\documentclass{article}
\usepackage{graphicx}
\usepackage{amsmath}
\usepackage{amssymb}
\usepackage{txfonts}
\usepackage{siunitx}
\usepackage{textgreek}
\usepackage{natbib}
\usepackage[letterpaper, left=.8in, right=.8in, top=1in, bottom=1in]{geometry}
\usepackage{hyperref}

\newcommand{\des}{DEST\-INY$^+$}
\newcommand{\amet}{\textalpha-meteoroid}
\newcommand{\bmet}{\textbeta-meteoroid}
\newcommand{\eg}{e.g.,}
\newcommand{\ie}{i.e.,}
\newcommand{\pion}{Pioneer~8/9}
\newcommand{\heos}{\mbox{HEOS-2}}
\newcommand{\qe}{$q$-$e$}

\DeclareSIUnit\solarradii{\text{\ensuremath{R_{\odot}}}}
\author{Maximilian~Sommer\thanks{University of Stuttgart, Germany. Email address: sommer@irs.uni-stuttgart.de}}

\usepackage{xcolor}
\definecolor{mylinkcolor}{RGB}{0, 51, 204}     
\definecolor{mycitecolor}{RGB}{0, 153, 0}      
\definecolor{myurlcolor}{RGB}{204, 0, 0}      
\definecolor{mytoccolor}{RGB}{0, 102, 204}    
\definecolor{myanchorcolor}{RGB}{0, 102, 204} 
\hypersetup{
    colorlinks=true,
    linkcolor=mylinkcolor,    
    citecolor=mylinkcolor,    
    urlcolor=myurlcolor,      
    filecolor=myurlcolor,     
    anchorcolor=myanchorcolor 
}

\begin{document}

\title{Alpha-Meteoroids then and now:\\ Unearthing an overlooked micrometeoroid population}

\date{July 2023}
\maketitle

\begin{abstract}
   The term `\amet{}' was introduced to describe a group of micrometeoroids 
   with certain dynamical properties, which---alongside the group of the \bmet{}s---had
   been identified by the first generation of reliable in-situ dust detectors in interplanetary space.
   In recent years, use of the term \amet{} has become more frequent again, 
   under a subtly but crucially altered definition.
   This work shall bring attention to the discrepancy between the term's original 
   and newly established meaning, and spotlight the now-overlooked group of particles that the
   term used to describe.
   We review past and present pertinent literature around the term \amet{},
   and assess the dynamics of the originally referred-to particles with
   respect to possible sources,
   showing that their formation is the expected consequence of 
   collisional grinding of the zodiacal cloud at short heliocentric distances.
   The abundance of the original \amet{}s, which are essentially `bound \bmet{}s', 
   makes them relevant to all in-situ dust experiments in the inner solar system.
   Due to the change of the term's meaning, however, they are not considered by contemporary studies.
   The characterization of this particle population could elucidate the processing of the innermost 
   zodiacal cloud, and should thus be objective of upcoming in-situ dust experiments.
   The attained ambiguity of the term \amet{} is not easily resolved,
   warranting great care and clarity going forward.
\end{abstract}

  \section{Introduction}
The first important in-situ measurements of cosmic dust were carried out 
by the dedicated dust sensors aboard the Pioneer~8 and~9 spacecraft.
Their most prominent discovery was a distinct group of dust particles that
appeared to be coming from the Sun on hyperbolic trajectories \citep{Berg1973evidence}.
In a seminal work, \citet{Zook1975source} coin the term \bmet{}s,
to refer to dust grains that `have their orbits significantly altered by radiation pressure',
including those affected strongly enough to assume escape trajectories.
The escaping \bmet{}s were subsequently confirmed by dust detectors onboard the Helios
\citep{Grun1980orbital} and later Ulysses spacecraft \citep{Baguhl1995dust,Wehry2004analysis}.
Since their discovery, the study of the \bmet{}s remains of high interest, 
being seen as a window of insight into the near-Sun physics that enable their creation. 
To what extent these miniscule particles are remnants of 
collisional grinding of the zodiacal cloud \citep[\eg][]{Zook1975source,Grun1985collisional}, 
of grains sublimating in the heat of the Sun \citep[\eg][]{Mukai1979model,Kobayashi2009dust},
or of rotational bursting of grains spun up by solar radiation \citep[\eg][]{Misconi1993spin,Herranen2020rotational} 
is still subject of scientific debate.

However, the data of these early dust detectors, onboard \pion, \heos, and Helios revealed another group of interplanetary dust particles,
which where characterized by approach directions centred around the heliocentric spacecraft apex,
as well as impact speeds that suggested the particles had large orbital eccentricities and
were encountered near their aphelion 
\citep{Zook1975source,Hoffmann1975first,Hoffmann1975temporal,Grun1980orbital}.
Initially referred to as `apex particles' by \citet[][]{Hoffmann1975first,Hoffmann1975temporal},
it became clear that they constituted a distinct dynamical class,
for which \citet{Grun1980dynamics} introduced the term `\amet{}s'---setting them apart
from the hyperbolic \bmet{}s, as well as from the sporadic micrometeoroids with higher angular momentum.
These \amet{}s, around \qty{1}{\um} in size, occupied a mass range in between the smaller \bmet{}s and the 
larger sporadic meteoroids ($\qty{e-13}{\gram} \le m \le \qty{e-11}{\gram}$).
\citet{Grun1980dynamics} argue that the dynamics of the \amet{}s speak against a 
Poynting-Robertson-drag-induced evolution from cometary or asteroidal orbits
and instead point to a common origin with the \bmet{}s 
(presumably, the collisional fragmentation of bigger grains closer to the Sun).
However, interest in the \amet{}s subsided with only few publications taking them up in the following years,
possibly owing to the growing popularity of in-situ cosmic dust
research on the outer solar system and the interstellar component, e.g., onboard
Galileo \citep{Grun1992galileo,Kruger1999detection},
Ulysses \citep{Grun1992ulysses,Kruger2007interstellar}, and 
Cassini \citep{Srama2004cassini,Altobelli2016flux}.

More recently, usage of the term \amet{}s has again become more frequent in pertinent literature.
Then again, this new usage of the term occurs generally under a different meaning than
originally defined by \citet{Grun1980dynamics}.
In this new context, all meteoroids that revolve on bound orbits are considered \amet{}s, 
explicitly including those that evolve from their cometary and asteroidal 
source orbits under Poynting-Robertson drag \citep[\eg][]{Krivov2000size,Szalay2021collisional}.

The goal of this work is to draw attention to this change of definition,
as well as to the group of particles originally named \amet{}s, which have been neglected as a result.
For in-situ dust research in the inner solar system, these grains may constitute 
the highest-flux interplanetary dust population besides the hyperbolic \bmet{}s.
In the prospect of large amounts of new data being gathered by highly sensitive dust detectors at and below 1~au, 
(\des \citep{Kruger2019modelling}, IMAP \citep{Sternovsky2022measuring}, MDM \citep{Kobayashi2020mercury}, and 
via antenna measurements with the already-in-service Solar Orbiter and Parker Solar Probe
\citep{Mann2019dust}), revisiting these grains is of high relevance.
To that end, we review the literature around the \amet{}s in Sect.~2 and
assess the dynamics and possible origins of the originally referred-to particles in Sect.~3.
Section~4 discusses the implications of the existence of the original \amet{}s and reconsiders
the naming convention.
  \section{Literature review: Alpha-Meteoroids}
The newly established meaning of the term \amet{}s (namely, all dust grains~/ meteoroids that are bound)
is straightforward and unambiguous. 
It may not be immediately clear how the original meaning deviates from that, given that a 
key feature of the originally referred-to particles is that they are in fact bound.
To highlight the subtlety of the original definition and to demonstrate its transformation, this section
takes a closer look at the literature and quotes the formative statements.
A list of all (traceable) publications using the term \amet{}s is given in Tab.~\ref{tab:amet_references}.

\subsection{Zook and Berg (1975)}
\citet{Zook1975source} give a comprehensive explanation for the dynamics of
micron and submicron-sized grains, detected by the \pion{} dust sensors.
Although the term \amet{} is not used here, \citet{Zook1975source} introduce the term \bmet{}
and discuss the group of particles \citet{Grun1980dynamics} will later call \amet{}s.
Nevertheless, this work is often credited for introducing the terms \bmet{} and \amet.

\citet{Zook1975source} show that the \pion{} data are consistent with the production
of hyperbolic particles in collisional breakups of meteoroids near the Sun, 
a mechanism previously proposed by \citet{Dohnanyi1971current}:
As micrometeoroids spiral inward under the Poynting-Robertson (PR) drag, their concentration
and relative velocities increase with decreasing solar distance, 
eventually causing them to suffer grain-grain collisions.
Submicron-sized fragments, having much larger area-to-mass ratios than the colliding parent grains,
suddenly receive a significant influence from solar radiation pressure,
which may sweep them away on unbound trajectories.
\citet{Zook1975source} express the effective strength of the radiation pressure for a given particle
as the quantity $\beta$ (which relates the radiation pressure force to the solar gravitational force),
leading them to introduce the term \bmet{}.
They write:
\begin{quote}
    `It will prove convenient to give a name to those meteoroids that are small enough to have their
    orbits significantly altered by radiation pressure. They will be called ``beta-meteoroids'' in this
    paper and will include those meteoroids that are in hyperbolic orbits.'
\citep[p.~186]{Zook1975source}
\end{quote}
Contrary to contemporary usage, the term \bmet{} was not introduced to describe just unbound dust particles,
but rather all grains whose orbits are significantly shaped by radiation pressure.
\citet{Zook1975source} explicitly discuss those \bmet{}s that upon creation 
would fail to become hyperbolic (e.g., due to their higher mass) but still assume highly eccentric orbits:
\begin{quote}
    `Beta-meteoroids that are created (by collisions or other means) with aphelia initially greater than
    1~au can be expected to complete a number of revolutions before their aphelia decay to within 1~au.
    Obviously, their probability of detection is enhanced over those particles put directly into hyperbolic trajectories.'
    \citep[p.~200]{Zook1975source}
\end{quote}
\citet{Zook1975source} further argue that these particles would be picked up by a sensor in a heliocentric orbit 
overtaking them around their low-momentum aphelion passage, and would thus be observed coming from 
near the spacecraft's apex direction.
Those that are created with aphelia near 1~au would be the most massive particles of this kind detectable by the Pioneers,
with an apparent approach most closely from the apex direction.
This is consistent with the fact that the average energy per impact recorded by \pion{}
reaches a maximum at the apex direction, as the authors show.

Adding to that, the \pion{} dust sensors also infrequently recorded so-called time-of-flight (TOF) events, 
where particles reached a second film sensor after penetrating the first, 
allowing for direct speed measurements.
In total, 20~TOF events were recorded over the course of both missions, that represented a particle mass \qty{\ge e-12}{\gram}
\citep{Berg1970orbital,Berg1971more,Wolf1976orbital}.
Of those, the prograde bound particles had impact directions centred around the spacecraft apex and impact velocities
between \qtylist{5;27}{\km\per\second} \citep[see also the review by][]{McDonnell1978microparticle}.\footnote{
    Orbit parameters of Pioneer~8: 
    $r_\mathrm{peri}\!=\!\qty{0.99}{\astronomicalunit}$, $r_\mathrm{apo}\!=\!\qty{1.09}{\astronomicalunit}$,
    mean heliocentric speed of \qty{29.3}{\km\per\second};
    and Pioneer~9:
    $r_\mathrm{peri}\!=\!\qty{0.75}{\astronomicalunit}$, $r_\mathrm{apo}\!=\!\qty{0.99}{\astronomicalunit}$,
    mean heliocentric speed of \qty{31.8}{\km\per\second}
    \citep[apsidal distances from][]{Dixon1975pioneer}.
}

\subsection{Hoffmann et al. (1975a,b)}
\citet{Hoffmann1975first} published results from the \heos{} dust experiment 
concurrently with the article by \citet{Zook1975source}.
Unlike the Pioneer sensors, the detector onboard the \heos{} satellite 
could not measure the hyperbolic \bmet{}s:
due to the spacecraft's power system and sensor accommodation, only viewing directions perpendicular
to the solar direction could be attained.\footnote{
    Note that \heos{} was in a highly elliptical Earth orbit with most of its time
    spent beyond \qty{100000}{km} from Earth. 
    For the purpose of this study, it can be considered as being in circular heliocentric orbit at \qty{1}{\astronomicalunit}.
}
However, like \citet{Zook1975source}, they characterized the group of particles later to be called \amet{}s.
Within the plane perpendicular to the Earth-Sun line, \citet{Hoffmann1975first} report
an anisotropy of dust flux in favour of the Earth's apex direction.
They introduce the term `apex particles' and point out their low impact velocity compared to particles 
approaching from other directions, as indicated by charge signal rise-time measurements
(see Sect.~\ref{Sect:accuracy} for remarks this method of speed determination).

In a subsequent analysis \citet{Hoffmann1975temporal} concretize their findings, stating an 
excess of apex flux of one order of magnitude over the anti-apex, ecliptic north, and south directions,
as well as an average impact velocity of the apex particles of \qty{10}{\km\per\s}
(around half of the impact speeds of particles approaching from the other directions).
This is in agreement with the conclusion of \citet{Zook1975source},
that these apex particles are encountered near their aphelion, where they are outpaced by the spacecraft.
\citet{Hoffmann1975temporal} also find that the apex particles are
predominantly larger grains of masses \qty{\ge e-12}{\gram}.

\subsection{Grün and Zook (1980)}
After \pion{} and \heos{}, the Helios spacecraft, orbiting the Sun on eccentric orbits between 0.3 and 1~au,
delivered more revealing in-situ dust data \citep{Grun1980orbital}.
Interpreting the consistent findings of \pion{}, \heos{}, and Helios, \citet{Grun1980dynamics} 
come to the conclusion that the apex particles constitute a distinct class of meteoroids 
for which they introduce the term \amet{}s:
\begin{quote}
    `Another class of meteoroids, intermediate in mass between the \bmet{}s described above and the larger
    ``sporadic'' meteoroids, had also been identified in the data from each of four separate experiments.
    These intermediate mass particles are observed to arrive from the heliocentric spacecraft apex
    direction in each case.
    As we shall show, these intermediate mass meteoroids appear to constitute a dynamical class of 
    meteoroids that is separate from both the larger sporadic meteoroids spiralling in toward the Sun
    under P-R drag and the smaller \bmet{}s on hyperbolic trajectories.
    We shall, for convenience, give this class of intermediate mass meteoroids a name: ``\amet{}s''.'
    \citep[p.~294]{Grun1980dynamics}
\end{quote}
\citet{Grun1980dynamics} interpret the \amet{}s as fragments
being generated in collisions alongside the smaller \bmet{}s 
(which, unlike \citet{Zook1975source}, they consider to be hyperbolic particles only),
writing:
\begin{quote}
    `Larger collisional fragments, with smaller $\beta$~values, will not be injected into 
    hyperbolic orbits after release from their parent bodies 
    but will have aphelia and eccentricities greatly increased over the parent objects. 
    This dynamical grouping is here called the \amet{} group.'
    \citep[p.~297]{Grun1980dynamics}
\end{quote}
In subsequent works, the Helios data are further analysed to characterize the \amet{}s, yielding
(broadly distributed) low semi-major axes and high eccentricities with averages of
\mbox{$\bar{a} \approx \qty{0.6}{\astronomicalunit}$} and $\bar{e} \approx 0.6$,
as well as a constraint on their average inclination of $\bar{i} < \qty{30}{\degree}$
\citep{Grun1981physikalische,Grun1985orbits}.

The classification scheme for in-situ detected dust particles as proposed by \citet{Grun1980dynamics}
considers three groups:
\begin{enumerate}
    \item \textbf{\bmet{}s}: Grains on hyperbolic trajectories; approach direction from the Sun; lowest detected masses. 
    \item \textbf{\amet{}s}: Grains on low-perihelion, highly eccentric orbits; encountered near their aphelion and thus 
    impacting from the apex direction; intermediate masses.
    \item \textbf{Sporadic meteoroids}: Grains on high-angular-momentum orbits 
    (\ie{} larger semi-major axis and lower eccentricity);
    no pronounced directionality; highest detected masses. 
\end{enumerate}
\citet{Grun1980dynamics} consider the third group to be the lower-mass-end of the sporadic 
meteoroid complex, whose particles may still be dynamically linked to their source body families 
and which may be observed as radar-meteors.

\citet{Grun1985collisional} iterate this classification, stating mass ranges of 
$m\!<\!\qty{e-13}{\gram}$ for \bmet{}s,
$\qty{e-13}{\gram}\!\lesssim\!m\!\lesssim\!\qty{e-11}{\gram}$ for \amet{}s, and
$m\!>\!\qty{e-11}{\gram}$ for the sporadic meteoroids.

\begin{figure}[ht]
    \centering
    \includegraphics[width=0.8\columnwidth,trim={0.5mm 0 0 0},clip]{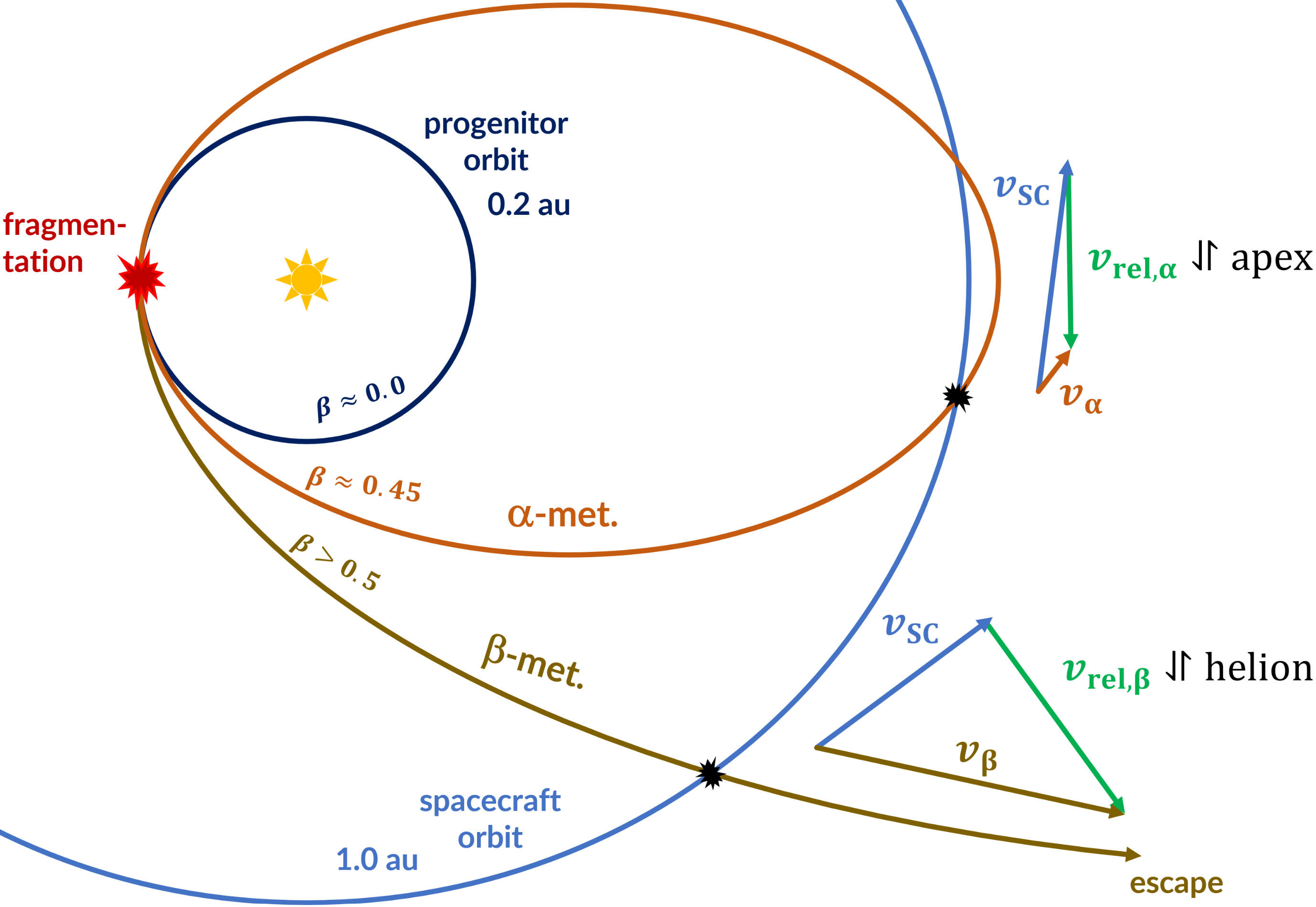}
    \caption{Orbits and detection scheme of \textalpha- \& \bmet{}s in the sense of \citet{Grun1980dynamics}.}
    \label{fig:orbit_scheme}
\end{figure}

\subsection{Transition to the new meaning}
Following its introduction, however, the term \amet{} does not seem to be adopted in the pertinent literature.
Only few publications take up the \amet{} term in the sense of \citet{Grun1980dynamics}, 
such as \citet{Shestakova1995dynamics} and \citet{Wehry1999identification}.
Some works adopt the term `apex particles' to refer to this class of in-situ-measured grains, having identified 
them in datasets of the Munich Dust Counter, as well as the GORID instrument
\citep[\eg][]{Igenbergs1991present,Iglseder1993cosmic,Iglseder1996cosmic,Svedhem2000new,McDonnell2001earth}.
In fact, Gr{\"u}n himself refrains from using the term \amet{}s in later works, 
resorting to the term `apex particles' instead \citep[\eg][]{Grun1985orbits,Grun1992ulysses}. 

\citet{Artymowicz1997dust} and \citet{Artymowicz1997beta} then take up the \textalpha-/\textbeta-meteoroid terminology 
to apply it to circumstellar discs in general, thereby extending the meaning of \amet{}s
to collisionally generated grains of all sizes that are released into bound orbits---instead 
of only those significantly affected by radiation pressure.
This usage is then adopted in other studies concerning circumstellar debris discs \citep[\eg][]{
Artymowicz2000beta,Krivova2000disk,Krivova2000size,Krivov2000size,Krivov2006dust,Mann2006dust,
Freistetter2007planets,Krivov2010debris,Kral2017exozodiacal},
where modelled discs typically consist of two-components, one of bound grains called \amet{}s, 
and one of unbound grains called \bmet{}s: 
\begin{quote}
    `We consider the dust disc as consisting of two dust populations which, borrowing the terminology
    from the Solar system studies, may be called \textalpha-~\& \bmet{}s \citep{Zook1975source}. 
    The former are larger grains that move round the star in bound orbits, 
    whereas the latter are smaller particles blown away from the star by the stellar radiation pressure.' 
    \citep[p.~1129]{Krivov2000size}
\end{quote}
In some contexts, the bound \amet{}s component may also be understood as to include fresh meteoroids, 
stemming directly from their cometary source bodies \citep{Krivova2000size}.
More recently, this usage is transpiring back into the context of in-situ dust detection 
\citep{Mann2021dust,Pusack2021dust,Szalay2021collisional}.
In the solar system's zodiacal cloud, the broad size regime covered by the adopted meaning 
arguably includes grains that evolve and circularize under PR drag, 
after having been released from their cometary or asteroidal source bodies 
(or from larger, collisionally evolving meteoroids 
whose orbital properties are still similar to those of their source bodies).
Therefore, under the new definition, the \amet{}s are generally considered to have circular orbits,
such as by \citet{Szalay2021collisional}, who use a two-component (\textalpha/\textbeta{})
zodiacal cloud model of bound (circularized) and unbound particles to simulate the influx onto the Parker Solar Probe.

The new usage is in contrast to the grains originally referred to as \amet{}s (or apex particles), 
which have a narrow size range and exhibit large eccentricities with low perihelia.
While it can be assumed that a fraction of the \amet{}s measured by the early dust detectors at \qty{1}{\astronomicalunit},
migrate further under PR drag and become circularized before undergoing further collisional grinding (or sublimation) to become hyperbolic \bmet{}s,
the deviation of the term \amet{} in contemporary literature from its original meaning is evident.

  \section{Alpha-meteoroids origin} \label{Sect:Dynamics}

The limited coverage of the original \amet{}s warrants a new and closer look 
at the dynamics and possible origins of the dust particles at hand, which we will take in this section.
In the following, we use the term \amet{} only under its original definition \citep{Grun1980dynamics},
referring to the low-angular-momentum and intermediate-mass dust grains,
identified by the early in-situ dust detectors in the form of apex particles.
The determined particle mass range of $\qty{e-13}{\gram}\!<\!m\!<\!\qty{e-11}{\gram}$ 
suggests that the motion of the \amet{}s is characterized by solar gravity,
the solar-radiation-induced force, and solar wind drag.
The radiation-induced force can be conceptually decomposed into a radial component (radiation pressure) 
and a drag component (Poynting-Robertson (PR) drag).
Based on the mass range, we can constrain the range of the \amet{}'s $\beta$-factor, 
which is the ratio of the radiation pressure force to solar gravity.
According to typical beta-curves obtained through Mie theory 
\citep[\eg][Fig.~6]{Gustafson2001interactions}, we find approximately $0.2\!<\!\beta\!<\!1$ for the \amet{}s.
The effect of solar wind drag is analogous to PR drag albeit at lower intensity \citep{Burns1979radiation}.
    
PR drag drains a particle's orbital energy, which causes it to spiral towards the Sun over long timescales.
It does so most effectively around perihelion, thus also causing a circularization of eccentric orbits.
Understanding how this connects to the low perihelia and high eccentricities present in the \amet{}s---which
are their defining qualities---is essential when investigating their nature and will be the focus here.
(The Lorentz force resulting from the interaction with the interplanetary magnetic field
may also be a relevant factor, especially for the smaller \amet{}s, and is discussed
separately in Sect.~\ref{Sect:EM-effects}).

Already with the first direct speed measurements of particles impacting the Pioneer sensors,
\citet{Berg1970orbital} note an incompatibility with a  direct asteroidal or cometary origin, and 
\citet{Zook1975source} as well as \citet{Grun1980dynamics} conclude that the derived dynamics
favour an origin close to the Sun, akin to the generation of hyperbolic \bmet{}s.
The underlying argument is this: Since the PR drag circularizes orbits before
substantially lowering perihelia, the low-perihelion and high-eccentricity orbits 
of the \amet{}s, could not have evolved from initially asteroidal or cometary orbits. 
To illustrate this proposition, we can analyse the evolution of perihelion and eccentricity 
of particles released from different sources.

\begin{figure}[!t]
    \centering
    \includegraphics[width=0.8\columnwidth,trim={0 0 0 0},clip]{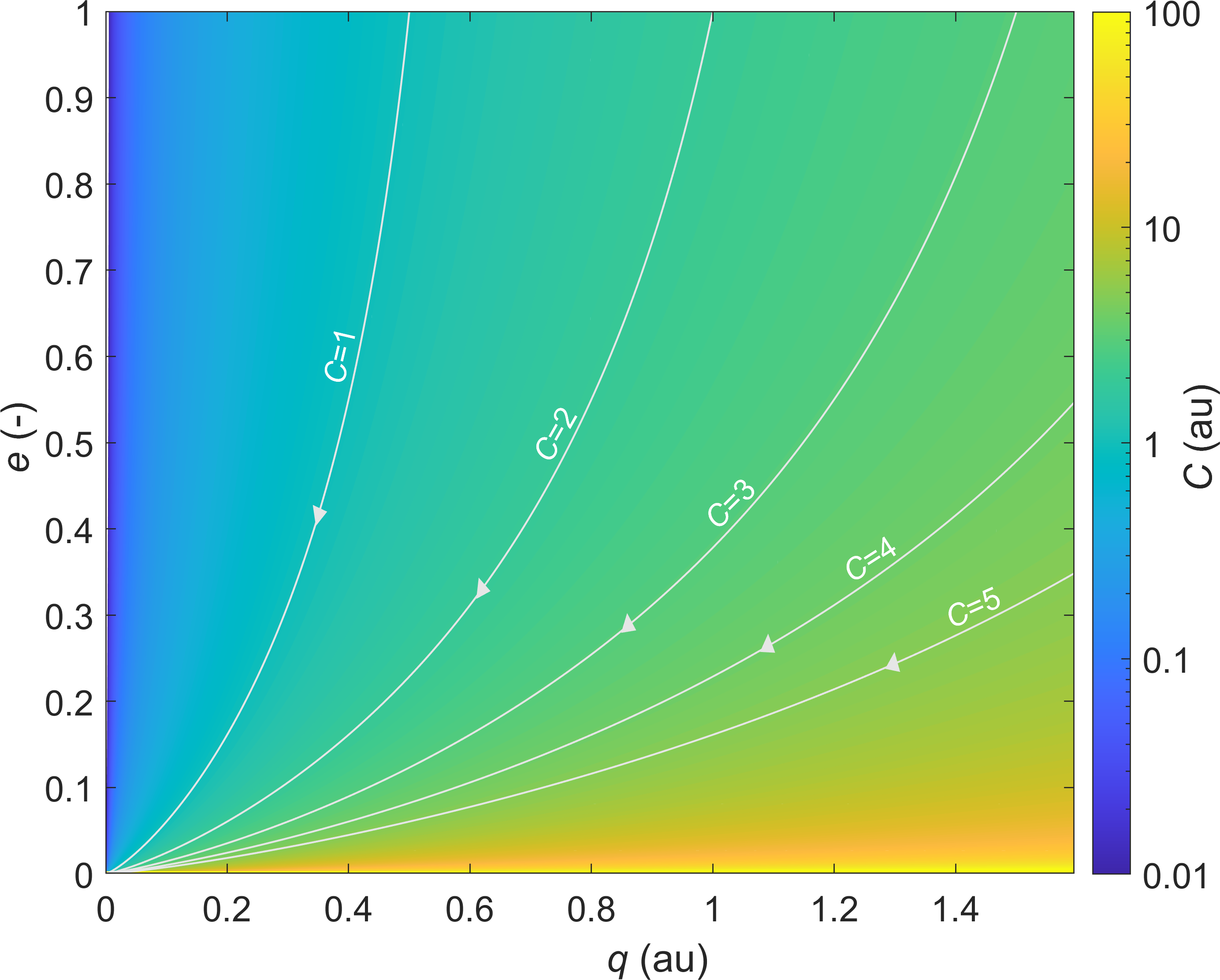}
    \caption{Value of the $C$-quantity in \qe{} space (Eq.~(\ref{eq:Cqe})). 
    Particles evolving solely under PR drag exhibit a continuous decrease in perihelion ($q$)
    and eccentricity ($e$) while preserving a constant $C$-value. 
    White contours illustrate their evolutionary tracks at representative integer values of $C$.}
    \label{fig:C_contourf}
\end{figure}

For a particle evolving under PR drag the semi-major axis and eccentricity decrease with time, 
which can be described by analytic solutions found by \citet{Wyatt1950poyntingrobertson}.
By incorporating the $\beta$-factor, these formulations can be expressed as:
\begin{align}
    & \frac{d a}{d t} = -\frac{ \beta \,\mu}{c} \cdot \frac{\left(2+3 e^2\right)}{a\left(1-e^2\right)^{3 / 2}} 
    \label{eq:PRdrag_sma} \\[1ex]
    & \frac{d e}{d t} = -\frac{5 \,\beta \,\mu}{2 \, c} \cdot \frac{e}{a^2\left(1-e^2\right)^{1 / 2}} 
    \label{eq:PRdrag_ecc} 
\end{align}
where $a$ is the semi-major axis of the particle, $e$ its eccentricity, 
$\mu$ the standard gravitational parameter of the Sun, and $c$ the speed of light.
By integrating the division of Eqs.~(\ref{eq:PRdrag_sma}) and~(\ref{eq:PRdrag_ecc}) 
(that is, $da/de$), \citet{Wyatt1950poyntingrobertson} derive further the quantity $C$,
which, for a given particle evolving under PR drag, remains constant at all times:
\begin{align}
    C = a \; e^{-4/5} \; \left(1 - e^2\right) = \rm{constant,}
\end{align}
where $a$ and $e$ are the semi-major axis and eccentricity of the particle at any arbitrary time.
Note that, although $C$ is in the dimension of $a$, it has no straightforward geometric representation.
By expressing $a\!=\!q/(1\!-\!e)$, where $q$ denotes the perihelion distance, 
we can rewrite Eq.~(\ref{eq:Cqe}) as:
\begin{align}
    C = q \; e^{-4/5} \; \left(1 + e\right) = \text{constant.}
    \label{eq:Cqe}
\end{align}
Figure~\ref{fig:C_contourf} illustrates the behaviour described by Eq.~(\ref{eq:Cqe}), 
showing contours of $C$ in \qe{} space along which particle orbits decay.
The curvature of the contours indicates that PR drag primarily reduces eccentricity
(\ie{} it tends to circularize orbits) before significantly lowering perihelion.
Note also that, for a given orbit, the RP-evolutionary track in \qe{}~space is 
independent of the particle's $\beta$-factor, 
which only determines the speed at which the particle progresses along the track.
\begin{figure}[h!t]
    \centering
    \includegraphics[width=0.8\columnwidth,trim={0 0 0 0},clip]{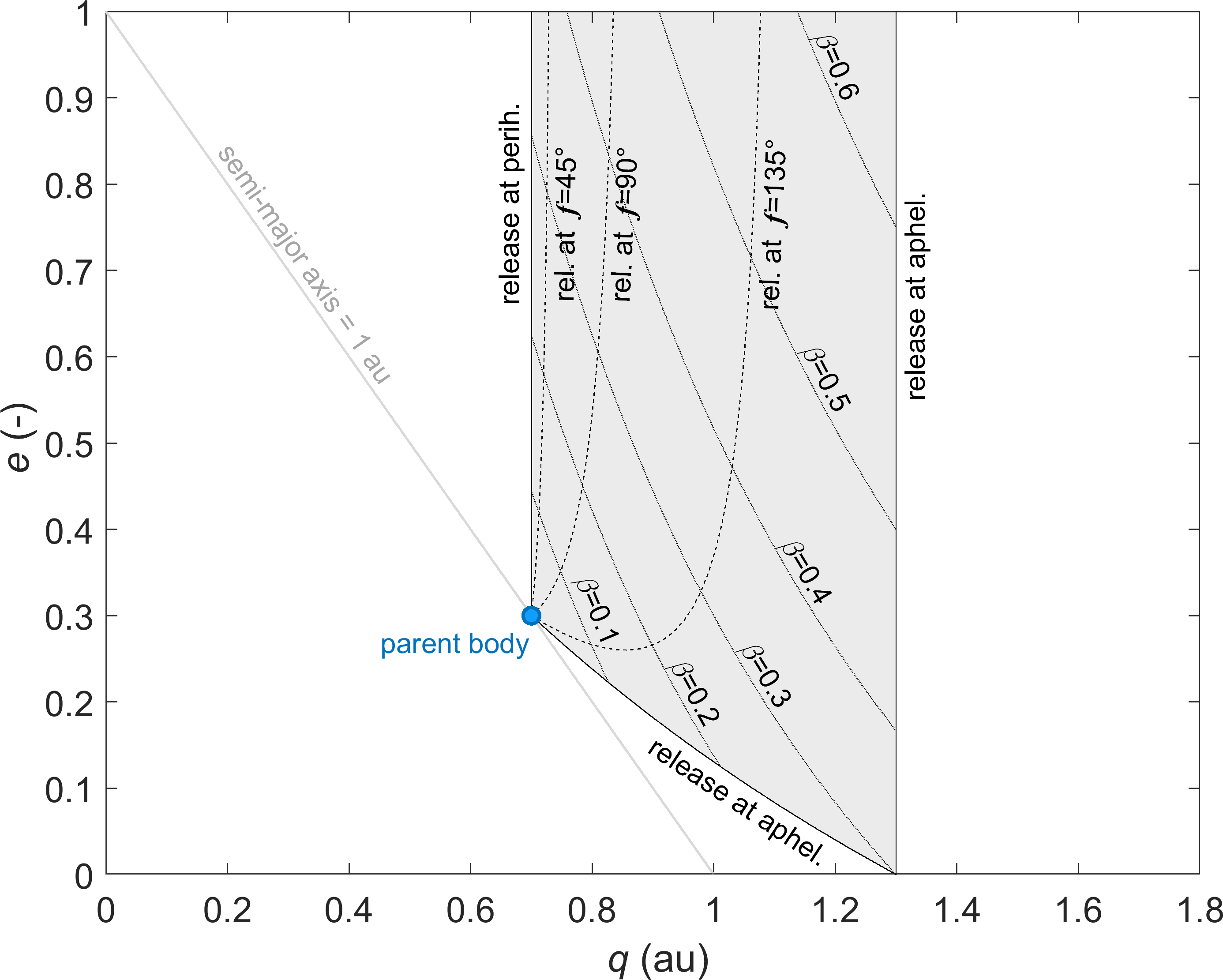}
    \caption{Release zone in \qe{}~space for dust released from a hypothetical source orbit
    with $q\!=\!0.7$ and $e\!=\!0.3$.
    The release zone, depicted by the grey area, represents the range of \qe{}~combinations
    that particles can exhibit upon release from the parent body
    depending on their $\beta$-factor and the release location along the source orbit.
    Contour lines within the release zone mark constant values of particle $\beta$ 
    and parent body true anomaly $f$ at release.
    Solid lines enclose the release zone and represent release at
    peri-~\& aphelion of the source body.
    We neglect the ejection velocity from the source body,
    the consideration of which would `blur' the release zone contours depicted here.
    Note that the release zone extends to unbound orbits ($e\!>\!1$),
    which are not relevant to the current study.}
    \label{fig:release_zone}
\end{figure}

Another crucial aspect to consider is that when a particle is released from a parent body,
it transitions into a new orbit upon exposure to solar radiation pressure.
This new orbit is characterized by an increased semi-major axis
and a potentially higher or lower eccentricity, 
depending on the release location within the source orbit and the magnitude of $\beta$.
The perihelion distance $q$ and eccentricity $e$ of the released particle may be determined from
the parent body perihelion distance $q_0$, its eccentricity $e_0$, its true anomaly $f$ at release,
as well as the $\beta$-factor of the particle as follows:
\begin{align}
    e^2=\frac{e_0^2+\beta^2+2 e_0 \, \beta \cos f}{(1-\beta)^2} \text{ ,}
\end{align}
and then
\begin{align}
    q=\frac{q_0 \, (1-\beta) \, (1-e)}{1 - e_0 - 2 \beta \, \dfrac{1 + e_0 \cos f}{1+e_0}}
\end{align}
\citep[equations modified from][]{Rigley2021comet}.
The range of possible \qe{} combinations that a particle can attain upon release is limited
and depends on the $q$ and $e$ of the source orbit.
We refer to this range as the `release zone'.
Figure~\ref{fig:release_zone} illustrates the release zone 
for a hypothetical parent body with $q\!=\!0.7$ and $e\!=\!0.3$.
Once released within the release zone, a particle's PR-evolution in \qe{} space 
now follows a constant value of $C$.

\begin{figure*}[!htp]
    \hspace{-4mm}\begin{tabular}{cc}
        \includegraphics[height=74mm,trim={12mm 2mm 0mm 6mm},clip]{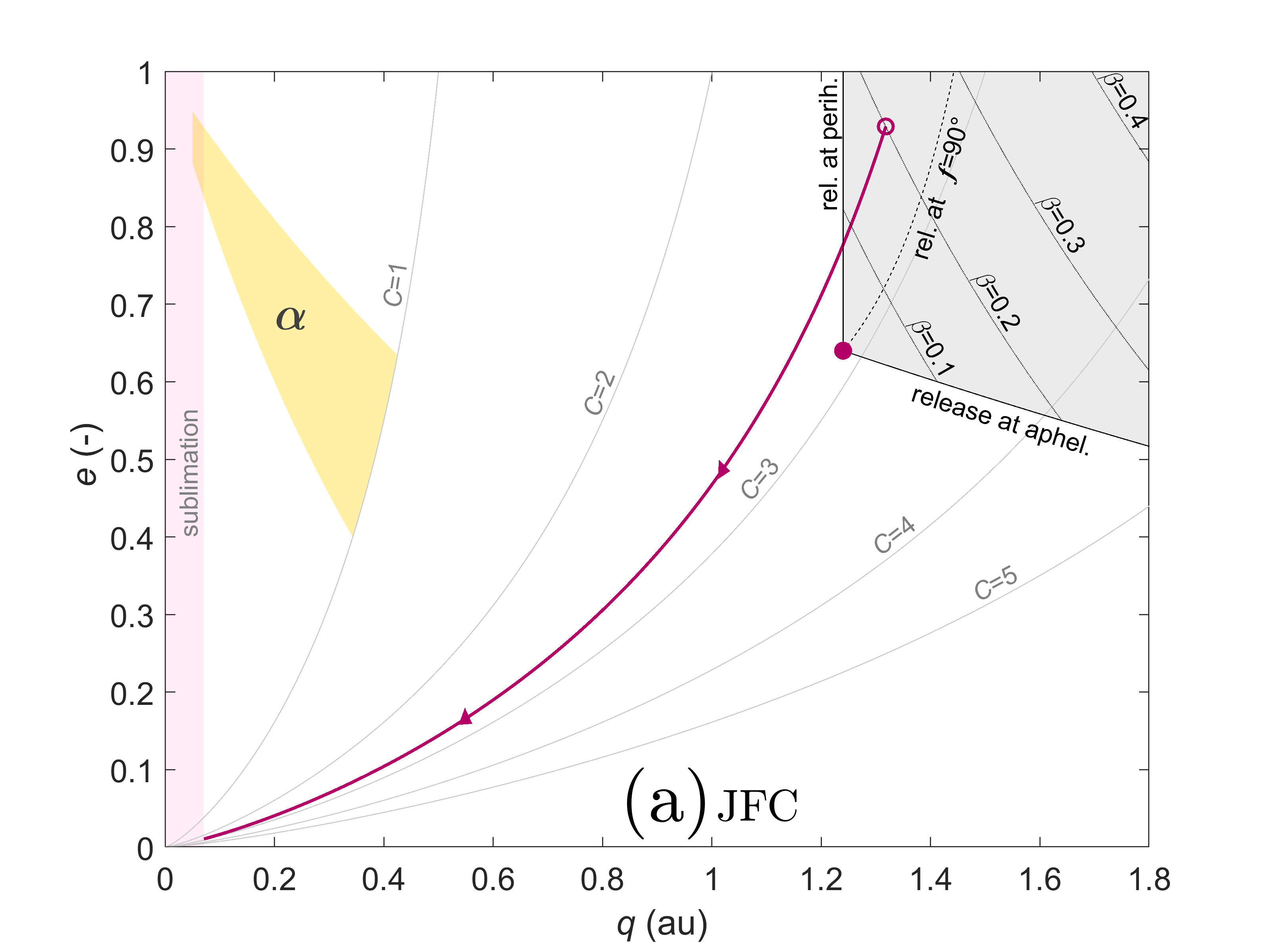} & \hspace{-13mm}
        \includegraphics[height=74mm,trim={17.5mm 2mm 16.5mm 6mm},clip]{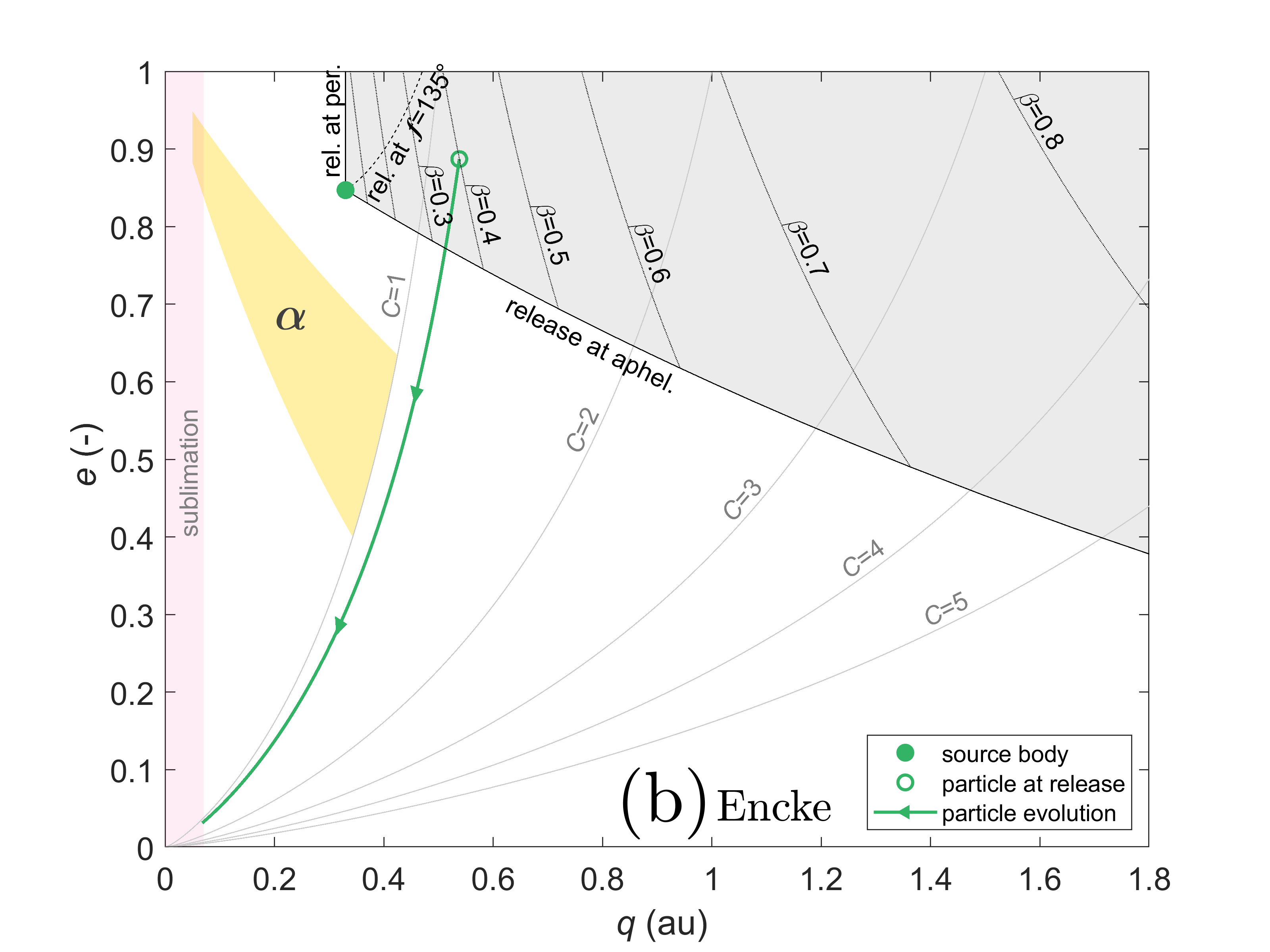} \\[-3ex]
        \includegraphics[height=74mm,trim={12mm 1mm 0mm 7mm},clip]{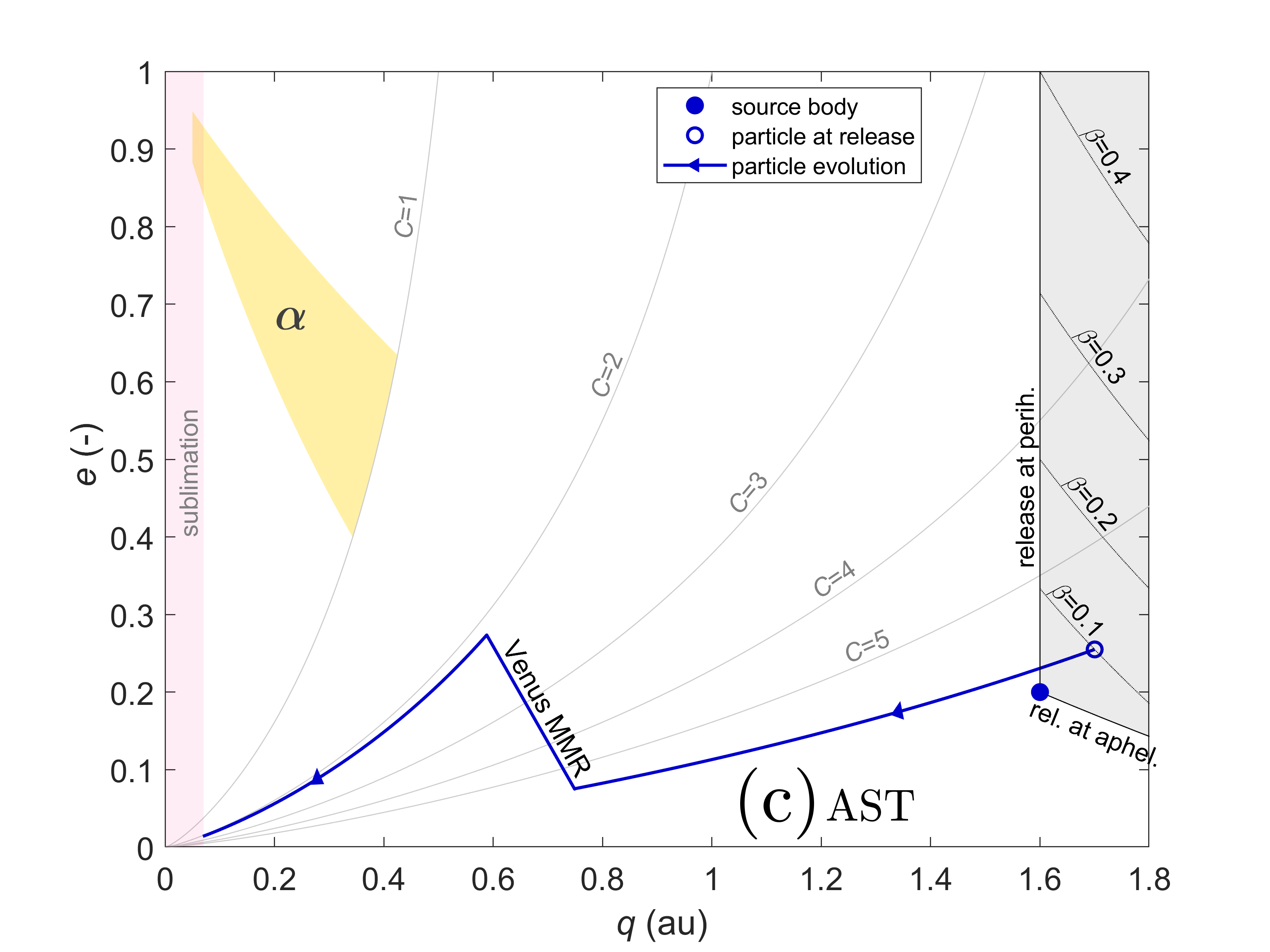} & \hspace{-13mm}
        \includegraphics[height=74mm,trim={17.5mm 1mm 16.5mm 7mm},clip]{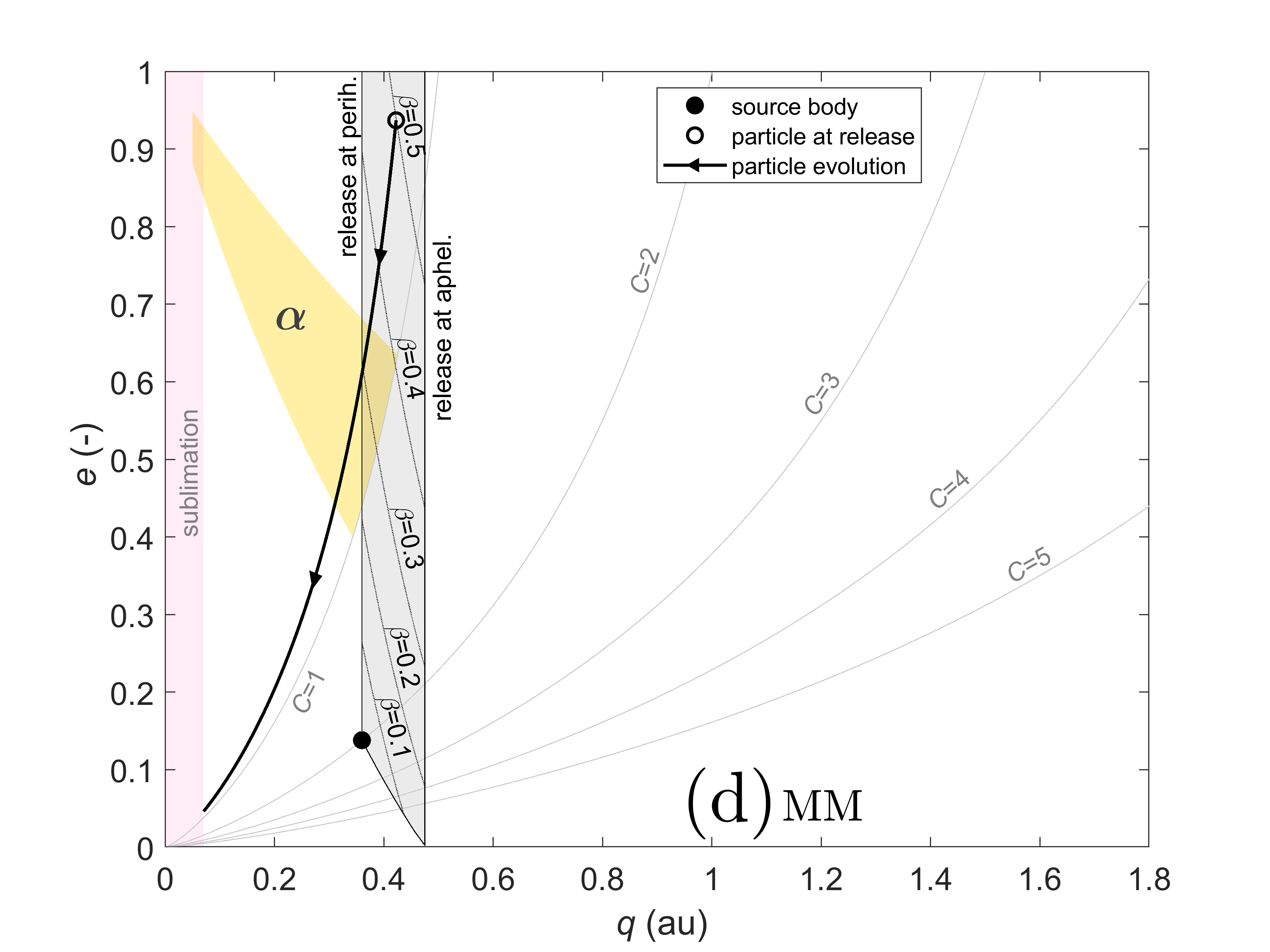}
    \end{tabular}
    \caption{Evolution of eccentricity and perihelion of exemplary particles released from different types of sources:
    (a) a JFC, in this example comet 67P,
    (b) Comet Encke, (c) an asteroid, and (d) a PR-evolved micrometeoroid (MM).
    Evolutionary tracks are idealized, \ie{} considering only PR evolution.
    Only case (b) considers planetary interaction.
    Particles may be released from the source body directly or generated from collisional fragmentation
    of larger released meteoroids that are still in the source body orbit.
    The grey area indicates the release zone, as explained in Fig.~\ref{fig:release_zone}.
    Evolutionary tracks of the exemplary particles starting from the release zone
    are displayed.
    Evolutionary tracks of particles starting elsewhere in the release zones can be inferred
    from the displayed contours of $C$ (units in~au).
    The yellow zone labelled `{\it{\textalpha}}' is set to span the \qe{}~combinations of the 
    low-angular-momentum apex particles registered in \pion{} TOF events
    (see Sect.~\ref{Sect:accuracy} for discussion).
    At $q\!<\!\qty{15}{\solarradii}$ grains may begin to sublimate indicated by the pink zone.}
    \label{fig:EQ_evolution}
\end{figure*}

The \amet{}s, with their high eccentricities and low perihelia,
are characterized by inherently low $C$-values.
Thus, for assessing whether dust particles released from a certain source 
can become \amet{}s, the minimum $C$-value that particles can attain 
without becoming hyperbolic ($e\!>\!1$) is critical.
In that way, we examine four different sources with respect to their ability to generate the \amet{}s:
Jupiter-family comets (JFC), Encke-type comets, asteroids, 
as well as fragmenting, PR-evolved micrometeoroids near the Sun.
Fig.~\ref{fig:EQ_evolution} shows the aforementioned release zones for dust released from 
representatives of these four sources in \qe{}~space, 
as well as the evolutionary tracks of exemplary released particles.
It also shows the \qe{}~region spanning typical \amet{}s as determined
by \pion{} TOF measurements, which notably exhibit $C\!<\!\qty{1}{\astronomicalunit}$.
In simple terms, in order for a source to be viable for \amet{} production,
the evolutionary tracks of released particles must be able to 
intersect the \amet{}~region (yellow area).
We further discuss the Lorentz force as well as sublimation near the Sun
as potential \amet{} creation mechanisms.

\subsection{Jupiter-family comets}
Fig.~\ref{fig:EQ_evolution}a illustrates the evolution of dust released from JFCs, which are
considered to be the primary dust source in the inner solar system \citep[\eg][]{Nesvorny2010cometary,Soja2019imem2}.
Here, we use comet 67P as a representative object of the JFC population 
(with $q\!=\!\qty{1.24}{\astronomicalunit}$ and $e\!=\!0.64$).
Grains may be released from the comet directly or from larger fragments released by the comet,
which evolve collisionally, rather than under PR drag, and still share the orbital characteristics of the comet.
Figure~\ref{fig:EQ_evolution}a illustrates the PR-evolutionary track of a single exemplary particle, 
but tracks originating from other points can be inferred from the displayed contours of $C$.
For release from 67P's orbit, the minimum $C$-value is approximately \qty{2.5}{\astronomicalunit}, 
occurring in particles released near perihelion with $\beta\!\approx\!0.2$.
This is far from the $C$-value required for particles to PR-evolve through the \amet{} \qe{}~region 
($C\!<\!\qty{1}{\astronomicalunit}$), making comets like 67P incapable of producing the \amet{}s.

\subsection{Comet Encke}
One might suspect comets that themselves exhibit a semi-major axis and eccentricity combination 
that corresponds to a lower $C$-value of being able to produce bound, `low-$C$' dust.
Figure~\ref{fig:EQ_evolution}b shows the dust release zone from Comet Encke's orbit 
($C_{Encke}\!\approx\!\qty{0.7}{\astronomicalunit}$), one of the lowest-perihelion periodic comets 
known to shed substantial amounts of material \citep[\eg][]{Sarugaku2015infrared}.
We see that the generation of dust from this type of orbit can yield low-$C$ particles
if their $\beta$-factor is sufficiently low ($\beta\!\lesssim\!0.3$), corresponding to 
the higher end of the \amet{} mass range (\qty{e-11}{\g} reported by \citet{Grun1985collisional}).
At $\beta=0.4$, particles can remain bound if released far from Encke's perihelion, 
with achievable values of $C\!\approx\!\qty{1}{\astronomicalunit}$. 
If release far from perihelion is acceptable, this creation pathway is borderline compatible 
with at least a fraction of the observed \amet{}s.

\subsection{Asteroidal dust in resonances}
Another conceivable scenario involves the phenomenon of eccentricity pumping by mean-motion resonances (MMRs).
Grains migrating towards the Sun, particularly from moderately eccentric asteroidal source orbits,
can become trapped in MMRs with the inner planets, causing them to maintain their semi-major axis
and thus halting their sunward progression.
This migration bottleneck is thought give rise to a circumsolar toroidal density enhancement, 
known as a `resonant ring'.
The presence of a discernable resonant ring associated with the Earth is supported
by the correlation of the predicted near-Earth ring structure 
(in particular, an Earth-centred gap and an Earth-trailing density enhancement) 
with zodiacal light brightness measurements \citep{Dermott1994circumsolar,Reach2010structure}.

When particles affected by PR drag are trapped in an external MMR, they maintain a quasi-constant
semi-major axis while their eccentricity steadily grows,
causing an effective lowering of their $C$-value.
The question is whether \amet{}-typical $C$-values can be reached via this mechanism.
For each resonance there is a $\beta$-independent maximum eccentricity that particles may attain
\citep[\eg][]{Weidenschilling1993orbital,Beauge1994capture},
which therefore also limits their minimum attainable $C$-value.
The evolution of an exemplary particle that has its eccentricity pumped
by an MMR is shown in Fig.~\ref{fig:EQ_evolution}c.
In this favourably idealized case, the particle (with $\beta\!=\!0.1$) gets trapped in the 4:5 external Venus resonance
and has its eccentricity increased to the maximum theoretically attainable value ($e_{\text{max}}\!=\!0.2736$),
resulting in $C_{min} \!\approx \!\qty{2.1}{\astronomicalunit}$.
(Note that MMRs of Venus allow for smaller $C$-values compared to those of Earth,
as they have the same limiting eccentricities at respectively smaller semi-major axes.
The same resonance of Earth yields $C_{min} \!\approx\! \qty{2.9}{\astronomicalunit}$.)
This is far from the demanded $C\!<\!\qty{1}{\astronomicalunit}$, 
notwithstanding the fact that trapping efficiency diminishes at such large $\beta$ 
\citep{Dermott2001orbital,Sommer2020effects}.
Theoretically, more remote resonances (\ie{} those of higher order or lower degree) 
allow for lower $C$-values to be attained 
(\eg{} $C_{min} \!\approx\! \qty{1.7}{\astronomicalunit}$ in the 3:5 Venus resonance).
However, those resonances have been shown to be effective in capturing particles in the absence of 
other planets and at larger particle sizes ($\beta \lesssim 0.01$) only, and are
entirely ineffective in the presence of a perturbing outer planet neighbour, such as Earth \citep{Sommer2020effects}.
Consequently, external MMRs can be ruled out as a mechanism to produce the \amet{}s.

On the other hand, the $\nu_6$ secular resonance and also various interior MMRs with Jupiter
have been shown by \citet{Morbidelli1998orbital} to be theoretically capable of pumping eccentricities 
of asteroidal meteoroids to nearly unity, lowering their perihelia to eventually reach the Sun.
For the last stage of their evolution, these would exhibit orbital elements
fairly similar to the \amet{}s discussed here.
It is doubtful, however, that these mechanisms can generate the observed dust grains.
That is because grains whose orbital energy is dissipated by PR drag---as is the case 
for the \amet{}s---can only be trapped in interior MMRs for limited amounts of time and 
experience a decrease of eccentricity rather than having it pumped \citep[\eg][]{Pastor2009motion,Klacka2020dust}.
It is also contrary to the expectation that smaller particles are less efficiently trapped
\citep{Gomes1995resonance,Dermott2001orbital}.
Similarly, the effect of secular resonances (such as the $\nu_6$) vanishes for grains less than several 10s of µm in size, due to the limited time
the PR-evolving grains spend near the resonance \citep{Kehoe2007effect,Espy2009dynamics}.
While larger meteoroids may be transferred by these effects onto orbits resembling those of the \amet{}s,
they may not be the progenitors of the small-size \amet{}s,
since the onset of radiation pressure onto smaller released particles would again alter their orbits substantially.

\subsection{Fragmentation of evolved micrometeoroids}
Fig.~\ref{fig:EQ_evolution}d shows the scenario put forward by \citet{Grun1980dynamics},
in which collisions among evolved micrometeoroids give rise to the \amet{}s.
In this example the progenitor is a meteoroid at $C\!\approx\!\qty{2}{\astronomicalunit}$, compatible with a direct PR decay
from a typical JFC orbit.
Assuming that velocity components added by the collision are negligible, the release zone 
for the fragments of this meteoroid exhibits a narrow range for possible perihelia
and wide range for possible eccentricities.
Large eccentricities and thus values of $C\!\lesssim\!\qty{1}{\astronomicalunit}$ may be assumed
by fragments with $0.3\!\lesssim\beta\lesssim\!0.5$.
In this case, if fragmentation occurs near aphelion, even bound low-$C$ grains with $\beta\!>\!0.5$ 
may be created, due to the eccentricity of the progenitor.
If not released directly into the \amet{} zone, fragments released at higher eccentricities 
may cross it upon further PR migration, as shown by the indicated evolutionary track.
Of the presented scenarios, release from a short heliocentric distance orbit thus poses
the most reliable pathway for generating grains of the \amet{} kind.
Of course, the releasing body does not have to be a fragmenting meteoroid necessarily 
and could theoretically also be a dust shedding asteroid with comparable orbital parameters,
that is, low heliocentric distance and moderate eccentricity.
However, to date there is only a single known member of the dynamical class of minor bodies that
revolve entirely within the orbit of Venus (Vatira asteroids) and predictions for its population size
are minuscule \citep{Greenstreet2012orbital,Sheppard2022deep}.

\subsection{Electromagnetic interaction} \label{Sect:EM-effects}
Thus far, we have neglected the influence of the Lorentz force, which results from the interaction of the
charge-accu\-mu\-lat\-ing dust grains with the interplanetary magnetic field (IMF), carried by the solar wind.
\citet{Morfill1986interaction} estimate that at a heliocentric distance of \qty{1}{\astronomicalunit} 
the Lorentz force becomes the dominant force for particles with masses $<\!\qty{e-14}{\gram}$.
The Lorentz force, as described by \citet{Parker1964perturbation},
acts largely perpendicular to the solar equatorial plane,
due to the dominant in-plane component of the IMF,
such that its principal effect on dust particles is a perturbation of their orbital 
inclination---a property where the \amet{}s show no strong characteristic.
Since the actual magnetic field fluctuates such that there is also a minor field component
normal to the solar equatorial plane present at all times,
there is also Lorentz force component within the orbital plane of affected particles, 
causing an analogous (yet less effective) perturbation of their
semi-major axis and eccentricity \citep{Parker1964perturbation}.

At low latitudes, where the rotating IMF exhibits a structure of variable-length sectors of alternating
field polarity, the Lorentz force acts effectively at random,
thus, inducing a stochastic dispersion of the orbital elements of particles in the micron size range,
also referred to as `Lorentz scattering'
\citep[\eg][]{Consolmagno1979lorentz,Consolmagno1980influence,Morfill1979motion,Barge1982diffusion,Mukai1984modification,Wallis1985stochastic,Fahr1985dustplasmagas,Morfill1986interaction}.
The derived magnitudes of this orbital dispersion, however, are inconsistent:
\citet{Consolmagno1979lorentz} finds an inclination dispersion about one order of magnitude
higher than \citet{Morfill1979motion} and two orders of magnitude higher than
\citet{Barge1982diffusion}, as pointed out by \citet{Leinert1985dynamics}.
As for effects within the orbital plane, 
Consolmagno concludes that Lorentz scattering in semi-major axis dominates over the drift resulting from PR drag at
particle radii $<\!\qty{2}{\um}$ at \qty{1}{\astronomicalunit} \citep[see][Fig.~5]{Consolmagno1979lorentz},
whereas \citet{Barge1982diffusion} find no meaningful dispersion in semi-major axis or eccentricities
for particles in the \amet{} size range.
Based on the former estimate, it could be considered that a stochastic spreading of eccentricities 
will drive some particles to higher eccentricities and thus potentially \amet{}-like orbits.
However, assessing the efficiency of this pathway is beyond the scope of this paper,
and can thus here be neither ruled out nor confirmed.

At high latitudes, on the other hand, the variable sector structure 
gives way to a unipolar magnetic field, which reverses polarity only once per 11-year solar cycle
and reaches down to solar equatorial latitudes of \qtyrange{15}{20}{\degree} 
during solar minimum \citep[\eg][]{Smith1993disappearance}.
Particles with considerable orbital inclination repeatedly enter this high-latitude field,
where, due to the less frequent polarity fluctuations and thus more consistent Lorentz 
perturbation, the spreading in orbital elements could be more effective 
\citep[see][Tab.~2, who find a dispersion in inclinations about two orders of magnitudes 
greater in particles with $i\!=\!\qty{30}{\degree}$ compared to $i\!=\!\qty{0}{\degree}$]
{Morfill1986interaction}.
Furthermore, at sub-micron particle sizes, the effect the steady Lorentz perturbation 
has within just one solar cycle becomes significant,
such that, rather than a random diffusion, it takes the character of a systematic drift.
Depending on the IMF polarity (also referred to as being in the `focussing' or `defocussing' configuration),
this secular change can direct particles either toward or away from the solar equatorial plane,
driving them preferentially to the low-~or high-latitude regime,
as described by \citet{Morfill1979motion} and \citet{Morfill1986interaction}.
(See also \citet{Hamilton1996electromagnetic}, who investigate the solar-cycle-dependent,
drift-induced escape of particles with radii $<\!\qty{100}{\nm}$, and \citet{Fahr1995evolution}, 
who conclude a negligible high-latitude drift in particles of several microns in size).
However, given that the \amet{}s have been found to exhibit an average inclination $<\!\qty{30}{\degree}$,
it seems unlikely that the systematic Lorentz perturbations at high latitude play a formative role
in their creation.

Furthermore, by including a hypothetical, non-zero average normal component of the IMF,
\citet{Lhotka2016charged} find that the Lorentz force could in principle 
induce a significant secular change of the semi-major axis,
which (depending on the value and sign of a particle's charge-to-mass ratio)
can accelerate, compensate, or even overcome orbital decay due to PR drag.
While they estimate that particles with radii as large as \qty{55}{\um} could withstand PR decay 
at \qty{1}{\astronomicalunit} via this systematic Lorentz perturbation,
a secular change of their eccentricity is not evident \citep[see][Fig.~2]{Lhotka2016charged}.
However, without a deeper understanding of the peculiarities of the interplanetary magnetic field,
such as a potential normal field component, we will not speculate on the relevance of this
effect for the \amet{}s, only that it adds to the uncertainty about the degree to which 
the electromagnetic interaction plays a role in their formation.

It could also be considered that `magnetic resonances' increase particle eccentricity and thus
transfer particles onto orbits with low $C$-value.
This type of resonance, as described by \citet{Morfill1979motion}, can occur at heliocentric distance 
$<\!\qty{0.3}{\astronomicalunit}$, where the short orbital period of particles may
enable a periodic interaction with the revolving IMF sectors.
However, \citet{Morfill1979motion} demonstrate the magnetic resonance only in a simplified model,
which does not consider variable sector lengths or solar-cycle-dependent magnetic polarity changes,
arguing that, under real conditions, the resonance could not act long enough to have a significant effect.
Thus, we rule out the magnetic resonance as an \amet{}-generating mechanism.

\subsection{Creation via sublimation} \label{Sect:sublimation}
Besides collisional grinding, sublimation of dust is considered an effective mechanism
for the creation of \bmet{}s \citep[see review by][]{Mann2004dust}.
As their orbits decay further toward the Sun, 
grains which survive the expulsion due to collisions are subject to an ever-higher solar thermal flux.
When sublimation sets in at a material-specific heliocentric distance 
(typically below 15~$R_\odot$, see \citet{Mann2004dust}, Tab.~2),
the mass loss of the particle steadily raises its $\beta$-factor.
The increasing effect of radiation pressure counteracts the decay due to PR drag and
eventually causes the particle to maintain a quasi-constant perihelion distance. 
Yet, as the increase in $\beta$ occurs around perihelion (where sublimation is strongest)
the eccentricity starts to grow, raising the aphelion with each revolution.
The further fate of the particle then depends on the material-specific maximum $\beta$-factor
that particles can assume \citep{Burns1979radiation,Shestakova1995dynamics,Krivov1998dynamics}:
Once $\beta$ surpasses a near-unity critical value
(typically around radii of \qtyrange[range-units=single]{100}{200}{\nm}),
the particle may become hyperbolic, that is, a \bmet{}.
On the other hand, particles made up of materials that reach a maximum $\beta$ below the threshold,
orbitally collapse and sublimate entirely after the maximum $\beta$ is passed.

One might think that, due to the incremental increase in eccentricity, 
particles will naturally surpass a stage with high aphelion before becoming hyperbolic,
such that they can be detected at \qty{1}{\astronomicalunit} in the manner of the apex particles,
as has been suggested by \citet{Shestakova1995dynamics}.
However, due to the low perihelion, this would correspond to a narrow range of near-unity orbital eccentricity.
If we demand an aphelion between \qtylist{1;1.5}{\astronomicalunit} the eccentricity range
for a grain sublimating at 10~$R_\odot$ becomes~0.91 to~0.94.
It is questionable whether an incremental increase in $\beta$ at perihelion 
can efficiently generate grains within such a high, yet narrow eccentricity range.
Moreover, a such created \amet{} may remain in this state for only one revolution,
given the sensitivity of the orbit to a changing $\beta$ at the next perihelion.
That is contrary to collisionally generated \amet{}s, which, once created,
may remain in such orbits for many more revolutions
(a typical \amet{} with $a\!=\!\qty{0.6}{\astronomicalunit}$, $e\!=\!0.6$, and $0.3\!<\!\beta\!<0.5$
has a PR lifetime of 200 to 300 years).
However, whether sublimation can be ruled out as an \amet{}-supplying mechanism must be left to further studies. 
If so, the presence of \amet{}s might be indicative of collisional grinding 
as the more effective loss-mechanism of the zodiacal cloud.

In this context it should be considered that recent findings of visible observations carried out by 
the Parker Solar Probe indicate a smooth decline in dust density starting from about 19~$R_\odot$ down to
5~$R_\odot$, as well as a dust free zone below 5~$R_\odot$ \citep{Howard2019nearsun,Stenborg2021psp,Stenborg2022psp}.
The observed absence of dust rings speaks against a sublimation loss occurring only 
at specific heliocentric distances \citep[see \eg][]{Kobayashi2009dust} 
and for a continuous process that gradually reduces the dust density 
\citep[or for a process that dissolves the dust rings effectively, \eg][]{Isobe1993effect}.

Other loss mechanisms, erosive sputtering by solar wind particles and rotational bursting,
may be also be relevant within a few $R_\odot$ \citep{Mann2004dust}
and could thus also play a role in the creation of \amet{}s.

  \section{Discussion} \label{Sect:Discn}

We examined the early (around 1980) literature around the term \amet{} and followed
its transition in meaning to its present-day usage.
We restated the characteristics of the originally described particles,
and reiterated the arguments put forward by the contemporaries of their discovery 
for a common origin with the hyperbolic \bmet{}s.
By analysing the evolutionary tracks of grains released from different source orbits,
we confirmed those arguments and rejected alternative formation pathways as being unable to yield the observed dynamics.
Among the alternative scenarios, the release of dust from Encke-type orbits is borderline compatible 
with the observed dynamics (which, however, could only account for the largest of the detected grains),
while a creation via sublimation or Lorentz force interaction cannot be definitively ruled out at this point.
Yet, a robust mechanism to produce the, as originally referred to, \amet{}s, is the 
release from progenitor-meteoroids in short-heliocentric-distance and low-eccentricity orbits,
which have migrated from their cometary and asteroidal source orbits under PR drag and eventually
suffer collisional fragmentation---equivalent to one of the widely accepted formation mechanisms of \bmet{}s
\citep[\eg][]{Wehry1999identification}.
If grinding of the zodiacal cloud at short heliocentric distances occurs with a size distribution for 
collisional products that spans $\beta$-factors less than 0.5 
\citep[that is, grains with radii roughly larger than \qtyrange{0.5}{0.8}{\um}, \eg][]{Wilck1996radiation},
the creation of these bound, high-eccentricity particles alongside the hyperbolic \bmet{}s is consequential.
Given that colliding progenitor meteoroids can be assumed to have typical sizes of \qtyrange{10}{100}{\um} 
(those which migrate most efficiently from their source orbits \citep{Grun1985collisional}), 
the creation of micron-sized fragments even in catastrophic collisions is not a far-fetched assumption.
This is also compatible with constraints on the largest fragment mass imposed by
typically assumed fragment mass distributions, which are based on the scaling of impact experiment results
\citep[\eg][]{Grun1985collisional,Ishimoto2000modeling,Krivov2005evolution}.

\subsection{Implications}

\subsubsection{Dust Models} \label{Sect:Models}
Some meteoroid engineering models allow analysis of flux and directionality
of micrometeoroids in the size range of the \amet{} along a user-specified spacecraft orbit,
though they are not equipped with a comparable particle population. 
One such model is the Interplanetary Meteoroid Engineering Model (IMEM), 
developed by \citet{Dikarev2005new,Dikarev2005update}.
Unlike the earlier empirical Divine-Staubach model, which introduces synthetic particle populations
derived from observations alone \citep{Divine1993five,Staubach1997meteoroid}, 
IMEM adopts a bottom-up physical approach, where the particle populations are constructed from the orbital evolution
(under PR drag) of meteoroids originating in source body populations (JFCs and asteroids).
Yet, as the source of the \amet{}s (presumably PR-drag-evolved progenitor meteoroids close to the Sun) 
is not considered, and since they cannot be created efficiently from cometary or asteroidal orbits
from PR drag alone (as shown in Sect.~\ref{Sect:Dynamics}), 
the \amet{}s cannot possibly be reproduced by IMEM's populations.
On the other hand, IMEM's representation of JFC and asteroidal dust is misguided, as its populations are also fitted
with datasets from in-situ dust detectors that were arguably exposed to the \amet{}s,
thus yielding unphysically high contributions of JFC and asteroidal dust at sizes that may be well below
the blowout limit for those sources \citep[\eg][]{Moorhead2021forbidden}.

IMEM is therefore an inadequate tool to make predictions for fluxes of grains smaller than a few microns,
although its usage may still be reasonable for analysing fluxes onto sensors whose detection threshold is
above the \amet{} size range \citep[\eg][]{Kobayashi2018situa}.
The same argument can be made for the successor IMEM2 \citep{Soja2019imem2}, where dust production
occurs only in cometary and asteroidal orbits, and which models particle radii down to \qty{1}{\um}.

This issue may be resolved in zodiacal cloud models by including an appropriate source population,
or, which is more physical, by treating grain-grain collisions not just as a loss but 
also as a productive mechanism \citep[\eg][]{Kral2013lidtdd,Rigley2021comet}.
Meteoroid environment models whose particle size range doesn't extend to the \amet{}s,
such as the MEM3 \citep{Moorhead2020nasa} or the \citet{Pokorny2019coorbital} model, 
are not affected by this issue.

\subsubsection{In-situ experiments}

As the \amet{}s have dominated the datasets of previous sub-micron-sensitive 
dust detectors deployed at and below 
\qty{1}{\astronomicalunit} (rivalled only by the hyperbolic \bmet{}s),
they can also be expected to play a prevalent role for the upcoming
generation of dust sensors going to the inner solar system,
\eg{} the impact plasma mass spectrometers of \des{} (\des{} Dust Analyzer (DDA)) and 
IMAP (Interstellar Dust Experiment (IDEX)), 
which will examine the properties of interstellar dust (ISD) at 
$\sim\!\qty{1}{\astronomicalunit}$ \citep{Kruger2019modelling,McComas2018interstellar}.
For a spacecraft in an ecliptic heliocentric orbit the influx of ISD particles,
which flows through the solar system with a predominant direction,
is modulated by its motion around the Sun.
When the spacecraft moves head-on to the flow of ISD,
the ISD flux is highly elevated due to the added relative velocity 
\citep[\eg][]{Altobelli2003cassini,Hervig2022decadal}, resulting in a kind of `interstellar season'.
This flux, however, occurs predominantly from the spacecraft apex,
posing a difficulty to distinguish the ISD particles from the, likewise apex-dominated, \amet{}s.
For a sensor deployed in a low-eccentricity orbit near \qty{1}{\astronomicalunit},
ISD grains will have relative velocities of \qtyrange[range-units=single]{40}{70}{\km\per\s} 
during the `interstellar season', whereas \amet{}s are encountered 
at velocities of \qtyrange[range-units=single]{5}{25}{\km\per\s} throughout the year.
Differentiation between interstellar and interplanetary dust 
could thus be achieved by determination of the impactor speeds
\citet[or, to some extent, by the impactor energy, as done by][]{Altobelli2003cassini,Altobelli2005interstellar}.

DDA may accomplish this task with its charge-sensing entrance grids that allow for a 
time-of-flight speed measurement of impacting grains larger than a few \qty{100}{\nm}.
However, smaller ISD particles, which are strongly modulated by the interplanetary magnetic field 
and whose occurrence in the inner solar system is thus dependent on the solar cycle 
\citep{Sterken2012flow,Strub2019heliospheric}, will not be registered by the charge-sensing entrance grids.
This hinders straightforward differentiation, yet it may be argued, 
that such small interplanetary dust particles will be on hyperbolic trajectories 
(due to their large $\beta$-factor) and are unlikely to approach from the apex direction.
(Although in the 10s of nm range, $\beta$ subsides again and, for certain materials, can assume
values compatible with bound orbits, see \eg{} \citet{Wilck1996radiation}.)

DDA also seeks to study the interplanetary dust environment at \qty{1}{\astronomicalunit}.
In that context, \citet{Kruger2019modelling} have used the IMEM tool to make predictions for flux, 
directionality, and relative contribution of JFC and asteroidal dust along the \des{} trajectory.
However, as discussed in Sect.~\ref{Sect:Models}, IMEM's representation of particles 
in the studied size range is misleading, in that it does not include the 
most relevant populations, and gives an unphysical account of the abundance of dust 
still dynamically linked to their source body families (JFC and asteroids).
The chemical analysis of interplanetary dust particles that have retained their dynamical affiliation is, of course, 
of high scientific value, yet, the incidence of such particles onto DDA may be much lower than estimated by IMEM.
The more abundant \amet{}s, on the other hand, lose their dynamical history in the 
(presumably collisional) processing at short heliocentric distances that drives their creation.
Nonetheless, their chemical characterization at high mass resolution 
(DDA: $m/dm\!\approx\!100\text{--}200$) is unprecedented
and may reveal valuable insights about the loss mechanisms of the zodiacal cloud.
It should be stressed that, while the \amet{}s may be predominant flux-wise,
there is a bias in sensors at \qty{1}{\astronomicalunit} to detect them 
\citep{Weidenschilling1978distribution,Altobelli2003cassini}.
Hence, they do not necessarily have to be the most abundant density-wise, too.

Another type of in-situ dust experiment are the electric field antennas onboard space missions,
which are in fact part of plasma science instruments, 
but can also sense dust impacts of micron and sub-micron-sized grains on the spacecraft surface \citep[\eg][]{Mann2019dust}.
Of relevance here are the inner heliosphere science missions,
Wind, STEREO, Solar Orbiter, and Parker Solar Probe, which carry such instruments.
While the datasets of STEREO and Solar Orbiter have so far not allowed a directional analysis of impacts 
\citep{Zaslavsky2012interplanetary,Zaslavsky2021first}---although an ISD component in the STEREO flux was identified 
by the modulation with ecliptic longitude \citep{Belheouane2012detection}---the Wind spacecraft,
as a spin-stabilized probe, has been able to retrieve directional information about impacting grains.
Stationed at the Sun-Earth L1 point, Wind observed an excess flux contribution from the spacecraft apex throughout the year
\citep{Malaspina2014interplanetary,Wood2015hypervelocity}, which is another indication of the prevalence of \amet{}s.

Likewise, The Parker Solar Probe (PSP) with its FIELDS instrument,
has recorded variable dust impacts rates along its highly eccentric orbit.
By comparing the flux modulation along the orbit to modelled impact rates generated by synthetic populations
of hyperbolic \bmet{}s as well as bound interplanetary dust (on circular orbits), 
\citet{Szalay2020nearsun,Szalay2021collisional} and \citet{Malaspina2020situ} conclude that most
of the influx onto PSP is in the form of hyperbolic \bmet{}s.
In addition, the PSP/FIELDS dataset allows directional analysis within the plane normal to the Sun direction (PSP's heat-shield plane),
due to the instruments four `planar antennas' \citep{Page2020examining,Malaspina2020situ,Pusack2021dust}.
Yet, the two-component model of dust on (i) hyperbolic and (2) circular orbits
does not predict certain features in flux rate and directionality
such as the consistent post-perihelion flux enhancement.
To match the post-perihelion flux enhancement \citep{Szalay2021collisional} introduce a third population (`\textbeta-stream')
of hyperbolic \bmet{}s emanating from the Geminids meteoroid stream, as products from collisions between
the meteoroid stream and the zodiacal cloud.
It is beyond the scope of this study to model the impact rate that would be produced along PSP's trajectory
by an \amet{}s population, yet, its qualitative effect may be considered.
On their outward orbital arc, the highly eccentric \amet{} resemble the hyperbolic \bmet{}s reasonably well 
at short heliocentric distances and introduce no new features, flux- or density-wise.
However, \amet{}s on their inward orbital arc would attain high relative speeds with respect
to PSP during its outbound arc, thus generating a post-perihelion flux 
arriving at the spacecraft from the anti-sunward direction.

At least during one orbit, the observed post-perihelion enhancement was accompanied by an 
increased fraction of impacts, for which the comparative signal strengths 
of the `planar antennas' could not give conclusive directional indications, 
implying impacts nearly perpendicular to PSP's heat-shield plane \citep{Pusack2021dust}.
Whether those impacts occurred predominantly on the sunward or on the anti-sunward side may be resolved
by further investigation of data of the fifth antenna of the FIELDS instrument, 
located at the back (anti-sunward side) of the spacecraft.
The pending results of the fifth antenna data analysis may thus confirm a post-perihelion contribution from
a population akin to the inward directed \amet{}s or the outward directed `\textbeta-stream', 
or a combination of both. 

It should be noted that \citet{Szalay2020nearsun} also consider a bound interplanetary dust population obtained
by extending the \citet{Pokorny2019coorbital} model from a minimal particle size of \qty{10}{\um} (diameter) to sub-micron sizes,
and conclude that it produces similar results for the PSP flux rate modulation as the idealized circular population.
However, the \citet{Pokorny2019coorbital} model considers grain-grain collisions only as a loss, but not as a productive
mechanism, and can therefore (as argued in Sect.~\ref{Sect:Models}) not account for the highly eccentric \amet{}s
present at these sizes.

\subsubsection{Radar Meteors}
Lastly, the relevance of the \amet{}s for radar meteor observations must be discussed.
Head-echo meteor observations with high-power large-aperture radars allow precise speed measurements of 
micrometeoroids entering the atmosphere.
In the case of the highly sensitive Arecibo Observatory (AO), detection of meteors produced by 
grains down to \qty{1}{\um} in diameter is feasible \citep[][who even report detection of \bmet{}s]{Janches2001orbital}.
It is thus conceivable that those instruments can also detect the \amet{}s, having sizes near the AO threshold.
Indeed, \citet{Janches2003geocentric,Sulzer2004meteoroid,Janches2005observed} report a population of slow
meteors at speeds of \qtyrange{10}{20}{\km\per\s} to emerge when AO observes near the Earth-apex
(alongside a stronger contribution from the retrograde north-/south-apex meteors).
The fact that this method detects predominantly meteors travelling along the radar beam axis \citep{Janches2004radiant},
indicates that these slow moving meteors indeed approach from the apex direction.
\citet{Janches2005observed} interpret this slow moving population as asteroidal grains on near circular heliocentric orbits,
whose atmospheric entry velocities are amplified by gravitational focussing.
One could argue, however, that grains with such low $v_\infty$, would not show a pronounced directionality when
entering the atmosphere. 
In addition, the slow apex meteors appear to be absent in head-echo data of the less sensitive 
Jicamarca Radio Observatory \citep{Janches2005observed}.
They could therefore be interpreted as being generated by meteoroids of smaller sizes that are
below the Jicamarca threshold, but above the detection threshold of AO.

However, if AO was in fact able to detect the \amet{}s, one would 
(due to the negative slope of the particle size distribution) 
expect them to dominate the dataset, which is certainly not the case.
A possible answer to this discrepancy could be an instrumental bias against low-velocity meteors,
as \citet{Hunt2004determination} and \citet{Close2007meteor} found to be present in head-echo observations.
\citet{Janches2008comparison} show that this velocity selection effect is not present in AO observations
at particles masses $>\!\qty{e-6}{\gram}$, yet, a bias at lower masses is not ruled out.
On the other hand, \citet{Fentzke2008semiempirical} find, that at AO's lower mass detection threshold of 
\qty{e-11}{\gram}---which is around the upper mass limit for \amet{}s---only grains of sufficient
speeds ablate and thus produce observable meteor phenomena.
Therefore, they conclude that such small particles are only detectable by AO at speeds $>\!\qty{30}{\km\per\s}$.
The fastest \amet{}s in the \pion{} TOF dataset exhibit $v_{\rm{impact}}\!\approx\!\qty{26}{\km\per\s}$, 
which (considering gravitational focussing) would amount to an atmospheric entry velocity of $\sim\!\qty{28}{\km\per\s}$. 
AO may thus not just be biased against the detection of \amet{}s, but rather entirely insensitive to them.
The conjecture that \amet{}s constitute the low-velocity apex meteors observed by AO is therefore speculative, at best.

\subsection{Accuracy of historic data} \label{Sect:accuracy}
This section reviews some aspects of the quality of the data retrieved from \pion{}, \heos{}, and Helios,
that led \citet{Grun1980dynamics} to point out the dynamical class of the \amet{}s.
Constraining the orbit of a particle requires knowledge about its velocity vector as well as $\beta$-factor,
which may be estimated from the derived impactor mass and assumed material properties.

With impact ionization dust sensors such as on \heos{} and Helios, the particle velocity is derived from 
calibration of the charge yield signal shape, which has inherently high uncertainties.
For slow impactors ($<\!\qty{25}{\km\per\s}$) onto the \heos{} sensor, 
\citet{Dietzel1973heos} state low uncertainties of \qtyrange{10}{20}{\percent}, 
whereas \citet{Grun1981physikalische} infer an uncertainty of a factor of \qtyrange{1.65}{2}{} for the Helios sensor
(which is more in line with uncertainties found in calibration studies 
with other sensors of this type \citep{Goller1989calibration,Igenbergs1991munich}).
The error in the velocity propagates to the mass determination, which, in the case of the Helios sensor,
has an uncertainty of a factor of~10 \citep{Grun1981physikalische}.
More recently, when revisiting impact experiments with the Cassini Cosmic Dust Analyzer, 
\citet{Hunziker2022impact} additionally found that particle porosity reduces the charge signal rise times,
which may cause an overestimate of impact velocities determined in this way, if particles are fluffy.

Although they are different designs, the \heos{} and the Helios (ecliptic) sensors have nearly equal
effective solid angles of \qty{1.03}{sr} and \qty{1.04}{sr} (around \qty{8}{\percent} of the entire sky),
which gives a reasonable limitation of the impactor direction.
While the constraint on the orbit of a single impactor may be considered insufficient,
given the uncertainties in the derived velocity, mass, and direction,
\citet{Grun1981physikalische} demonstrates that by ascribing probability distributions to these quantities
and adding up the resulting orbital elements distributions of all registered particles,
a conclusive characterization of their dynamic properties is possible
\citep[see also][]{Schmidt1979distribution,Schmidt1980orbital,Schmidt1980bahnelemente,Grun1985orbits}.

On the other hand, the direct speed measurements stemming from the \pion{} time-of-flight events
certainly provide a more reliable velocity vector determination for single impactors than the \heos{} and Helios sensors.
Nonetheless, this method may underestimate speeds to some extent, 
due to the deceleration of grains upon penetration of the front film sensors.
According to \citet{Berg1969pioneer}, the particle velocity is reduced by \qty{40}{\percent} for 
grains of the minimum kinetic energy (and \qty{5}{\percent} at ten times the threshold energy).
Two of the ten bound and prograde TOF-measured particles have kinetic energies close to the threshold level of \qty{100}{\nano\joule},
whereas the remaining 8 particles have energies considerably higher than the threshold with a logarithmic average of \qty{2300}{\nano\joule} 
(calculated from velocities and masses given by \citet{McDonnell1978microparticle}).
Considering deceleration for those two particles would lead to a lowering of their perihelia (an increase of their $C$-value).
One of the remaining 8 TOF-measured particles exhibits a less eccentric orbit
($q\!=\!\qty{0.61}{\astronomicalunit}$, $e\!=\!0.26$, and $C\!\approx\!\qty{2}{\astronomicalunit}$),
compatible with release from a cometary orbit and subsequent PR evolution.
The \amet{} zone illustrated in Fig.~\ref{fig:EQ_evolution} has been set to span only the remaining 
7~particles, which all have $C\!<\!\qty{1}{\astronomicalunit}$.
The direct TOF speed measurements also allow for a more reliable determination of the impactor mass,
thus arguably allowing a more accurate estimate of the $\beta$-factor, compared to pure impact ionization detectors.
Furthermore, the effective solid angle for the TOF events ($\lesssim\!\qty{0.23}{sr}$ 
as estimated from the angular sensitivity given by \citet{Grun1973reliability})
is significantly below that of the \heos{} and Helios sensors, 
providing a better constraint on the impactor direction.

\subsection{Terminology}
The confusion around the terminology of the particles discussed here suggests rethinking
of the prevalent naming convention.
A continued double use of the term \amet{} can, of course, only be discouraged,
and one might be drawn to use a different name for the originally referred-to particles.

The term `apex particles' seems straightforward and descriptive,
but carries meaning only in the context of in-situ detection from certain orbits.
The Helios probe, itself on a high-eccentricity orbit (with peri-~and aphelion of roughly 
\qtylist[range-units=single]{0.3;1}{\astronomicalunit}), observed this population 
to impact from the apex direction only when sufficiently far from aphelion.
Around its low-momentum aphelion, relative velocities with respect to the apex particles diminishes,
causing them to lose their directional signature \citep{Grun1985orbits}.
This term also bears the risk of confusing the particles with the meteoroids that approach the Earth from the apex
direction, generating the north-/south-apex radiants seen in radar meteors.
These grains move on retrograde orbits, stemming from Halley-type and Oort cloud comets \citep[\eg][]{Nesvorny2011dynamics,Pokorny2014dynamical}, 
and have no dynamical relation with the apex particles discussed here.
A small fraction of these apex meteors may also be supplied by JFCs with low perihelia \citep{Nesvorny2011dynamical},
which can have similar orbits to the particles discussed here albeit at much higher particle masses
(necessary to withstand expulsion by radiation pressure upon release).
 
A suitable replacement for the original \amet{} term may be `bound \bmet{}s', 
which is descriptive and conveys the relatedness to hyperbolic \bmet{}s.
This is also in the original sense of the \bmet{} term, 
which refers to grains with substantial $\beta$-value, bound or unbound \citep{Zook1975source}.
However, also the term \bmet{}s has undergone a change of definition since then
and has commonly become to be understood as to refer to unbound particles only.
One is therefore faced with the dilemma of either returning to the original definition of the term
`\amet{}' or returning to the original definition of `\bmet{}'---accepting an ambiguity in literature in either case.
This may be resolved by resorting to a term such as `failed \bmet{}s', which upholds the relatedness to \bmet{}s,
yet also alludes to the fact that they are not \textit{true} (\ie{} hyperbolic) \bmet{}s.
However, for the sake of brevity, this author still favours a return to the original \amet{} meaning.
  \newpage
\section{Conclusion}

The literature review presented in this work has shown, that the term \amet{} has,
since its inception by \citet{Grun1980dynamics}, undergone a change of definition.
Originally, the term referred to grains in a narrow size range around roughly \qty{1}{\um},
which revolved on bound, yet high-eccentricity and low-perihelion orbits.
Due to their dynamics, these were seen as `siblings' of the \bmet{}s,
presumably stemming from the same formative process near the Sun.
Yet, with the adoption of the term in the circumstellar disc modelling domain, 
\amet{}s became to be understood as merely the bound component of circumstellar dust discs 
(in addition to the unbound \bmet{}-component),
which has pushed the originally referred-to class of particles out of sight of contemporary research.

Our assessment of the dynamics of the original \amet{}s suggests that the particles,
do not evolve directly from cometary or asteroidal source orbits, and that they are 
generated effectively (and inevitably) in collisions among evolved grains near the Sun.
The \amet{}s generated this way are in essence `bound \bmet{}s',
having failed to attain unbound trajectories due to being slightly too massive.
Therefore, we reinforce the argument made by \citet{Grun1980dynamics}---although
creation pathways via release from Encke-type comets, sublimation, or Lorentz perturbation
are not definitively ruled out.

This population is not accounted for in most physical meteoroid environment models (such as IMEM) 
which typically do not consider the production of \bmet{}s and, thus, can also not include the \amet{}s.
Therefore, we stress the importance of considering collisional production in models,
that aim to describe micron- and submicron-sized dust populations.

Recently, due to a renewed interest in inner-solar-system interplanetary dust research 
fuelled by new missions,
the altered definition of the \amet{}s has transpired back into in-situ dust detection research.
The now-overlooked original \amet{}s, however, which are characterized by their high eccentricities,
had been the predominantly detected bound dust population
by the early generations of sensitive dust detectors at \qty{1}{\astronomicalunit}.
Considering them is therefore vital for current and upcoming in-situ dust experiments,
such as onboard \des{}, IMAP, and the Parker Solar Probe
(as well as, potentially, on the Lunar Gateway \citep{Wozniakiewicz2021cosmic}
or the Dolphin mission \citep{Sterken2022dolphin}).

Physical, chemical, and dynamical characterization to the degree enabled by the latest 
sophisticated instruments is unprecedented for the \amet{}s and may elucidate
their formation and the processing of the zodiacal cloud at short heliocentric distances.
Signatures of sublimation and melting \citep{Belton1966dynamics},
high sensitivity to electromagnetic interaction (high charge-to-mass ratio), 
or a flux dependence on the solar cycle
are just examples of what new-generation dust detectors might look out for
to further our understanding of the peculiar \amet{}s.

\section*{Acknowledgements}

I thank Russel Howard for a stimulating discussion at COS\-PAR 2022;
Eberhard Grün, Ralf Srama, and Harald Krüger for their assistance in acquiring valuable literature;
Veerle Sterken for providing helpful feedback;
as well as two anonymous reviewers for their constructive suggestions.

Partial support provided by the German Research Foundation (DFG, grant no. SR77/6-1) 
and the German Aerospace Center (DLR, grant no. 50OO2101) is gratefully acknowledged.
\renewcommand{\thetable}{A.\arabic{table}}
  \begin{appendix}
    \renewcommand{\thetable}{A.\arabic{table}}
    \onecolumn
    \section{Alpha-meteoroids reference list}
    \begin{table*}[ht]
        \renewcommand*{\arraystretch}{1.1}
        \centering

        \begin{tabular}{ p{4.84cm} b{1.79cm} p{10cm} }
            Reference & Definition of \amet{} & Comment \\ 
            \hline
            \citet{Zook1975source} & - &
            Introduce the term `\bmet{}' for grains whose orbits are significantly shaped by radiation pressure, including those that are bound.\\
            \citet{Hoffmann1975first,Hoffmann1975temporal} & - &
            Introduce the term `apex particles', referring to the group of grains measured by \heos{} 
            appearing to come from the heliocentric apex.\\
            \citet{Grun1980dynamics} & original &
            Introduce the term `\amet{}' for the apex particles, due to their apparent relatedness to the hyperbolic \bmet{}s\\
            \citet{Grun1981physikalische} & original &
            Research report (in German) on Helios results. Characterizes the \amet{}s.\\
            \citet{Grun1980orbital,Grun1985orbits} & - &
            Characterize apex particles measured by Helios, not using the term \amet{}.\\
            \citet{Grun1985collisional} & original & 
            Summarizes findings about \amet{}s. \\
            \citet{Grun1992ulysses} & - &
            Mention apex particles, not using the term \amet{}s.\\
            \citet{Iglseder1993cosmic,Iglseder1996cosmic} & - & 
            Characterize apex particles measured by Hiten, not using the term \amet{}.\\
            \citet{Shestakova1995dynamics} & original & 
            -\\
            \citet{Artymowicz1997dust,Artymowicz1997beta} & new &
            Apply the \textalpha-/\bmet{} terminology to circumstellar discs in general, thereby extending
            the meaning of \amet s to all grains that are released from parent bodies into bound orbits.\\
            \citet{Wehry1999identification} & original &
            -\\
            \citet{Krivov2000size,Krivov2006dust} & new & 
            -\\
            \citet{Krivova2000disk,Krivova2000size} & new & 
            -\\
            \citet{Artymowicz2000beta} & new &
            -\\
            \citet{Fechtig2001historical} & original &
            -\\
            \citet{Mann2006dust} & new &
            -\\
            \citet{Freistetter2007planets} & new &
            -\\
            \citet{Soja2010dynamics} & original & 
            -\\ 
            \citet{Krivov2010debris} & new &
            -\\
            \citet{Kral2017exozodiacal} & new &
            -\\
            \citet{Szalay2021collisional} & new &
            -\\
            \citet{Mann2021dust} & new &
            -\\ 
            \citet{Pusack2021dust} & new &
            -\\ 
        \end{tabular}
        \caption{\label{tab:literature} A collection of published articles discussing \amet s or using the term.
        The list may be incomplete, \ie{} there may be published articles not listed here using the term \amet.
        Also listed are a few connected works using the term `apex particles'.} 
        \label{tab:amet_references}
    \end{table*}

\end{appendix}

  \newpage
	\bibliographystyle{elsarticle-harv}
	\bibliography{references}

\begin{thebibliography}{146}
\expandafter\ifx\csname natexlab\endcsname\relax\def\natexlab#1{#1}\fi
\providecommand{\url}[1]{\texttt{#1}}
\providecommand{\href}[2]{#2}
\providecommand{\path}[1]{#1}
\providecommand{\DOIprefix}{doi:}
\providecommand{\ArXivprefix}{arXiv:}
\providecommand{\URLprefix}{URL: }
\providecommand{\Pubmedprefix}{pmid:}
\providecommand{\doi}[1]{\href{http://dx.doi.org/#1}{\path{#1}}}
\providecommand{\Pubmed}[1]{\href{pmid:#1}{\path{#1}}}
\providecommand{\bibinfo}[2]{#2}
\ifx\xfnm\relax \def\xfnm[#1]{\unskip,\space#1}\fi
\bibitem[{Altobelli et~al.(2005)Altobelli, Kempf, Kr{\"u}ger, Landgraf, Roy and
  Gr{\"u}n}]{Altobelli2005interstellar}
\bibinfo{author}{Altobelli, N.}, \bibinfo{author}{Kempf, S.},
  \bibinfo{author}{Kr{\"u}ger, H.}, \bibinfo{author}{Landgraf, M.},
  \bibinfo{author}{Roy, M.}, \bibinfo{author}{Gr{\"u}n, E.},
  \bibinfo{year}{2005}.
\newblock \bibinfo{title}{Interstellar dust flux measurements by the
  {{Galileo}} dust instrument between the orbits of {{Venus}} and {{Mars}}}.
\newblock \bibinfo{journal}{Journal of Geophysical Research: Space Physics}
  \bibinfo{volume}{110}.
\newblock \DOIprefix\doi{10.1029/2004JA010772}.
\bibitem[{Altobelli et~al.(2003)Altobelli, Kempf, Landgraf, Srama, Dikarev,
  Kr{\"u}ger, {Moragas-Klostermeyer} and Gr{\"u}n}]{Altobelli2003cassini}
\bibinfo{author}{Altobelli, N.}, \bibinfo{author}{Kempf, S.},
  \bibinfo{author}{Landgraf, M.}, \bibinfo{author}{Srama, R.},
  \bibinfo{author}{Dikarev, V.}, \bibinfo{author}{Kr{\"u}ger, H.},
  \bibinfo{author}{{Moragas-Klostermeyer}, G.}, \bibinfo{author}{Gr{\"u}n, E.},
  \bibinfo{year}{2003}.
\newblock \bibinfo{title}{Cassini between {{Venus}} and {{Earth}}:
  {{Detection}} of interstellar dust}.
\newblock \bibinfo{journal}{Journal of Geophysical Research: Space Physics}
  \bibinfo{volume}{108}.
\newblock \DOIprefix\doi{10.1029/2003JA009874}.
\bibitem[{Altobelli et~al.(2016)Altobelli, Postberg, Fiege, Trieloff, Kimura,
  Sterken, Hsu, Hillier, Khawaja, {Moragas-Klostermeyer}, Blum, Burton, Srama,
  Kempf and Gr{\"u}n}]{Altobelli2016flux}
\bibinfo{author}{Altobelli, N.}, \bibinfo{author}{Postberg, F.},
  \bibinfo{author}{Fiege, K.}, \bibinfo{author}{Trieloff, M.},
  \bibinfo{author}{Kimura, H.}, \bibinfo{author}{Sterken, V.J.},
  \bibinfo{author}{Hsu, H.W.}, \bibinfo{author}{Hillier, J.},
  \bibinfo{author}{Khawaja, N.}, \bibinfo{author}{{Moragas-Klostermeyer}, G.},
  \bibinfo{author}{Blum, J.}, \bibinfo{author}{Burton, M.},
  \bibinfo{author}{Srama, R.}, \bibinfo{author}{Kempf, S.},
  \bibinfo{author}{Gr{\"u}n, E.}, \bibinfo{year}{2016}.
\newblock \bibinfo{title}{Flux and composition of interstellar dust at
  {{Saturn}} from {{Cassini}}'s {{Cosmic Dust Analyzer}}}.
\newblock \bibinfo{journal}{Science} \bibinfo{volume}{352},
  \bibinfo{pages}{312--318}.
\newblock \DOIprefix\doi{10.1126/science.aac6397}.
\bibitem[{Artymowicz(1997)}]{Artymowicz1997beta}
\bibinfo{author}{Artymowicz, P.}, \bibinfo{year}{1997}.
\newblock \bibinfo{title}{{{BETA PICTORIS}}: {{An Early Solar System}}?}
\newblock \bibinfo{journal}{Annual Review of Earth and Planetary Sciences}
  \bibinfo{volume}{25}, \bibinfo{pages}{175--219}.
\newblock \DOIprefix\doi{10.1146/annurev.earth.25.1.175}.
\bibitem[{Artymowicz(2000)}]{Artymowicz2000beta}
\bibinfo{author}{Artymowicz, P.}, \bibinfo{year}{2000}.
\newblock \bibinfo{title}{Beta {{Pictoris}} and {{Other Solar Systems}}}.
\newblock \bibinfo{journal}{Space Science Reviews} \bibinfo{volume}{92},
  \bibinfo{pages}{69--86}.
\newblock \DOIprefix\doi{10.1023/A:1005297300878}.
\bibitem[{Artymowicz and Clampin(1997)}]{Artymowicz1997dust}
\bibinfo{author}{Artymowicz, P.}, \bibinfo{author}{Clampin, M.},
  \bibinfo{year}{1997}.
\newblock \bibinfo{title}{Dust around {{Main}}-{{Sequence Stars}}: {{Nature}}
  or {{Nurture}} by the {{Interstellar Medium}}?}
\newblock \bibinfo{journal}{The Astrophysical Journal} \bibinfo{volume}{490},
  \bibinfo{pages}{863--878}.
\newblock \DOIprefix\doi{10.1086/304889}.
\bibitem[{Baguhl et~al.(1995)Baguhl, Hamilton, Gr{\"u}n, Fechtig, Kissel,
  Linkert, Linkert, Riemann, Staubach, Dermott, Hanner, Polanskey, Lindblad,
  Mann, McDonnell, Morfill, Schwehm and Zook}]{Baguhl1995dust}
\bibinfo{author}{Baguhl, M.}, \bibinfo{author}{Hamilton, D.P.},
  \bibinfo{author}{Gr{\"u}n, E.}, \bibinfo{author}{Fechtig, H.},
  \bibinfo{author}{Kissel, J.}, \bibinfo{author}{Linkert, D.},
  \bibinfo{author}{Linkert, G.}, \bibinfo{author}{Riemann, R.},
  \bibinfo{author}{Staubach, P.}, \bibinfo{author}{Dermott, S.F.},
  \bibinfo{author}{Hanner, M.S.}, \bibinfo{author}{Polanskey, C.},
  \bibinfo{author}{Lindblad, B.A.}, \bibinfo{author}{Mann, .},
  \bibinfo{author}{McDonnell, J.A.M.}, \bibinfo{author}{Morfill, G.E.},
  \bibinfo{author}{Schwehm, G.}, \bibinfo{author}{Zook, H.A.},
  \bibinfo{year}{1995}.
\newblock \bibinfo{title}{Dust {{Measurements}} at {{High Ecliptic
  Latitudes}}}.
\newblock \bibinfo{journal}{Science} \bibinfo{volume}{268},
  \bibinfo{pages}{1016--1019}.
\newblock \DOIprefix\doi{10.1126/science.268.5213.1016}.
\bibitem[{Barge et~al.(1982)Barge, Pellat and Millet}]{Barge1982diffusion}
\bibinfo{author}{Barge, P.}, \bibinfo{author}{Pellat, R.},
  \bibinfo{author}{Millet, J.}, \bibinfo{year}{1982}.
\newblock \bibinfo{title}{Diffusion of {{Keplerian}} motions by a stochastic
  force. {{II}} - {{Lorentz}} scattering of interplanetary dusts}.
\newblock \bibinfo{journal}{Astronomy and Astrophysics} \bibinfo{volume}{115},
  \bibinfo{pages}{8--19}.
\bibitem[{Beaug{\'e} and {Ferraz-Mello}(1994)}]{Beauge1994capture}
\bibinfo{author}{Beaug{\'e}, C.}, \bibinfo{author}{{Ferraz-Mello}, S.},
  \bibinfo{year}{1994}.
\newblock \bibinfo{title}{Capture in {{Exterior Mean-Motion Resonances Due}} to
  {{Poynting-Robertson Drag}}}.
\newblock \bibinfo{journal}{Icarus} \bibinfo{volume}{110},
  \bibinfo{pages}{239--260}.
\newblock \DOIprefix\doi{10.1006/icar.1994.1119}.
\bibitem[{Belheouane et~al.(2012)Belheouane, Zaslavsky, {Meyer-Vernet},
  Issautier, Mann and Maksimovic}]{Belheouane2012detection}
\bibinfo{author}{Belheouane, S.}, \bibinfo{author}{Zaslavsky, A.},
  \bibinfo{author}{{Meyer-Vernet}, N.}, \bibinfo{author}{Issautier, K.},
  \bibinfo{author}{Mann, I.}, \bibinfo{author}{Maksimovic, M.},
  \bibinfo{year}{2012}.
\newblock \bibinfo{title}{Detection of {{Interstellar Dust}} with
  {{STEREO}}/{{WAVES}} at 1 {{AU}}}.
\newblock \bibinfo{journal}{Solar Physics} \bibinfo{volume}{281},
  \bibinfo{pages}{501--506}.
\newblock \DOIprefix\doi{10.1007/s11207-012-9995-7}.
\bibitem[{Belton(1966)}]{Belton1966dynamics}
\bibinfo{author}{Belton, M.J.S.}, \bibinfo{year}{1966}.
\newblock \bibinfo{title}{Dynamics of {{Interplanetary Dust}}}.
\newblock \bibinfo{journal}{Science} \bibinfo{volume}{151},
  \bibinfo{pages}{35--44}.
\newblock \DOIprefix\doi{10.1126/science.151.3706.35}.
\bibitem[{Berg and Gerloff(1970)}]{Berg1970orbital}
\bibinfo{author}{Berg, O.E.}, \bibinfo{author}{Gerloff, U.},
  \bibinfo{year}{1970}.
\newblock \bibinfo{title}{Orbital elements of micrometeorites derived from
  {{Pioneer}} 8 measurements}.
\newblock \bibinfo{journal}{Journal of Geophysical Research}
  \bibinfo{volume}{75}, \bibinfo{pages}{6932--6939}.
\newblock \DOIprefix\doi{10.1029/JA075i034p06932}.
\bibitem[{Berg and Gerloff(1971)}]{Berg1971more}
\bibinfo{author}{Berg, O.E.}, \bibinfo{author}{Gerloff, U.},
  \bibinfo{year}{1971}.
\newblock \bibinfo{title}{More than two years of micrometeorite data from two
  {{Pioneer}} satellites}, in: \bibinfo{booktitle}{Space {{Research XI}}}, pp.
  \bibinfo{pages}{225--235}.
\bibitem[{Berg and Gr{\"u}n(1973)}]{Berg1973evidence}
\bibinfo{author}{Berg, O.E.}, \bibinfo{author}{Gr{\"u}n, E.},
  \bibinfo{year}{1973}.
\newblock \bibinfo{title}{Evidence of hyperbolic cosmic dust particles.}, in:
  \bibinfo{booktitle}{Space {{Research XIII}}}, pp.
  \bibinfo{pages}{1047--1055}.
\bibitem[{Berg and Richardson(1969)}]{Berg1969pioneer}
\bibinfo{author}{Berg, O.E.}, \bibinfo{author}{Richardson, F.F.},
  \bibinfo{year}{1969}.
\newblock \bibinfo{title}{The {{Pioneer}} 8 {{Cosmic Dust Experiment}}}.
\newblock \bibinfo{journal}{Review of Scientific Instruments}
  \bibinfo{volume}{40}, \bibinfo{pages}{1333--1337}.
\newblock \DOIprefix\doi{10.1063/1.1683778}.
\bibitem[{Burns et~al.(1979)Burns, Lamy and Soter}]{Burns1979radiation}
\bibinfo{author}{Burns, J.A.}, \bibinfo{author}{Lamy, P.L.},
  \bibinfo{author}{Soter, S.}, \bibinfo{year}{1979}.
\newblock \bibinfo{title}{Radiation forces on small particles in the solar
  system}.
\newblock \bibinfo{journal}{Icarus} \bibinfo{volume}{40},
  \bibinfo{pages}{1--48}.
\newblock \DOIprefix\doi{10.1016/0019-1035(79)90050-2}.
\bibitem[{Close et~al.(2007)Close, Brown, {Campbell-Brown}, Oppenheim and
  Colestock}]{Close2007meteor}
\bibinfo{author}{Close, S.}, \bibinfo{author}{Brown, P.},
  \bibinfo{author}{{Campbell-Brown}, M.}, \bibinfo{author}{Oppenheim, M.},
  \bibinfo{author}{Colestock, P.}, \bibinfo{year}{2007}.
\newblock \bibinfo{title}{Meteor head echo radar data: {{Mass}}\textendash
  velocity selection effects}.
\newblock \bibinfo{journal}{Icarus} \bibinfo{volume}{186},
  \bibinfo{pages}{547--556}.
\newblock \DOIprefix\doi{10.1016/j.icarus.2006.09.007}.
\bibitem[{Consolmagno(1979)}]{Consolmagno1979lorentz}
\bibinfo{author}{Consolmagno, G.}, \bibinfo{year}{1979}.
\newblock \bibinfo{title}{Lorentz scattering of interplanetary dust}.
\newblock \bibinfo{journal}{Icarus} \bibinfo{volume}{38},
  \bibinfo{pages}{398--410}.
\newblock \DOIprefix\doi{10.1016/0019-1035(79)90195-7}.
\bibitem[{Consolmagno(1980)}]{Consolmagno1980influence}
\bibinfo{author}{Consolmagno, G.J.}, \bibinfo{year}{1980}.
\newblock \bibinfo{title}{Influence of the interplanetary magnetic field on
  cometary and primordial dust orbits: {{Applications}} of {{Lorentz}}
  scattering}.
\newblock \bibinfo{journal}{Icarus} \bibinfo{volume}{43},
  \bibinfo{pages}{203--214}.
\newblock \DOIprefix\doi{10.1016/0019-1035(80)90121-9}.
\bibitem[{Dermott et~al.(1994)Dermott, Jayaraman, Xu, Gustafson and
  Liou}]{Dermott1994circumsolar}
\bibinfo{author}{Dermott, S.F.}, \bibinfo{author}{Jayaraman, S.},
  \bibinfo{author}{Xu, Y.L.}, \bibinfo{author}{Gustafson, B.{\AA}.S.},
  \bibinfo{author}{Liou, J.C.}, \bibinfo{year}{1994}.
\newblock \bibinfo{title}{A circumsolar ring of asteroidal dust in resonant
  lock with the {{Earth}}}.
\newblock \bibinfo{journal}{Nature} \bibinfo{volume}{369},
  \bibinfo{pages}{719--723}.
\newblock \URLprefix \url{http://www.nature.com/articles/369719a0},
  \DOIprefix\doi{10.1038/369719a0}.
\bibitem[{Dermott et~al.(2001)Dermott, Kehoe, Grogan, Durda, Jayaraman,
  Kortenkamp and Wyatt}]{Dermott2001orbital}
\bibinfo{author}{Dermott, S.F.}, \bibinfo{author}{Kehoe, T.J.J.},
  \bibinfo{author}{Grogan, K.}, \bibinfo{author}{Durda, D.D.},
  \bibinfo{author}{Jayaraman, S.}, \bibinfo{author}{Kortenkamp, S.J.},
  \bibinfo{author}{Wyatt, M.C.}, \bibinfo{year}{2001}.
\newblock \bibinfo{title}{Orbital {{Evolution}} of {{Interplanetary Dust}}},
  in: \bibinfo{editor}{Appenzeller, I.}, \bibinfo{editor}{G{\"o}rner, G.},
  \bibinfo{editor}{Harwit, M.}, \bibinfo{editor}{Kippenhahn, R.},
  \bibinfo{editor}{Lequeux, J.}, \bibinfo{editor}{Strittmatter, P.A.},
  \bibinfo{editor}{Trimble, V.}, \bibinfo{editor}{Gr{\"u}n, E.},
  \bibinfo{editor}{Gustafson, B.{\AA}.S.}, \bibinfo{editor}{Dermott, S.},
  \bibinfo{editor}{Fechtig, H.} (Eds.), \bibinfo{booktitle}{Interplanetary
  {{Dust}}}. \bibinfo{publisher}{Springer Berlin Heidelberg}, pp.
  \bibinfo{pages}{569--639}.
\newblock \DOIprefix\doi{10.1007/978-3-642-56428-4_12}.
\bibitem[{Dietzel et~al.(1973)Dietzel, Eichhorn, Fechtig, Grun, Hoffmann and
  Kissel}]{Dietzel1973heos}
\bibinfo{author}{Dietzel, H.}, \bibinfo{author}{Eichhorn, G.},
  \bibinfo{author}{Fechtig, H.}, \bibinfo{author}{Grun, E.},
  \bibinfo{author}{Hoffmann, H.J.}, \bibinfo{author}{Kissel, J.},
  \bibinfo{year}{1973}.
\newblock \bibinfo{title}{The {{HEOS}} 2 and {{HELIOS}} micrometeoroid
  experiments}.
\newblock \bibinfo{journal}{Journal of Physics E: Scientific Instruments}
  \bibinfo{volume}{6}, \bibinfo{pages}{209--217}.
\newblock \DOIprefix\doi{10.1088/0022-3735/6/3/008}.
\bibitem[{Dikarev et~al.(2005a)Dikarev, Gr{\"u}n, Baggaley, Galligan, Landgraf
  and Jehn}]{Dikarev2005new}
\bibinfo{author}{Dikarev, V.}, \bibinfo{author}{Gr{\"u}n, E.},
  \bibinfo{author}{Baggaley, J.}, \bibinfo{author}{Galligan, D.},
  \bibinfo{author}{Landgraf, M.}, \bibinfo{author}{Jehn, R.},
  \bibinfo{year}{2005}a.
\newblock \bibinfo{title}{The new {{ESA}} meteoroid model}.
\newblock \bibinfo{journal}{Advances in Space Research} \bibinfo{volume}{35},
  \bibinfo{pages}{1282--1289}.
\newblock \DOIprefix\doi{10.1016/j.asr.2005.05.014}.
\bibitem[{Dikarev et~al.(2005b)Dikarev, Gr{\"u}n, Landgraf and
  Jehn}]{Dikarev2005update}
\bibinfo{author}{Dikarev, V.}, \bibinfo{author}{Gr{\"u}n, E.},
  \bibinfo{author}{Landgraf, M.}, \bibinfo{author}{Jehn, R.},
  \bibinfo{year}{2005}b.
\newblock \bibinfo{title}{{Update of the ESA Meteoroid Model}}, in:
  \bibinfo{booktitle}{{4th European Conference on Space Debris, ESA SP-587}}.
\bibitem[{Divine(1993)}]{Divine1993five}
\bibinfo{author}{Divine, N.}, \bibinfo{year}{1993}.
\newblock \bibinfo{title}{Five populations of interplanetary meteoroids}.
\newblock \bibinfo{journal}{Journal of Geophysical Research: Planets}
  \bibinfo{volume}{98}, \bibinfo{pages}{17029--17048}.
\newblock \DOIprefix\doi{10.1029/93JE01203}.
\bibitem[{Dixon(1975)}]{Dixon1975pioneer}
\bibinfo{author}{Dixon, W.J.}, \bibinfo{year}{1975}.
\newblock \bibinfo{title}{Pioneer spacecraft reliability and performance}.
\newblock \bibinfo{journal}{Acta Astronautica} \bibinfo{volume}{2},
  \bibinfo{pages}{801--817}.
\newblock \URLprefix
  \url{https://www.sciencedirect.com/science/article/pii/0094576575900223},
  \DOIprefix\doi{10.1016/0094-5765(75)90022-3}.
\bibitem[{Dohnanyi(1971)}]{Dohnanyi1971current}
\bibinfo{author}{Dohnanyi, J.S.}, \bibinfo{year}{1971}.
\newblock \bibinfo{title}{Current {{Evolution}} of {{Meteoroids}}}.
\newblock \bibinfo{journal}{International Astronomical Union Colloquium}
  \bibinfo{volume}{13}, \bibinfo{pages}{363--374}.
\newblock \DOIprefix\doi{10.1017/S0252921100049289}.
\bibitem[{Espy(2009)}]{Espy2009dynamics}
\bibinfo{author}{Espy, A.J.}, \bibinfo{year}{2009}.
\newblock \bibinfo{title}{The Dynamics of Asteroidal Dust and Structure of the
  Zodiacal Cloud}.
\newblock Ph.D. thesis. University of Florida.
\bibitem[{Fahr and Ripken(1985)}]{Fahr1985dustplasmagas}
\bibinfo{author}{Fahr, H.J.}, \bibinfo{author}{Ripken, H.W.},
  \bibinfo{year}{1985}.
\newblock \bibinfo{title}{Dust-{{Plasma-Gas Interactions}} in the
  {{Heliosphere}}}.
\newblock \bibinfo{journal}{International Astronomical Union Colloquium}
  \bibinfo{volume}{85}, \bibinfo{pages}{305--323}.
\newblock \DOIprefix\doi{10.1017/S0252921100084840}.
\bibitem[{Fahr et~al.(1995)Fahr, Scherer and Banaszkiewicz}]{Fahr1995evolution}
\bibinfo{author}{Fahr, H.J.}, \bibinfo{author}{Scherer, K.},
  \bibinfo{author}{Banaszkiewicz, M.}, \bibinfo{year}{1995}.
\newblock \bibinfo{title}{The evolution of the zodiacal dust cloud under plasma
  drag and lorentz forces in the latitudinally asymmetric solar wind}.
\newblock \bibinfo{journal}{Planetary and Space Science} \bibinfo{volume}{43},
  \bibinfo{pages}{301--312}.
\newblock \DOIprefix\doi{10.1016/0032-0633(94)00174-P}.
\bibitem[{Fechtig et~al.(2001)Fechtig, Leinert and
  Berg}]{Fechtig2001historical}
\bibinfo{author}{Fechtig, H.}, \bibinfo{author}{Leinert, C.},
  \bibinfo{author}{Berg, O.E.}, \bibinfo{year}{2001}.
\newblock \bibinfo{title}{Historical {{Perspectives}}}, in:
  \bibinfo{booktitle}{Interplanetary {{Dust}}}. Astronomy and {{Astrophysics
  Library}}, pp. \bibinfo{pages}{1--55}.
\newblock \DOIprefix\doi{10.1007/978-3-642-56428-4_1}.
\bibitem[{Fentzke and Janches(2008)}]{Fentzke2008semiempirical}
\bibinfo{author}{Fentzke, J.T.}, \bibinfo{author}{Janches, D.},
  \bibinfo{year}{2008}.
\newblock \bibinfo{title}{A semi-empirical model of the contribution from
  sporadic meteoroid sources on the meteor input function in the {{MLT}}
  observed at {{Arecibo}}}.
\newblock \bibinfo{journal}{Journal of Geophysical Research: Space Physics}
  \bibinfo{volume}{113}.
\newblock \DOIprefix\doi{10.1029/2007JA012531}.
\bibitem[{Freistetter et~al.(2007)Freistetter, Krivov and
  L{\"o}hne}]{Freistetter2007planets}
\bibinfo{author}{Freistetter, F.}, \bibinfo{author}{Krivov, A.V.},
  \bibinfo{author}{L{\"o}hne, T.}, \bibinfo{year}{2007}.
\newblock \bibinfo{title}{Planets of {$\beta$} {{Pictoris}} revisited}.
\newblock \bibinfo{journal}{Astronomy \& Astrophysics} \bibinfo{volume}{466},
  \bibinfo{pages}{389--393}.
\newblock \DOIprefix\doi{10.1051/0004-6361:20066746}.
\bibitem[{G{\"o}ller and Gr{\"u}n(1989)}]{Goller1989calibration}
\bibinfo{author}{G{\"o}ller, J.}, \bibinfo{author}{Gr{\"u}n, E.},
  \bibinfo{year}{1989}.
\newblock \bibinfo{title}{Calibration of the {{Galileo}}/{{Ulysses}} dust
  detectors with different projectile materials and at varying impact angles}.
\newblock \bibinfo{journal}{Planetary and Space Science} \bibinfo{volume}{37},
  \bibinfo{pages}{1197--1206}.
\newblock \DOIprefix\doi{10.1016/0032-0633(89)90014-7}.
\bibitem[{Gomes(1995)}]{Gomes1995resonance}
\bibinfo{author}{Gomes, R.}, \bibinfo{year}{1995}.
\newblock \bibinfo{title}{Resonance trapping and evolution of particles subject
  to poynting-robertson drag: {{Adiabatic}} and non-adiabatic approaches}.
\newblock \bibinfo{journal}{Celestial Mechanics and Dynamical Astronomy}
  \bibinfo{volume}{61}, \bibinfo{pages}{97--113}.
\newblock \DOIprefix\doi{10.1007/BF00051690}.
\bibitem[{Greenstreet et~al.(2012)Greenstreet, Ngo and
  Gladman}]{Greenstreet2012orbital}
\bibinfo{author}{Greenstreet, S.}, \bibinfo{author}{Ngo, H.},
  \bibinfo{author}{Gladman, B.}, \bibinfo{year}{2012}.
\newblock \bibinfo{title}{The orbital distribution of {{Near-Earth Objects}}
  inside {{Earth}}'s orbit}.
\newblock \bibinfo{journal}{Icarus} \bibinfo{volume}{217},
  \bibinfo{pages}{355--366}.
\newblock \DOIprefix\doi{10.1016/j.icarus.2011.11.010}.
\bibitem[{Gr{\"u}n(1981)}]{Grun1981physikalische}
\bibinfo{author}{Gr{\"u}n, E.}, \bibinfo{year}{1981}.
\newblock \bibinfo{title}{Physikalische Und Chemische {{Eigenschaften}} Des
  Interplanetaren {{Staubes}} - {{Messungen}} Des
  {{Mikrometeoritenexperimentes}} Auf {{Helios}}}.
\newblock \bibinfo{type}{Bundesministerium F\"ur {{Forschung}} Und
  {{Technologie}}, {{Forschungsbreicht}}} \bibinfo{number}{W81-034}. {BMFT}.
\bibitem[{Gr{\"u}n et~al.(1973)Gr{\"u}n, Berg and
  Dohnanyi}]{Grun1973reliability}
\bibinfo{author}{Gr{\"u}n, E.}, \bibinfo{author}{Berg, O.E.},
  \bibinfo{author}{Dohnanyi, J.S.}, \bibinfo{year}{1973}.
\newblock \bibinfo{title}{Reliability of cosmic dust data from {{Pioneers}} 8
  and 9}, in: \bibinfo{booktitle}{Space Research {{XVIII}}, {{Proceedings}} of
  the {{Open Meetings}} of the {{Working Groups}} on {{Physical Sciences}} of
  the 15th {{Plenary Meeting}} of {{COSPAR}}}, pp. \bibinfo{pages}{1057--1062}.
\bibitem[{Gr{\"u}n et~al.(1992a)Gr{\"u}n, Fechtig, Hanner, Kissel, Lindblad,
  Linkert, Maas, Morfill and Zook}]{Grun1992galileo}
\bibinfo{author}{Gr{\"u}n, E.}, \bibinfo{author}{Fechtig, H.},
  \bibinfo{author}{Hanner, M.S.}, \bibinfo{author}{Kissel, J.},
  \bibinfo{author}{Lindblad, B.A.}, \bibinfo{author}{Linkert, D.},
  \bibinfo{author}{Maas, D.}, \bibinfo{author}{Morfill, G.E.},
  \bibinfo{author}{Zook, H.A.}, \bibinfo{year}{1992}a.
\newblock \bibinfo{title}{The {{Galileo Dust Detector}}}.
\newblock \bibinfo{journal}{Space Science Reviews} \bibinfo{volume}{60},
  \bibinfo{pages}{317--340}.
\newblock \DOIprefix\doi{10.1007/BF00216860}.
\bibitem[{Gr{\"u}n et~al.(1985a)Gr{\"u}n, Fechtig and Kissel}]{Grun1985orbits}
\bibinfo{author}{Gr{\"u}n, E.}, \bibinfo{author}{Fechtig, H.},
  \bibinfo{author}{Kissel, J.}, \bibinfo{year}{1985}a.
\newblock \bibinfo{title}{Orbits of {{Interplanetary Dust Particles Inside}} 1
  {{AU}} as {{Observed}} by {{Helios}}}, in: \bibinfo{booktitle}{Properties and
  {{Interactions}} of {{Interplanetary Dust}}}, \bibinfo{address}{{Dordrecht}}.
  pp. \bibinfo{pages}{105--111}.
\newblock \DOIprefix\doi{10.1017/S0252921100084438}.
\bibitem[{Gr{\"u}n et~al.(1992b)Gr{\"u}n, Fechtig, Kissel, Linkert, Maas,
  McDonnell, Morfill, Schwehm, Zook and Giese}]{Grun1992ulysses}
\bibinfo{author}{Gr{\"u}n, E.}, \bibinfo{author}{Fechtig, H.},
  \bibinfo{author}{Kissel, J.}, \bibinfo{author}{Linkert, D.},
  \bibinfo{author}{Maas, D.}, \bibinfo{author}{McDonnell, J.a.M.},
  \bibinfo{author}{Morfill, G.E.}, \bibinfo{author}{Schwehm, G.},
  \bibinfo{author}{Zook, H.A.}, \bibinfo{author}{Giese, R.H.},
  \bibinfo{year}{1992}b.
\newblock \bibinfo{title}{The {{Ulysses}} dust experiment}.
\newblock \bibinfo{journal}{Astronomy \& Astrophysics Supplement Series}
  \bibinfo{volume}{92}, \bibinfo{pages}{411--423}.
\bibitem[{Gr{\"u}n et~al.(1980)Gr{\"u}n, Pailer, Fechtig and
  Kissel}]{Grun1980orbital}
\bibinfo{author}{Gr{\"u}n, E.}, \bibinfo{author}{Pailer, N.},
  \bibinfo{author}{Fechtig, H.}, \bibinfo{author}{Kissel, J.},
  \bibinfo{year}{1980}.
\newblock \bibinfo{title}{Orbital and physical characteristics of
  micrometeoroids in the inner solar system as observed by {{Helios}} 1}.
\newblock \bibinfo{journal}{Planetary and Space Science} \bibinfo{volume}{28},
  \bibinfo{pages}{333--349}.
\newblock \DOIprefix\doi{10.1016/0032-0633(80)90022-7}.
\bibitem[{Gr{\"u}n et~al.(1985b)Gr{\"u}n, Zook, Fechtig and
  Giese}]{Grun1985collisional}
\bibinfo{author}{Gr{\"u}n, E.}, \bibinfo{author}{Zook, H.},
  \bibinfo{author}{Fechtig, H.}, \bibinfo{author}{Giese, R.},
  \bibinfo{year}{1985}b.
\newblock \bibinfo{title}{Collisional balance of the meteoritic complex}.
\newblock \bibinfo{journal}{Icarus} \bibinfo{volume}{62},
  \bibinfo{pages}{244--272}.
\newblock \DOIprefix\doi{10.1016/0019-1035(85)90121-6}.
\bibitem[{Gr{\"u}n and Zook(1980)}]{Grun1980dynamics}
\bibinfo{author}{Gr{\"u}n, E.}, \bibinfo{author}{Zook, H.A.},
  \bibinfo{year}{1980}.
\newblock \bibinfo{title}{Dynamics of {{Micrometeoroids}}}.
\newblock \bibinfo{journal}{Symposium - International Astronomical Union}
  \bibinfo{volume}{90}, \bibinfo{pages}{293--298}.
\newblock \DOIprefix\doi{10.1017/S0074180900066912}.
\bibitem[{Gustafson et~al.(2001)Gustafson, Kolokolova, Xu, Greenberg and
  Stognienko}]{Gustafson2001interactions}
\bibinfo{author}{Gustafson, B.{\AA}.S.}, \bibinfo{author}{Kolokolova, L.},
  \bibinfo{author}{Xu, Y.l.}, \bibinfo{author}{Greenberg, J.M.},
  \bibinfo{author}{Stognienko, R.}, \bibinfo{year}{2001}.
\newblock \bibinfo{title}{Interactions with {{Electromagnetic Radiation}}:
  {{Theory}} and {{Laboratory Simulations}}}, in:
  \bibinfo{booktitle}{Interplanetary {{Dust}}}. Astronomy and {{Astrophysics
  Library}}, pp. \bibinfo{pages}{509--567}.
\newblock \URLprefix \url{https://doi.org/10.1007/978-3-642-56428-4_11},
  \DOIprefix\doi{10.1007/978-3-642-56428-4_11}.
\bibitem[{Hamilton et~al.(1996)Hamilton, Gr{\"u}n and
  Baguhl}]{Hamilton1996electromagnetic}
\bibinfo{author}{Hamilton, D.P.}, \bibinfo{author}{Gr{\"u}n, E.},
  \bibinfo{author}{Baguhl, M.}, \bibinfo{year}{1996}.
\newblock \bibinfo{title}{Electromagnetic {{Escape}} of {{Dust}} from the
  {{Solar System}}}, in: \bibinfo{booktitle}{{{IAU Colloquium}}}, pp.
  \bibinfo{pages}{31--34}.
\newblock \DOIprefix\doi{10.1017/S0252921100501225}.
\bibitem[{Herranen(2020)}]{Herranen2020rotational}
\bibinfo{author}{Herranen, J.}, \bibinfo{year}{2020}.
\newblock \bibinfo{title}{Rotational {{Disruption}} of {{Nonspherical Cometary
  Dust Particles}} by {{Radiative Torques}}}.
\newblock \bibinfo{journal}{The Astrophysical Journal} \bibinfo{volume}{893},
  \bibinfo{pages}{109}.
\newblock \DOIprefix\doi{10.3847/1538-4357/ab8009}.
\bibitem[{Hervig et~al.(2022)Hervig, Malaspina, Sterken, Wilson~III, Hunziker
  and Bailey}]{Hervig2022decadal}
\bibinfo{author}{Hervig, M.E.}, \bibinfo{author}{Malaspina, D.},
  \bibinfo{author}{Sterken, V.}, \bibinfo{author}{Wilson~III, L.B.},
  \bibinfo{author}{Hunziker, S.}, \bibinfo{author}{Bailey, S.M.},
  \bibinfo{year}{2022}.
\newblock \bibinfo{title}{Decadal and {{Annual Variations}} in {{Meteoric Flux
  From Ulysses}}, {{Wind}}, and {{SOFIE Observations}}}.
\newblock \bibinfo{journal}{Journal of Geophysical Research: Space Physics}
  \bibinfo{volume}{127}, \bibinfo{pages}{e2022JA030749}.
\newblock \DOIprefix\doi{10.1029/2022JA030749}.
\bibitem[{Hoffmann et~al.(1975a)Hoffmann, Fechtig, Gr{\"u}n and
  Kissel}]{Hoffmann1975first}
\bibinfo{author}{Hoffmann, H.J.}, \bibinfo{author}{Fechtig, H.},
  \bibinfo{author}{Gr{\"u}n, E.}, \bibinfo{author}{Kissel, J.},
  \bibinfo{year}{1975}a.
\newblock \bibinfo{title}{First results of the micrometeoroid experiment s 215
  on the {{HEOS}} 2 satellite}.
\newblock \bibinfo{journal}{Planetary and Space Science} \bibinfo{volume}{23},
  \bibinfo{pages}{215--224}.
\newblock \DOIprefix\doi{10.1016/0032-0633(75)90080-X}.
\bibitem[{Hoffmann et~al.(1975b)Hoffmann, Fechtig, Gr{\"u}n and
  Kissel}]{Hoffmann1975temporal}
\bibinfo{author}{Hoffmann, H.J.}, \bibinfo{author}{Fechtig, H.},
  \bibinfo{author}{Gr{\"u}n, E.}, \bibinfo{author}{Kissel, J.},
  \bibinfo{year}{1975}b.
\newblock \bibinfo{title}{Temporal fluctuations and anisotropy of the
  micrometeoroid flux in the {{Earth-Moon}} system measured by {{HEOS}} 2}.
\newblock \bibinfo{journal}{Planetary and Space Science} \bibinfo{volume}{23},
  \bibinfo{pages}{985--991}.
\newblock \DOIprefix\doi{10.1016/0032-0633(75)90186-5}.
\bibitem[{Howard et~al.(2019)Howard, Vourlidas, Bothmer, Colaninno, DeForest,
  Gallagher, Hall, Hess, Higginson, Korendyke, Kouloumvakos, Lamy, Liewer,
  Linker, Linton, Penteado, Plunkett, Poirier, Raouafi, Rich, Rochus,
  Rouillard, Socker, Stenborg, Thernisien and Viall}]{Howard2019nearsun}
\bibinfo{author}{Howard, R.A.}, \bibinfo{author}{Vourlidas, A.},
  \bibinfo{author}{Bothmer, V.}, \bibinfo{author}{Colaninno, R.C.},
  \bibinfo{author}{DeForest, C.E.}, \bibinfo{author}{Gallagher, B.},
  \bibinfo{author}{Hall, J.R.}, \bibinfo{author}{Hess, P.},
  \bibinfo{author}{Higginson, A.K.}, \bibinfo{author}{Korendyke, C.M.},
  \bibinfo{author}{Kouloumvakos, A.}, \bibinfo{author}{Lamy, P.L.},
  \bibinfo{author}{Liewer, P.C.}, \bibinfo{author}{Linker, J.},
  \bibinfo{author}{Linton, M.}, \bibinfo{author}{Penteado, P.},
  \bibinfo{author}{Plunkett, S.P.}, \bibinfo{author}{Poirier, N.},
  \bibinfo{author}{Raouafi, N.E.}, \bibinfo{author}{Rich, N.},
  \bibinfo{author}{Rochus, P.}, \bibinfo{author}{Rouillard, A.P.},
  \bibinfo{author}{Socker, D.G.}, \bibinfo{author}{Stenborg, G.},
  \bibinfo{author}{Thernisien, A.F.}, \bibinfo{author}{Viall, N.M.},
  \bibinfo{year}{2019}.
\newblock \bibinfo{title}{Near-{{Sun}} observations of an {{F-corona}} decrease
  and {{K-corona}} fine structure}.
\newblock \bibinfo{journal}{Nature} \bibinfo{volume}{576},
  \bibinfo{pages}{232--236}.
\newblock \DOIprefix\doi{10.1038/s41586-019-1807-x}.
\bibitem[{Hunt et~al.(2004)Hunt, Oppenheim, Close, Brown, McKeen and
  Minardi}]{Hunt2004determination}
\bibinfo{author}{Hunt, S.M.}, \bibinfo{author}{Oppenheim, M.},
  \bibinfo{author}{Close, S.}, \bibinfo{author}{Brown, P.G.},
  \bibinfo{author}{McKeen, F.}, \bibinfo{author}{Minardi, M.},
  \bibinfo{year}{2004}.
\newblock \bibinfo{title}{Determination of the meteoroid velocity distribution
  at the {{Earth}} using high-gain radar}.
\newblock \bibinfo{journal}{Icarus} \bibinfo{volume}{168},
  \bibinfo{pages}{34--42}.
\newblock \DOIprefix\doi{10.1016/j.icarus.2003.08.006}.
\bibitem[{Hunziker et~al.(2022)Hunziker, {Moragas-Klostermeyer}, Hillier,
  Fielding, Hornung, Lovett, Armes, Fontanese, James, Hsu, Herrmann, Fechler,
  Poch, Pommerol, Srama, Malaspina and Sterken}]{Hunziker2022impact}
\bibinfo{author}{Hunziker, S.}, \bibinfo{author}{{Moragas-Klostermeyer}, G.},
  \bibinfo{author}{Hillier, J.K.}, \bibinfo{author}{Fielding, L.A.},
  \bibinfo{author}{Hornung, K.}, \bibinfo{author}{Lovett, J.R.},
  \bibinfo{author}{Armes, S.P.}, \bibinfo{author}{Fontanese, J.},
  \bibinfo{author}{James, D.}, \bibinfo{author}{Hsu, H.W.},
  \bibinfo{author}{Herrmann, I.}, \bibinfo{author}{Fechler, N.},
  \bibinfo{author}{Poch, O.}, \bibinfo{author}{Pommerol, A.},
  \bibinfo{author}{Srama, R.}, \bibinfo{author}{Malaspina, D.},
  \bibinfo{author}{Sterken, V.J.}, \bibinfo{year}{2022}.
\newblock \bibinfo{title}{Impact ionization dust detection with compact, hollow
  and fluffy dust analogs}.
\newblock \bibinfo{journal}{Planetary and Space Science} \bibinfo{volume}{220},
  \bibinfo{pages}{105536}.
\newblock \DOIprefix\doi{10.1016/j.pss.2022.105536}.
\bibitem[{Igenbergs et~al.(1991a)Igenbergs, H{\"u}dephol, Uesugi, Hayashi,
  Svedhem, Iglseder, Koller, Glasmachers, Gr{\"u}n, Schwehm, Mizutani,
  Yamamoto, Fujimura, Ishii, Araki, Yamakoshi and
  Nogami}]{Igenbergs1991present}
\bibinfo{author}{Igenbergs, E.}, \bibinfo{author}{H{\"u}dephol, A.},
  \bibinfo{author}{Uesugi, K.}, \bibinfo{author}{Hayashi, T.},
  \bibinfo{author}{Svedhem, H.}, \bibinfo{author}{Iglseder, H.},
  \bibinfo{author}{Koller, G.}, \bibinfo{author}{Glasmachers, A.},
  \bibinfo{author}{Gr{\"u}n, E.}, \bibinfo{author}{Schwehm, G.},
  \bibinfo{author}{Mizutani, H.}, \bibinfo{author}{Yamamoto, T.},
  \bibinfo{author}{Fujimura, A.}, \bibinfo{author}{Ishii, N.},
  \bibinfo{author}{Araki, H.}, \bibinfo{author}{Yamakoshi, K.},
  \bibinfo{author}{Nogami, K.}, \bibinfo{year}{1991}a.
\newblock \bibinfo{title}{The {{Present Status}} of the {{Munich Dust Counter
  Experiment}} on {{Board}} of the {{Hiten Spacecraft}}}, in:
  \bibinfo{editor}{{Levasseur-Regourd}, A.C.}, \bibinfo{editor}{Hasegawa, H.}
  (Eds.), \bibinfo{booktitle}{Origin and {{Evolution}} of {{Interplanetary
  Dust}}}, \bibinfo{address}{{Dordrecht}}. pp. \bibinfo{pages}{15--20}.
\newblock \DOIprefix\doi{10.1007/978-94-011-3640-2_3}.
\bibitem[{Igenbergs et~al.(1991b)Igenbergs, H{\"u}depohl, Uesugi, Hayashi,
  Svedhem, Iglseder, Koller, Glasmachers, Gr{\"u}n, Schwehm, Mizutani,
  Yamamoto, Fujimura, Ishii, Araki, Yamakoshi and Nogami}]{Igenbergs1991munich}
\bibinfo{author}{Igenbergs, E.}, \bibinfo{author}{H{\"u}depohl, A.},
  \bibinfo{author}{Uesugi, K.}, \bibinfo{author}{Hayashi, T.},
  \bibinfo{author}{Svedhem, H.}, \bibinfo{author}{Iglseder, H.},
  \bibinfo{author}{Koller, G.}, \bibinfo{author}{Glasmachers, A.},
  \bibinfo{author}{Gr{\"u}n, E.}, \bibinfo{author}{Schwehm, G.},
  \bibinfo{author}{Mizutani, H.}, \bibinfo{author}{Yamamoto, T.},
  \bibinfo{author}{Fujimura, A.}, \bibinfo{author}{Ishii, N.},
  \bibinfo{author}{Araki, H.}, \bibinfo{author}{Yamakoshi, K.},
  \bibinfo{author}{Nogami, K.}, \bibinfo{year}{1991}b.
\newblock \bibinfo{title}{The {{Munich Dust Counter}} \textemdash{} {{A Cosmic
  Dust Experiment}} on {{Board}} of the {{Muses-A Mission}} of {{Japan}}}, in:
  \bibinfo{booktitle}{Origin and {{Evolution}} of {{Interplanetary Dust}}}, pp.
  \bibinfo{pages}{45--48}.
\newblock \DOIprefix\doi{10.1007/978-94-011-3640-2_9}.
\bibitem[{Iglseder et~al.(1993)Iglseder, M{\"u}nzenmayer, Svedhem and
  Gr{\"u}n}]{Iglseder1993cosmic}
\bibinfo{author}{Iglseder, H.}, \bibinfo{author}{M{\"u}nzenmayer, R.},
  \bibinfo{author}{Svedhem, H.}, \bibinfo{author}{Gr{\"u}n, E.},
  \bibinfo{year}{1993}.
\newblock \bibinfo{title}{Cosmic dust and space debris measurements with the
  {{Munich}} dust counter on board the satellites hiten and brem-sat}.
\newblock \bibinfo{journal}{Advances in Space Research} \bibinfo{volume}{13},
  \bibinfo{pages}{129--132}.
\newblock \DOIprefix\doi{10.1016/0273-1177(93)90579-Z}.
\bibitem[{Iglseder et~al.(1996)Iglseder, Uesugi and
  Svedhem}]{Iglseder1996cosmic}
\bibinfo{author}{Iglseder, H.}, \bibinfo{author}{Uesugi, K.},
  \bibinfo{author}{Svedhem, H.}, \bibinfo{year}{1996}.
\newblock \bibinfo{title}{Cosmic dust measurements in lunar orbit}.
\newblock \bibinfo{journal}{Advances in Space Research} \bibinfo{volume}{17},
  \bibinfo{pages}{177--182}.
\newblock \DOIprefix\doi{10.1016/0273-1177(95)00777-C}.
\bibitem[{Ishimoto(2000)}]{Ishimoto2000modeling}
\bibinfo{author}{Ishimoto, H.}, \bibinfo{year}{2000}.
\newblock \bibinfo{title}{Modeling the number density distribution of
  interplanetary dust on the ecliptic plane within {{5AU}} of the {{Sun}}}.
\newblock \bibinfo{journal}{Astronomy and Astrophysics} \bibinfo{volume}{362},
  \bibinfo{pages}{1158--1173}.
\bibitem[{Isobe and {Sateesh-Kumar}(1993)}]{Isobe1993effect}
\bibinfo{author}{Isobe, S.}, \bibinfo{author}{{Sateesh-Kumar}, A.},
  \bibinfo{year}{1993}.
\newblock \bibinfo{title}{An effect of {{Lorenz}} force on interplanetary
  dust}, in: \bibinfo{booktitle}{Meteoroids and Their Parent Bodies}, p.
  \bibinfo{pages}{381}.
\bibitem[{Janches and Chau(2005)}]{Janches2005observed}
\bibinfo{author}{Janches, D.}, \bibinfo{author}{Chau, J.L.},
  \bibinfo{year}{2005}.
\newblock \bibinfo{title}{Observed diurnal and seasonal behavior of the
  micrometeor flux using the {{Arecibo}} and {{Jicamarca}} radars}.
\newblock \bibinfo{journal}{Journal of Atmospheric and Solar-Terrestrial
  Physics} \bibinfo{volume}{67}, \bibinfo{pages}{1196--1210}.
\newblock \DOIprefix\doi{10.1016/j.jastp.2005.06.011}.
\bibitem[{Janches et~al.(2008)Janches, Close and
  Fentzke}]{Janches2008comparison}
\bibinfo{author}{Janches, D.}, \bibinfo{author}{Close, S.},
  \bibinfo{author}{Fentzke, J.T.}, \bibinfo{year}{2008}.
\newblock \bibinfo{title}{A comparison of detection sensitivity between
  {{ALTAIR}} and {{Arecibo}} meteor observations: {{Can}} high power and large
  aperture radars detect low velocity meteor head-echoes}.
\newblock \bibinfo{journal}{Icarus} \bibinfo{volume}{193},
  \bibinfo{pages}{105--111}.
\newblock \DOIprefix\doi{10.1016/j.icarus.2007.08.022}.
\bibitem[{Janches et~al.(2001)Janches, Meisel and Mathews}]{Janches2001orbital}
\bibinfo{author}{Janches, D.}, \bibinfo{author}{Meisel, D.D.},
  \bibinfo{author}{Mathews, J.D.}, \bibinfo{year}{2001}.
\newblock \bibinfo{title}{Orbital {{Properties}} of the {{Arecibo
  Micrometeoroids}} at {{Earth Interception}}}.
\newblock \bibinfo{journal}{Icarus} \bibinfo{volume}{150},
  \bibinfo{pages}{206--218}.
\newblock \DOIprefix\doi{10.1006/icar.2000.6575}.
\bibitem[{Janches et~al.(2003)Janches, Nolan, Meisel, Mathews, Zhou and
  Moser}]{Janches2003geocentric}
\bibinfo{author}{Janches, D.}, \bibinfo{author}{Nolan, M.C.},
  \bibinfo{author}{Meisel, D.D.}, \bibinfo{author}{Mathews, J.D.},
  \bibinfo{author}{Zhou, Q.H.}, \bibinfo{author}{Moser, D.E.},
  \bibinfo{year}{2003}.
\newblock \bibinfo{title}{On the geocentric micrometeor velocity distribution}.
\newblock \bibinfo{journal}{Journal of Geophysical Research: Space Physics}
  \bibinfo{volume}{108}.
\newblock \DOIprefix\doi{10.1029/2002JA009789}.
\bibitem[{Janches et~al.(2004)Janches, Nolan and Sulzer}]{Janches2004radiant}
\bibinfo{author}{Janches, D.}, \bibinfo{author}{Nolan, M.C.},
  \bibinfo{author}{Sulzer, M.}, \bibinfo{year}{2004}.
\newblock \bibinfo{title}{Radiant measurement accuracy of micrometeors detected
  by the {{Arecibo}} 430 {{MHz Dual-Beam Radar}}}.
\newblock \bibinfo{journal}{Atmospheric Chemistry and Physics}
  \bibinfo{volume}{4}, \bibinfo{pages}{621--626}.
\newblock \DOIprefix\doi{10.5194/acp-4-621-2004}.
\bibitem[{Kehoe et~al.(2007)Kehoe, Dermott and
  {Mahoney-Hopping}}]{Kehoe2007effect}
\bibinfo{author}{Kehoe, T.J.J.}, \bibinfo{author}{Dermott, S.F.},
  \bibinfo{author}{{Mahoney-Hopping}, L.M.}, \bibinfo{year}{2007}.
\newblock \bibinfo{title}{The {{Effect}} of {{Inter-Particle Collisions}} on
  the {{Dynamical Evolution}} of {{Asteroidal Dust}} and the {{Structure}} of
  the {{Zodiacal Cloud}}}.
\newblock \bibinfo{journal}{ESA-SP} \bibinfo{volume}{643},
  \bibinfo{pages}{81--85}.
\bibitem[{Kla{\v c}ka et~al.(2020)Kla{\v c}ka, Nagy and Jur{\v
  c}i}]{Klacka2020dust}
\bibinfo{author}{Kla{\v c}ka, J.}, \bibinfo{author}{Nagy, R.},
  \bibinfo{author}{Jur{\v c}i, M.}, \bibinfo{year}{2020}.
\newblock \bibinfo{title}{Dust grains in mean motion orbital resonances with a
  planet}.
\newblock \bibinfo{journal}{Planetary and Space Science} \bibinfo{volume}{182},
  \bibinfo{pages}{104852}.
\newblock \DOIprefix\doi{10.1016/j.pss.2020.104852}.
\bibitem[{Kobayashi et~al.(2009)Kobayashi, Watanabe, Kimura and
  Yamamoto}]{Kobayashi2009dust}
\bibinfo{author}{Kobayashi, H.}, \bibinfo{author}{Watanabe, S.i.},
  \bibinfo{author}{Kimura, H.}, \bibinfo{author}{Yamamoto, T.},
  \bibinfo{year}{2009}.
\newblock \bibinfo{title}{Dust ring formation due to sublimation of dust grains
  drifting radially inward by the {{Poynting}}\textendash{{Robertson}} drag:
  {{An}} analytical model}.
\newblock \bibinfo{journal}{Icarus} \bibinfo{volume}{201},
  \bibinfo{pages}{395--405}.
\newblock \DOIprefix\doi{10.1016/j.icarus.2009.01.002}.
\bibitem[{Kobayashi et~al.(2018)Kobayashi, Kr{\"u}ger, Senshu, Wada, Okudaira,
  Sasaki and Kimura}]{Kobayashi2018situa}
\bibinfo{author}{Kobayashi, M.}, \bibinfo{author}{Kr{\"u}ger, H.},
  \bibinfo{author}{Senshu, H.}, \bibinfo{author}{Wada, K.},
  \bibinfo{author}{Okudaira, O.}, \bibinfo{author}{Sasaki, S.},
  \bibinfo{author}{Kimura, H.}, \bibinfo{year}{2018}.
\newblock \bibinfo{title}{In situ observations of dust particles in {{Martian}}
  dust belts using a large-sensitive-area dust sensor}.
\newblock \bibinfo{journal}{Planetary and Space Science} \bibinfo{volume}{156},
  \bibinfo{pages}{41--46}.
\newblock \DOIprefix\doi{10.1016/j.pss.2017.12.011}.
\bibitem[{Kobayashi et~al.(2020)Kobayashi, Shibata, Nogami, Fujii, Hasegawa,
  Hirabayashi, Hirai, Iwai, Kimura, Miyachi, Nakamura, Ohashi, Sasaki, Takechi,
  Yano, Kr{\"u}ger, Lohse, Srama, Strub and Gr{\"u}n}]{Kobayashi2020mercury}
\bibinfo{author}{Kobayashi, M.}, \bibinfo{author}{Shibata, H.},
  \bibinfo{author}{Nogami, K.}, \bibinfo{author}{Fujii, M.},
  \bibinfo{author}{Hasegawa, S.}, \bibinfo{author}{Hirabayashi, M.},
  \bibinfo{author}{Hirai, T.}, \bibinfo{author}{Iwai, T.},
  \bibinfo{author}{Kimura, H.}, \bibinfo{author}{Miyachi, T.},
  \bibinfo{author}{Nakamura, M.}, \bibinfo{author}{Ohashi, H.},
  \bibinfo{author}{Sasaki, S.}, \bibinfo{author}{Takechi, S.},
  \bibinfo{author}{Yano, H.}, \bibinfo{author}{Kr{\"u}ger, H.},
  \bibinfo{author}{Lohse, A.K.}, \bibinfo{author}{Srama, R.},
  \bibinfo{author}{Strub, P.}, \bibinfo{author}{Gr{\"u}n, E.},
  \bibinfo{year}{2020}.
\newblock \bibinfo{title}{Mercury {{Dust Monitor}} ({{MDM}}) {{Onboard}} the
  {{Mio Orbiter}} of the {{BepiColombo Mission}}}.
\newblock \bibinfo{journal}{Space Science Reviews} \bibinfo{volume}{216},
  \bibinfo{pages}{144}.
\newblock \DOIprefix\doi{10.1007/s11214-020-00775-7}.
\bibitem[{Kral et~al.(2017)Kral, Krivov, Defr{\`e}re, {van Lieshout}, Bonsor,
  Augereau, Th{\'e}bault, Ertel, Lebreton and Absil}]{Kral2017exozodiacal}
\bibinfo{author}{Kral, Q.}, \bibinfo{author}{Krivov, A.V.},
  \bibinfo{author}{Defr{\`e}re, D.}, \bibinfo{author}{{van Lieshout}, R.},
  \bibinfo{author}{Bonsor, A.}, \bibinfo{author}{Augereau, J.C.},
  \bibinfo{author}{Th{\'e}bault, P.}, \bibinfo{author}{Ertel, S.},
  \bibinfo{author}{Lebreton, J.}, \bibinfo{author}{Absil, O.},
  \bibinfo{year}{2017}.
\newblock \bibinfo{title}{Exozodiacal clouds: Hot and warm dust around main
  sequence stars}.
\newblock \bibinfo{journal}{Astronomical Review} \bibinfo{volume}{13},
  \bibinfo{pages}{69--111}.
\newblock \DOIprefix\doi{10.1080/21672857.2017.1353202}.
\bibitem[{Kral et~al.(2013)Kral, Th{\'e}bault and Charnoz}]{Kral2013lidtdd}
\bibinfo{author}{Kral, Q.}, \bibinfo{author}{Th{\'e}bault, P.},
  \bibinfo{author}{Charnoz, S.}, \bibinfo{year}{2013}.
\newblock \bibinfo{title}{{{LIDT-DD}}: {{A}} new self-consistent debris disc
  model that includes radiation pressure and couples dynamical and collisional
  evolution}.
\newblock \bibinfo{journal}{Astronomy \& Astrophysics} \bibinfo{volume}{558},
  \bibinfo{pages}{A121}.
\newblock \URLprefix
  \url{https://www.aanda.org/articles/aa/abs/2013/10/aa21398-13/aa21398-13.html},
  \DOIprefix\doi{10.1051/0004-6361/201321398}.
\bibitem[{Krivov et~al.(1998)Krivov, Kimura and Mann}]{Krivov1998dynamics}
\bibinfo{author}{Krivov, A.}, \bibinfo{author}{Kimura, H.},
  \bibinfo{author}{Mann, I.}, \bibinfo{year}{1998}.
\newblock \bibinfo{title}{Dynamics of {{Dust}} near the {{Sun}}}.
\newblock \bibinfo{journal}{Icarus} \bibinfo{volume}{134},
  \bibinfo{pages}{311--327}.
\newblock \DOIprefix\doi{10.1006/icar.1998.5949}.
\bibitem[{Krivov(2010)}]{Krivov2010debris}
\bibinfo{author}{Krivov, A.V.}, \bibinfo{year}{2010}.
\newblock \bibinfo{title}{Debris disks: Seeing dust, thinking of planetesimals
  and planets}.
\newblock \bibinfo{journal}{Research in Astronomy and Astrophysics}
  \bibinfo{volume}{10}, \bibinfo{pages}{383}.
\newblock \DOIprefix\doi{10.1088/1674-4527/10/5/001}.
\bibitem[{Krivov et~al.(2006)Krivov, L{\"o}hne and Srem{\v
  c}evi{\'c}}]{Krivov2006dust}
\bibinfo{author}{Krivov, A.V.}, \bibinfo{author}{L{\"o}hne, T.},
  \bibinfo{author}{Srem{\v c}evi{\'c}, M.}, \bibinfo{year}{2006}.
\newblock \bibinfo{title}{Dust distributions in debris disks: Effects of
  gravity, radiation pressure and collisions}.
\newblock \bibinfo{journal}{Astronomy \& Astrophysics} \bibinfo{volume}{455},
  \bibinfo{pages}{509--519}.
\newblock \DOIprefix\doi{10.1051/0004-6361:20064907}.
\bibitem[{Krivov et~al.(2000)Krivov, Mann and Krivova}]{Krivov2000size}
\bibinfo{author}{Krivov, A.V.}, \bibinfo{author}{Mann, I.},
  \bibinfo{author}{Krivova, N.A.}, \bibinfo{year}{2000}.
\newblock \bibinfo{title}{Size distributions of dust in circumstellar debris
  discs}.
\newblock \bibinfo{journal}{Astronomy \& Astrophysics} \bibinfo{volume}{362},
  \bibinfo{pages}{1121--1137}.
\bibitem[{Krivov et~al.(2005)Krivov, Srem{\v c}evi{\'c} and
  Spahn}]{Krivov2005evolution}
\bibinfo{author}{Krivov, A.V.}, \bibinfo{author}{Srem{\v c}evi{\'c}, M.},
  \bibinfo{author}{Spahn, F.}, \bibinfo{year}{2005}.
\newblock \bibinfo{title}{Evolution of a {{Keplerian}} disk of colliding and
  fragmenting particles: A kinetic model with application to the
  {{Edgeworth}}\textendash{{Kuiper}} belt}.
\newblock \bibinfo{journal}{Icarus} \bibinfo{volume}{174},
  \bibinfo{pages}{105--134}.
\newblock \DOIprefix\doi{10.1016/j.icarus.2004.10.003}.
\bibitem[{Krivova et~al.(2000a)Krivova, Krivov and Mann}]{Krivova2000disk}
\bibinfo{author}{Krivova, N.A.}, \bibinfo{author}{Krivov, A.V.},
  \bibinfo{author}{Mann, I.}, \bibinfo{year}{2000}a.
\newblock \bibinfo{title}{The {{Disk}} of {$\beta$} {{Pictoris}} in the
  {{Light}} of {{Polarimetric Data}}}.
\newblock \bibinfo{journal}{The Astrophysical Journal} \bibinfo{volume}{539},
  \bibinfo{pages}{424--434}.
\newblock \DOIprefix\doi{10.1086/309214}.
\bibitem[{Krivova et~al.(2000b)Krivova, Krivov and Mann}]{Krivova2000size}
\bibinfo{author}{Krivova, N.A.}, \bibinfo{author}{Krivov, A.V.},
  \bibinfo{author}{Mann, I.}, \bibinfo{year}{2000}b.
\newblock \bibinfo{title}{Size {{Distribution}} of {{Dust}} in the {{Disk}} of
  {$\beta$} {{Pictoris}}}, in: \bibinfo{booktitle}{Disks, {{Planetesimals}},
  and {{Planets}}, {{ASP Conference Series}}}, p. \bibinfo{pages}{387}.
\newblock \URLprefix
  \url{https://ui.adsabs.harvard.edu/abs/2000ASPC..219..387K}.
\bibitem[{Kr{\"u}ger et~al.(1999)Kr{\"u}ger, Krivov, Hamilton and
  Gr{\"u}n}]{Kruger1999detection}
\bibinfo{author}{Kr{\"u}ger, H.}, \bibinfo{author}{Krivov, A.V.},
  \bibinfo{author}{Hamilton, D.P.}, \bibinfo{author}{Gr{\"u}n, E.},
  \bibinfo{year}{1999}.
\newblock \bibinfo{title}{Detection of an impact-generated dust cloud around
  {{Ganymede}}}.
\newblock \bibinfo{journal}{Nature} \bibinfo{volume}{399},
  \bibinfo{pages}{558--560}.
\newblock \DOIprefix\doi{10.1038/21136}.
\bibitem[{Kr{\"u}ger et~al.(2007)Kr{\"u}ger, Landgraf, Altobelli and
  Gr{\"u}n}]{Kruger2007interstellar}
\bibinfo{author}{Kr{\"u}ger, H.}, \bibinfo{author}{Landgraf, M.},
  \bibinfo{author}{Altobelli, N.}, \bibinfo{author}{Gr{\"u}n, E.},
  \bibinfo{year}{2007}.
\newblock \bibinfo{title}{Interstellar {{Dust}} in the {{Solar System}}}.
\newblock \bibinfo{journal}{Space Science Reviews} \bibinfo{volume}{130},
  \bibinfo{pages}{401--408}.
\newblock \DOIprefix\doi{10.1007/s11214-007-9181-7}.
\bibitem[{Kr{\"u}ger et~al.(2019)Kr{\"u}ger, Strub, Srama, Kobayashi, Arai,
  Kimura, Hirai, {Moragas-Klostermeyer}, Altobelli, Sterken, Agarwal, Sommer
  and Gr{\"u}n}]{Kruger2019modelling}
\bibinfo{author}{Kr{\"u}ger, H.}, \bibinfo{author}{Strub, P.},
  \bibinfo{author}{Srama, R.}, \bibinfo{author}{Kobayashi, M.},
  \bibinfo{author}{Arai, T.}, \bibinfo{author}{Kimura, H.},
  \bibinfo{author}{Hirai, T.}, \bibinfo{author}{{Moragas-Klostermeyer}, G.},
  \bibinfo{author}{Altobelli, N.}, \bibinfo{author}{Sterken, V.J.},
  \bibinfo{author}{Agarwal, J.}, \bibinfo{author}{Sommer, M.},
  \bibinfo{author}{Gr{\"u}n, E.}, \bibinfo{year}{2019}.
\newblock \bibinfo{title}{Modelling {{DESTINY}}+ interplanetary and
  interstellar dust measurements en route to the active asteroid (3200)
  {{Phaethon}}}.
\newblock \bibinfo{journal}{Planetary and Space Science} \bibinfo{volume}{172},
  \bibinfo{pages}{22--42}.
\newblock \DOIprefix\doi{10.1016/j.pss.2019.04.005}.
\bibitem[{Leinert(1985)}]{Leinert1985dynamics}
\bibinfo{author}{Leinert, C.}, \bibinfo{year}{1985}.
\newblock \bibinfo{title}{Dynamics and {{Spatial Distribution}} of
  {{Interplanetary Dust}}}, in: \bibinfo{booktitle}{Properties and
  {{Interactions}} of {{Interplanetary Dust}}, {{Proc}}. of the {{IAU Colloq}}.
  85}, pp. \bibinfo{pages}{369--375}.
\newblock \DOIprefix\doi{10.1017/S0252921100084931}.
\bibitem[{Lhotka et~al.(2016)Lhotka, Bourdin and Narita}]{Lhotka2016charged}
\bibinfo{author}{Lhotka, C.}, \bibinfo{author}{Bourdin, P.},
  \bibinfo{author}{Narita, Y.}, \bibinfo{year}{2016}.
\newblock \bibinfo{title}{{{CHARGED DUST GRAIN DYNAMICS SUBJECT TO SOLAR
  WIND}}, {{POYNTING}}\textendash{{ROBERTSON DRAG}}, {{AND THE INTERPLANETARY
  MAGNETIC FIELD}}}.
\newblock \bibinfo{journal}{The Astrophysical Journal} \bibinfo{volume}{828},
  \bibinfo{pages}{10}.
\newblock \DOIprefix\doi{10.3847/0004-637X/828/1/10}.
\bibitem[{Malaspina et~al.(2014)Malaspina, Hor{\'a}nyi, Zaslavsky, Goetz,
  Wilson~III and Kersten}]{Malaspina2014interplanetary}
\bibinfo{author}{Malaspina, D.M.}, \bibinfo{author}{Hor{\'a}nyi, M.},
  \bibinfo{author}{Zaslavsky, A.}, \bibinfo{author}{Goetz, K.},
  \bibinfo{author}{Wilson~III, L.B.}, \bibinfo{author}{Kersten, K.},
  \bibinfo{year}{2014}.
\newblock \bibinfo{title}{Interplanetary and interstellar dust observed by the
  {{Wind}}/{{WAVES}} electric field instrument}.
\newblock \bibinfo{journal}{Geophysical Research Letters} \bibinfo{volume}{41},
  \bibinfo{pages}{266--272}.
\newblock \DOIprefix\doi{10.1002/2013GL058786}.
\bibitem[{Malaspina et~al.(2020)Malaspina, Szalay, Pokorn{\'y}, Page, Bale,
  Bonnell, {de Wit}, Goetz, Goodrich, Harvey, MacDowall and
  Pulupa}]{Malaspina2020situ}
\bibinfo{author}{Malaspina, D.M.}, \bibinfo{author}{Szalay, J.R.},
  \bibinfo{author}{Pokorn{\'y}, P.}, \bibinfo{author}{Page, B.},
  \bibinfo{author}{Bale, S.D.}, \bibinfo{author}{Bonnell, J.W.},
  \bibinfo{author}{{de Wit}, T.D.}, \bibinfo{author}{Goetz, K.},
  \bibinfo{author}{Goodrich, K.}, \bibinfo{author}{Harvey, P.R.},
  \bibinfo{author}{MacDowall, R.J.}, \bibinfo{author}{Pulupa, M.},
  \bibinfo{year}{2020}.
\newblock \bibinfo{title}{In {{Situ Observations}} of {{Interplanetary Dust
  Variability}} in the {{Inner Heliosphere}}}.
\newblock \bibinfo{journal}{The Astrophysical Journal} \bibinfo{volume}{892},
  \bibinfo{pages}{115}.
\newblock \DOIprefix\doi{10.3847/1538-4357/ab799b}.
\bibitem[{Mann and Czechowski(2021)}]{Mann2021dust}
\bibinfo{author}{Mann, I.}, \bibinfo{author}{Czechowski, A.},
  \bibinfo{year}{2021}.
\newblock \bibinfo{title}{Dust observations from {{Parker Solar Probe}}: Dust
  ejection from the inner {{Solar System}}}.
\newblock \bibinfo{journal}{Astronomy \& Astrophysics} \bibinfo{volume}{650},
  \bibinfo{pages}{A29}.
\newblock \DOIprefix\doi{10.1051/0004-6361/202039362}.
\bibitem[{Mann et~al.(2004)Mann, Kimura, Biesecker, Tsurutani, Gr{\"u}n,
  McKibben, Liou, MacQueen, Mukai, Guhathakurta and Lamy}]{Mann2004dust}
\bibinfo{author}{Mann, I.}, \bibinfo{author}{Kimura, H.},
  \bibinfo{author}{Biesecker, D.A.}, \bibinfo{author}{Tsurutani, B.T.},
  \bibinfo{author}{Gr{\"u}n, E.}, \bibinfo{author}{McKibben, R.B.},
  \bibinfo{author}{Liou, J.C.}, \bibinfo{author}{MacQueen, R.M.},
  \bibinfo{author}{Mukai, T.}, \bibinfo{author}{Guhathakurta, M.},
  \bibinfo{author}{Lamy, P.}, \bibinfo{year}{2004}.
\newblock \bibinfo{title}{Dust {{Near The Sun}}}.
\newblock \bibinfo{journal}{Space Science Reviews} \bibinfo{volume}{110},
  \bibinfo{pages}{269--305}.
\newblock \DOIprefix\doi{10.1023/B:SPAC.0000023440.82735.ba}.
\bibitem[{Mann et~al.(2006)Mann, K{\"o}hler, Kimura, Cechowski and
  Minato}]{Mann2006dust}
\bibinfo{author}{Mann, I.}, \bibinfo{author}{K{\"o}hler, M.},
  \bibinfo{author}{Kimura, H.}, \bibinfo{author}{Cechowski, A.},
  \bibinfo{author}{Minato, T.}, \bibinfo{year}{2006}.
\newblock \bibinfo{title}{Dust in the solar system and in extra-solar planetary
  systems}.
\newblock \bibinfo{journal}{The Astronomy and Astrophysics Review}
  \bibinfo{volume}{13}, \bibinfo{pages}{159--228}.
\newblock \DOIprefix\doi{10.1007/s00159-006-0028-0}.
\bibitem[{Mann et~al.(2019)Mann, Nouz{\'a}k, Vaverka, Antonsen, Fredriksen,
  Issautier, Malaspina, {Meyer-Vernet}, Pavl{\r{u}}, Sternovsky, Stude, Ye and
  Zaslavsky}]{Mann2019dust}
\bibinfo{author}{Mann, I.}, \bibinfo{author}{Nouz{\'a}k, L.},
  \bibinfo{author}{Vaverka, J.}, \bibinfo{author}{Antonsen, T.},
  \bibinfo{author}{Fredriksen, {\AA}.}, \bibinfo{author}{Issautier, K.},
  \bibinfo{author}{Malaspina, D.}, \bibinfo{author}{{Meyer-Vernet}, N.},
  \bibinfo{author}{Pavl{\r{u}}, J.}, \bibinfo{author}{Sternovsky, Z.},
  \bibinfo{author}{Stude, J.}, \bibinfo{author}{Ye, S.},
  \bibinfo{author}{Zaslavsky, A.}, \bibinfo{year}{2019}.
\newblock \bibinfo{title}{Dust observations with antenna measurements and its
  prospects for observations with {{Parker Solar Probe}} and {{Solar
  Orbiter}}}.
\newblock \bibinfo{journal}{Annales Geophysicae} \bibinfo{volume}{37},
  \bibinfo{pages}{1121--1140}.
\newblock \DOIprefix\doi{10.5194/angeo-37-1121-2019}.
\bibitem[{McComas et~al.(2018)McComas, Christian, Schwadron, Fox, Westlake,
  Allegrini, Baker, Biesecker, Bzowski, Clark, Cohen, Cohen, Dayeh, Decker, {de
  Nolfo}, Desai, Ebert, Elliott, Fahr, Frisch, Funsten, Fuselier, Galli,
  Galvin, Giacalone, Gkioulidou, Guo, Horanyi, Isenberg, Janzen, Kistler,
  Korreck, Kubiak, Kucharek, Larsen, Leske, Lugaz, Luhmann, Matthaeus,
  Mitchell, Moebius, Ogasawara, Reisenfeld, Richardson, Russell, Sok{\'o}{\l},
  Spence, Skoug, Sternovsky, Swaczyna, Szalay, Tokumaru, Wiedenbeck, Wurz, Zank
  and Zirnstein}]{McComas2018interstellar}
\bibinfo{author}{McComas, D.J.}, \bibinfo{author}{Christian, E.R.},
  \bibinfo{author}{Schwadron, N.A.}, \bibinfo{author}{Fox, N.},
  \bibinfo{author}{Westlake, J.}, \bibinfo{author}{Allegrini, F.},
  \bibinfo{author}{Baker, D.N.}, \bibinfo{author}{Biesecker, D.},
  \bibinfo{author}{Bzowski, M.}, \bibinfo{author}{Clark, G.},
  \bibinfo{author}{Cohen, C.M.S.}, \bibinfo{author}{Cohen, I.},
  \bibinfo{author}{Dayeh, M.A.}, \bibinfo{author}{Decker, R.},
  \bibinfo{author}{{de Nolfo}, G.A.}, \bibinfo{author}{Desai, M.I.},
  \bibinfo{author}{Ebert, R.W.}, \bibinfo{author}{Elliott, H.A.},
  \bibinfo{author}{Fahr, H.}, \bibinfo{author}{Frisch, P.C.},
  \bibinfo{author}{Funsten, H.O.}, \bibinfo{author}{Fuselier, S.A.},
  \bibinfo{author}{Galli, A.}, \bibinfo{author}{Galvin, A.B.},
  \bibinfo{author}{Giacalone, J.}, \bibinfo{author}{Gkioulidou, M.},
  \bibinfo{author}{Guo, F.}, \bibinfo{author}{Horanyi, M.},
  \bibinfo{author}{Isenberg, P.}, \bibinfo{author}{Janzen, P.},
  \bibinfo{author}{Kistler, L.M.}, \bibinfo{author}{Korreck, K.},
  \bibinfo{author}{Kubiak, M.A.}, \bibinfo{author}{Kucharek, H.},
  \bibinfo{author}{Larsen, B.A.}, \bibinfo{author}{Leske, R.A.},
  \bibinfo{author}{Lugaz, N.}, \bibinfo{author}{Luhmann, J.},
  \bibinfo{author}{Matthaeus, W.}, \bibinfo{author}{Mitchell, D.},
  \bibinfo{author}{Moebius, E.}, \bibinfo{author}{Ogasawara, K.},
  \bibinfo{author}{Reisenfeld, D.B.}, \bibinfo{author}{Richardson, J.D.},
  \bibinfo{author}{Russell, C.T.}, \bibinfo{author}{Sok{\'o}{\l}, J.M.},
  \bibinfo{author}{Spence, H.E.}, \bibinfo{author}{Skoug, R.},
  \bibinfo{author}{Sternovsky, Z.}, \bibinfo{author}{Swaczyna, P.},
  \bibinfo{author}{Szalay, J.R.}, \bibinfo{author}{Tokumaru, M.},
  \bibinfo{author}{Wiedenbeck, M.E.}, \bibinfo{author}{Wurz, P.},
  \bibinfo{author}{Zank, G.P.}, \bibinfo{author}{Zirnstein, E.J.},
  \bibinfo{year}{2018}.
\newblock \bibinfo{title}{Interstellar {{Mapping}} and {{Acceleration Probe}}
  ({{IMAP}}): {{A New NASA Mission}}}.
\newblock \bibinfo{journal}{Space Science Reviews} \bibinfo{volume}{214},
  \bibinfo{pages}{116}.
\newblock \DOIprefix\doi{10.1007/s11214-018-0550-1}.
\bibitem[{McDonnell(1978)}]{McDonnell1978microparticle}
\bibinfo{author}{McDonnell, J.A.M.}, \bibinfo{year}{1978}.
\newblock \bibinfo{title}{Microparticle studies by space instrumentation}, in:
  \bibinfo{booktitle}{Cosmic {{Dust}}}, pp. \bibinfo{pages}{337--426}.
\bibitem[{McDonnell et~al.(2001)McDonnell, McBride, Green, Ratcliff, Gardner
  and Griffiths}]{McDonnell2001earth}
\bibinfo{author}{McDonnell, T.}, \bibinfo{author}{McBride, N.},
  \bibinfo{author}{Green, S.F.}, \bibinfo{author}{Ratcliff, P.R.},
  \bibinfo{author}{Gardner, D.J.}, \bibinfo{author}{Griffiths, A.D.},
  \bibinfo{year}{2001}.
\newblock \bibinfo{title}{Near {{Earth Environment}}}, in:
  \bibinfo{editor}{Gr{\"u}n, E.}, \bibinfo{editor}{Gustafson, B.{\AA}.S.},
  \bibinfo{editor}{Dermott, S.}, \bibinfo{editor}{Fechtig, H.} (Eds.),
  \bibinfo{booktitle}{Interplanetary {{Dust}}}. \bibinfo{address}{{Berlin,
  Heidelberg}}. Astronomy and {{Astrophysics Library}}, pp.
  \bibinfo{pages}{163--231}.
\newblock \DOIprefix\doi{10.1007/978-3-642-56428-4_4}.
\bibitem[{Misconi(1993)}]{Misconi1993spin}
\bibinfo{author}{Misconi, N.Y.}, \bibinfo{year}{1993}.
\newblock \bibinfo{title}{The spin of cosmic dust: Rotational bursting of
  circumsolar dust in the {{F}} corona}.
\newblock \bibinfo{journal}{Journal of Geophysical Research: Space Physics}
  \bibinfo{volume}{98}, \bibinfo{pages}{18951--18961}.
\newblock \DOIprefix\doi{10.1029/93JA01521}.
\bibitem[{Moorhead(2021)}]{Moorhead2021forbidden}
\bibinfo{author}{Moorhead, A.V.}, \bibinfo{year}{2021}.
\newblock \bibinfo{title}{Forbidden mass ranges for shower meteoroids}.
\newblock \bibinfo{journal}{Icarus} \bibinfo{volume}{354},
  \bibinfo{pages}{113949}.
\newblock \URLprefix \url{http://dx.doi.org/10.1016/j.icarus.2020.113949},
  \DOIprefix\doi{10.1016/j.icarus.2020.113949}.
\bibitem[{Moorhead et~al.(2020)Moorhead, Kingery and Ehlert}]{Moorhead2020nasa}
\bibinfo{author}{Moorhead, A.V.}, \bibinfo{author}{Kingery, A.},
  \bibinfo{author}{Ehlert, S.}, \bibinfo{year}{2020}.
\newblock \bibinfo{title}{{{NASA}}'s {{Meteoroid Engineering Model}} 3 and
  {{Its Ability}} to {{Replicate Spacecraft Impact Rates}}}.
\newblock \bibinfo{journal}{Journal of Spacecraft and Rockets}
  \bibinfo{volume}{57}, \bibinfo{pages}{160--176}.
\newblock \URLprefix \url{https://arc.aiaa.org/doi/10.2514/1.A34561},
  \DOIprefix\doi{10.2514/1.A34561}.
\bibitem[{Morbidelli and Gladman(1998)}]{Morbidelli1998orbital}
\bibinfo{author}{Morbidelli, A.}, \bibinfo{author}{Gladman, B.},
  \bibinfo{year}{1998}.
\newblock \bibinfo{title}{Orbital and temporal distributions of meteorites
  originating in the asteroid belt}.
\newblock \bibinfo{journal}{Meteoritics \& Planetary Science}
  \bibinfo{volume}{33}, \bibinfo{pages}{999--1016}.
\newblock \DOIprefix\doi{10.1111/j.1945-5100.1998.tb01707.x}.
\bibitem[{Morfill and Gr{\"u}n(1979)}]{Morfill1979motion}
\bibinfo{author}{Morfill, G.}, \bibinfo{author}{Gr{\"u}n, E.},
  \bibinfo{year}{1979}.
\newblock \bibinfo{title}{The motion of charged dust particles in
  interplanetary space\textemdash{{I}}. {{The}} zodiacal dust cloud}.
\newblock \bibinfo{journal}{Planetary and Space Science} \bibinfo{volume}{27},
  \bibinfo{pages}{1269--1282}.
\newblock \DOIprefix\doi{10.1016/0032-0633(79)90105-3}.
\bibitem[{Morfill et~al.(1986)Morfill, Gr{\"u}n and
  Leinert}]{Morfill1986interaction}
\bibinfo{author}{Morfill, G.E.}, \bibinfo{author}{Gr{\"u}n, E.},
  \bibinfo{author}{Leinert, C.}, \bibinfo{year}{1986}.
\newblock \bibinfo{title}{The {{Interaction}} of {{Solid Particles}} with the
  {{Interplanetary Medium}}}, in: \bibinfo{booktitle}{The {{Sun}} and the
  {{Heliosphere}} in {{Three Dimensions}}, {{Proceedings}} of the 19th {{ESLAB
  Symposium}}, {{Astrophysics}} and {{Space Science Library}}}, pp.
  \bibinfo{pages}{455--474}.
\newblock \DOIprefix\doi{10.1007/978-94-009-4612-5_53}.
\bibitem[{Mukai and Giese(1984)}]{Mukai1984modification}
\bibinfo{author}{Mukai, T.}, \bibinfo{author}{Giese, R.H.},
  \bibinfo{year}{1984}.
\newblock \bibinfo{title}{Modification of the spatial distribution of
  interplanetary dust grains by {{Lorentz}} forces}.
\newblock \bibinfo{journal}{Astronomy and Astrophysics} \bibinfo{volume}{131},
  \bibinfo{pages}{355--363}.
\bibitem[{Mukai and Yamamoto(1979)}]{Mukai1979model}
\bibinfo{author}{Mukai, T.}, \bibinfo{author}{Yamamoto, T.},
  \bibinfo{year}{1979}.
\newblock \bibinfo{title}{A {{Model}} of the {{Circumsolar Dust Cloud}}}.
\newblock \bibinfo{journal}{Publications of the Astronomical Society of Japan}
  \bibinfo{volume}{31}, \bibinfo{pages}{585--596}.
\bibitem[{Nesvorn{\'y} et~al.(2011a)Nesvorn{\'y}, Janches, Vokrouhlick{\'y},
  Pokorn{\'y}, Bottke and Jenniskens}]{Nesvorny2011dynamical}
\bibinfo{author}{Nesvorn{\'y}, D.}, \bibinfo{author}{Janches, D.},
  \bibinfo{author}{Vokrouhlick{\'y}, D.}, \bibinfo{author}{Pokorn{\'y}, P.},
  \bibinfo{author}{Bottke, W.F.}, \bibinfo{author}{Jenniskens, P.},
  \bibinfo{year}{2011}a.
\newblock \bibinfo{title}{{{DYNAMICAL MODEL FOR THE ZODIACAL CLOUD AND SPORADIC
  METEORS}}}.
\newblock \bibinfo{journal}{The Astrophysical Journal} \bibinfo{volume}{743},
  \bibinfo{pages}{129}.
\newblock \DOIprefix\doi{10.1088/0004-637X/743/2/129}.
\bibitem[{Nesvorn{\'y} et~al.(2010)Nesvorn{\'y}, Jenniskens, Levison, Bottke,
  Vokrouhlick{\'y} and Gounelle}]{Nesvorny2010cometary}
\bibinfo{author}{Nesvorn{\'y}, D.}, \bibinfo{author}{Jenniskens, P.},
  \bibinfo{author}{Levison, H.F.}, \bibinfo{author}{Bottke, W.F.},
  \bibinfo{author}{Vokrouhlick{\'y}, D.}, \bibinfo{author}{Gounelle, M.},
  \bibinfo{year}{2010}.
\newblock \bibinfo{title}{{{COMETARY ORIGIN OF THE ZODIACAL CLOUD AND
  CARBONACEOUS MICROMETEORITES}}. {{IMPLICATIONS FOR HOT DEBRIS DISKS}}}.
\newblock \bibinfo{journal}{The Astrophysical Journal} \bibinfo{volume}{713},
  \bibinfo{pages}{816--836}.
\newblock \DOIprefix\doi{10.1088/0004-637X/713/2/816}.
\bibitem[{Nesvorn{\'y} et~al.(2011b)Nesvorn{\'y}, Vokrouhlick{\'y}, Pokorn{\'y}
  and Janches}]{Nesvorny2011dynamics}
\bibinfo{author}{Nesvorn{\'y}, D.}, \bibinfo{author}{Vokrouhlick{\'y}, D.},
  \bibinfo{author}{Pokorn{\'y}, P.}, \bibinfo{author}{Janches, D.},
  \bibinfo{year}{2011}b.
\newblock \bibinfo{title}{{{DYNAMICS OF DUST PARTICLES RELEASED FROM OORT CLOUD
  COMETS AND THEIR CONTRIBUTION TO RADAR METEORS}}}.
\newblock \bibinfo{journal}{The Astrophysical Journal} \bibinfo{volume}{743},
  \bibinfo{pages}{37}.
\newblock \DOIprefix\doi{10.1088/0004-637X/743/1/37}.
\bibitem[{Page et~al.(2020)Page, Bale, Bonnell, Goetz, Goodrich, Harvey,
  Larsen, MacDowall, Malaspina, Pokorn{\'y}, Pulupa and
  Szalay}]{Page2020examining}
\bibinfo{author}{Page, B.}, \bibinfo{author}{Bale, S.D.},
  \bibinfo{author}{Bonnell, J.W.}, \bibinfo{author}{Goetz, K.},
  \bibinfo{author}{Goodrich, K.}, \bibinfo{author}{Harvey, P.R.},
  \bibinfo{author}{Larsen, R.}, \bibinfo{author}{MacDowall, R.J.},
  \bibinfo{author}{Malaspina, D.M.}, \bibinfo{author}{Pokorn{\'y}, P.},
  \bibinfo{author}{Pulupa, M.}, \bibinfo{author}{Szalay, J.R.},
  \bibinfo{year}{2020}.
\newblock \bibinfo{title}{Examining {{Dust Directionality}} with the
  {{{\emph{Parker Solar Probe}}}} {{FIELDS Instrument}}}.
\newblock \bibinfo{journal}{The Astrophysical Journal Supplement Series}
  \bibinfo{volume}{246}, \bibinfo{pages}{51}.
\newblock \DOIprefix\doi{10.3847/1538-4365/ab5f6a}.
\bibitem[{Parker(1964)}]{Parker1964perturbation}
\bibinfo{author}{Parker, E.N.}, \bibinfo{year}{1964}.
\newblock \bibinfo{title}{The {{Perturbation}} of {{Interplanetary Dust
  Grains}} by the {{Solar Wind}}.}
\newblock \bibinfo{journal}{The Astrophysical Journal} \bibinfo{volume}{139},
  \bibinfo{pages}{951}.
\newblock \DOIprefix\doi{10.1086/147829}.
\bibitem[{P{\'a}stor et~al.(2009)P{\'a}stor, Kla{\v c}ka and
  K{\'o}mar}]{Pastor2009motion}
\bibinfo{author}{P{\'a}stor, P.}, \bibinfo{author}{Kla{\v c}ka, J.},
  \bibinfo{author}{K{\'o}mar, L.}, \bibinfo{year}{2009}.
\newblock \bibinfo{title}{Motion of dust in mean motion resonances with
  planets}.
\newblock \bibinfo{journal}{Celestial Mechanics and Dynamical Astronomy}
  \bibinfo{volume}{103}, \bibinfo{pages}{343--364}.
\newblock \DOIprefix\doi{10.1007/s10569-009-9202-9}.
\bibitem[{Pokorn{\'y} and Kuchner(2019)}]{Pokorny2019coorbital}
\bibinfo{author}{Pokorn{\'y}, P.}, \bibinfo{author}{Kuchner, M.},
  \bibinfo{year}{2019}.
\newblock \bibinfo{title}{Co-orbital {{Asteroids}} as the {{Source}} of
  {{Venus}}'s {{Zodiacal Dust Ring}}}.
\newblock \bibinfo{journal}{The Astrophysical Journal} \bibinfo{volume}{873},
  \bibinfo{pages}{L16}.
\newblock \DOIprefix\doi{10.3847/2041-8213/ab0827}.
\bibitem[{Pokorn{\'y} et~al.(2014)Pokorn{\'y}, Vokrouhlick{\'y}, Nesvorn{\'y},
  {Campbell-Brown} and Brown}]{Pokorny2014dynamical}
\bibinfo{author}{Pokorn{\'y}, P.}, \bibinfo{author}{Vokrouhlick{\'y}, D.},
  \bibinfo{author}{Nesvorn{\'y}, D.}, \bibinfo{author}{{Campbell-Brown}, M.},
  \bibinfo{author}{Brown, P.}, \bibinfo{year}{2014}.
\newblock \bibinfo{title}{{{DYNAMICAL MODEL FOR THE TOROIDAL SPORADIC
  METEORS}}}.
\newblock \bibinfo{journal}{The Astrophysical Journal} \bibinfo{volume}{789},
  \bibinfo{pages}{25}.
\newblock \DOIprefix\doi{10.1088/0004-637X/789/1/25}.
\bibitem[{Pusack et~al.(2021)Pusack, Malaspina, Szalay, Bale, Goetz, MacDowall
  and Pulupa}]{Pusack2021dust}
\bibinfo{author}{Pusack, A.}, \bibinfo{author}{Malaspina, D.M.},
  \bibinfo{author}{Szalay, J.R.}, \bibinfo{author}{Bale, S.D.},
  \bibinfo{author}{Goetz, K.}, \bibinfo{author}{MacDowall, R.J.},
  \bibinfo{author}{Pulupa, M.}, \bibinfo{year}{2021}.
\newblock \bibinfo{title}{Dust {{Directionality}} and an {{Anomalous
  Interplanetary Dust Population Detected}} by the {{Parker Solar Probe}}}.
\newblock \bibinfo{journal}{The Planetary Science Journal} \bibinfo{volume}{2},
  \bibinfo{pages}{186}.
\newblock \DOIprefix\doi{10.3847/PSJ/ac0bb9}.
\bibitem[{Reach(2010)}]{Reach2010structure}
\bibinfo{author}{Reach, W.T.}, \bibinfo{year}{2010}.
\newblock \bibinfo{title}{Structure of the {{Earth}}'s circumsolar dust ring}.
\newblock \bibinfo{journal}{Icarus} \bibinfo{volume}{209},
  \bibinfo{pages}{848--850}.
\newblock \URLprefix
  \url{https://www.sciencedirect.com/science/article/pii/S0019103510002563},
  \DOIprefix\doi{10.1016/j.icarus.2010.06.034}.
\bibitem[{Rigley and Wyatt(2021)}]{Rigley2021comet}
\bibinfo{author}{Rigley, J.K.}, \bibinfo{author}{Wyatt, M.C.},
  \bibinfo{year}{2021}.
\newblock \bibinfo{title}{Comet fragmentation as a source of the zodiacal
  cloud}.
\newblock \bibinfo{journal}{Monthly Notices of the Royal Astronomical Society}
  \bibinfo{volume}{510}, \bibinfo{pages}{834--857}.
\newblock \DOIprefix\doi{10.1093/mnras/stab3482}.
\bibitem[{Sarugaku et~al.(2015)Sarugaku, Ishiguro, Ueno, Usui and
  Reach}]{Sarugaku2015infrared}
\bibinfo{author}{Sarugaku, Y.}, \bibinfo{author}{Ishiguro, M.},
  \bibinfo{author}{Ueno, M.}, \bibinfo{author}{Usui, F.},
  \bibinfo{author}{Reach, W.T.}, \bibinfo{year}{2015}.
\newblock \bibinfo{title}{{{INFRARED AND OPTICAL IMAGINGS OF THE COMET
  2P}}/{{ENCKE DUST CLOUD IN THE}} 2003 {{RETURN}}}.
\newblock \bibinfo{journal}{The Astrophysical Journal} \bibinfo{volume}{804},
  \bibinfo{pages}{127}.
\newblock \DOIprefix\doi{10.1088/0004-637X/804/2/127}.
\bibitem[{Schmidt(1980)}]{Schmidt1980bahnelemente}
\bibinfo{author}{Schmidt, K.D.}, \bibinfo{year}{1980}.
\newblock \bibinfo{title}{Bahnelemente von {{Mikrometeoriten}}: {{Analyse}} von
  {{Messungen}} Der {{Sonnensonde Helios}} 1}.
\newblock \bibinfo{type}{Bundesministerium F\"ur {{Forschung}} Und
  {{Technologie}}, {{Forschungsbreicht}}} \bibinfo{number}{W 80-036}. {BMFT}.
\bibitem[{Schmidt and Gr{\"u}n(1979)}]{Schmidt1979distribution}
\bibinfo{author}{Schmidt, K.D.}, \bibinfo{author}{Gr{\"u}n, E.},
  \bibinfo{year}{1979}.
\newblock \bibinfo{title}{The distribution of orbital elements of
  interplanetary dust in the inner solar system as detected by the {{Helios}}
  spaceprobe.}, in: \bibinfo{booktitle}{Space Research {{XIX}}, {{Proceedings}}
  of the {{Open Meetings}} of the {{Working Groups}} on {{Physical Sciences}}
  of the 21st {{Plenary Meeting}} of {{COSPAR}}}, pp.
  \bibinfo{pages}{439--442}.
\bibitem[{Schmidt and Gr{\"u}n(1980)}]{Schmidt1980orbital}
\bibinfo{author}{Schmidt, K.D.}, \bibinfo{author}{Gr{\"u}n, E.},
  \bibinfo{year}{1980}.
\newblock \bibinfo{title}{Orbital {{Elements}} of {{Micrometeoroids Detected}}
  by the {{Helios}} 1 {{Space Probe}} in the {{Inner Solar System}}}, in:
  \bibinfo{booktitle}{{{IAU Symposium}} No. 90}, pp. \bibinfo{pages}{321--324}.
\newblock \DOIprefix\doi{10.1017/S0074180900066997}.
\bibitem[{Sheppard et~al.(2022)Sheppard, Tholen, Pokorn{\'y}, Micheli,
  Dell'Antonio, Fu, Trujillo, Beaton, Carlsten, {Drlica-Wagner},
  {Mart{\'i}nez-V{\'a}zquez}, Mau, {Santana-Ros}, {Santana-Silva}, Sif{\'o}n,
  Simha, Thirouin, Trilling, Vivas and Zenteno}]{Sheppard2022deep}
\bibinfo{author}{Sheppard, S.S.}, \bibinfo{author}{Tholen, D.J.},
  \bibinfo{author}{Pokorn{\'y}, P.}, \bibinfo{author}{Micheli, M.},
  \bibinfo{author}{Dell'Antonio, I.}, \bibinfo{author}{Fu, S.},
  \bibinfo{author}{Trujillo, C.A.}, \bibinfo{author}{Beaton, R.},
  \bibinfo{author}{Carlsten, S.}, \bibinfo{author}{{Drlica-Wagner}, A.},
  \bibinfo{author}{{Mart{\'i}nez-V{\'a}zquez}, C.}, \bibinfo{author}{Mau, S.},
  \bibinfo{author}{{Santana-Ros}, T.}, \bibinfo{author}{{Santana-Silva}, L.},
  \bibinfo{author}{Sif{\'o}n, C.}, \bibinfo{author}{Simha, S.},
  \bibinfo{author}{Thirouin, A.}, \bibinfo{author}{Trilling, D.},
  \bibinfo{author}{Vivas, A.K.}, \bibinfo{author}{Zenteno, A.},
  \bibinfo{year}{2022}.
\newblock \bibinfo{title}{A {{Deep}} and {{Wide Twilight Survey}} for
  {{Asteroids Interior}} to {{Earth}} and {{Venus}}}.
\newblock \bibinfo{journal}{The Astronomical Journal} \bibinfo{volume}{164},
  \bibinfo{pages}{168}.
\newblock \DOIprefix\doi{10.3847/1538-3881/ac8cff}.
\bibitem[{Shestakova and Tambovtseva(1995)}]{Shestakova1995dynamics}
\bibinfo{author}{Shestakova, L.I.}, \bibinfo{author}{Tambovtseva, L.V.},
  \bibinfo{year}{1995}.
\newblock \bibinfo{title}{Dynamics of dust grains near the {{Sun}}}.
\newblock \bibinfo{journal}{Astronomical \& Astrophysical Transactions}
  \bibinfo{volume}{8}, \bibinfo{pages}{59--81}.
\newblock \DOIprefix\doi{10.1080/10556799508203297}.
\bibitem[{Smith et~al.(1993)Smith, Neugebauer, Balogh, Bame, Erd{\"o}s,
  Forsyth, Goldstein, Phillips and Tsurutani}]{Smith1993disappearance}
\bibinfo{author}{Smith, E.J.}, \bibinfo{author}{Neugebauer, M.},
  \bibinfo{author}{Balogh, A.}, \bibinfo{author}{Bame, S.J.},
  \bibinfo{author}{Erd{\"o}s, G.}, \bibinfo{author}{Forsyth, R.J.},
  \bibinfo{author}{Goldstein, B.E.}, \bibinfo{author}{Phillips, J.L.},
  \bibinfo{author}{Tsurutani, B.T.}, \bibinfo{year}{1993}.
\newblock \bibinfo{title}{Disappearance of the heliospheric sector structure at
  {{Ulysses}}}.
\newblock \bibinfo{journal}{Geophysical Research Letters} \bibinfo{volume}{20},
  \bibinfo{pages}{2327--2330}.
\newblock \DOIprefix\doi{10.1029/93GL02632}.
\bibitem[{Soja(2010)}]{Soja2010dynamics}
\bibinfo{author}{Soja, R.H.}, \bibinfo{year}{2010}.
\newblock \bibinfo{title}{Dynamics of the {{Solar System Meteoroid
  Population}}}.
\newblock Ph.D. thesis. University of Canterbury.
\newblock \URLprefix \url{http://dx.doi.org/10.26021/7541}.
\bibitem[{Soja et~al.(2019)Soja, Gr{\"u}n, Strub, Sommer, Millinger,
  Vaubaillon, Alius, Camodeca, Hein, Laskar, Gastineau, Fienga, Schwarzkopf,
  Herzog, Gutsche, Skuppin and Srama}]{Soja2019imem2}
\bibinfo{author}{Soja, R.H.}, \bibinfo{author}{Gr{\"u}n, E.},
  \bibinfo{author}{Strub, P.}, \bibinfo{author}{Sommer, M.},
  \bibinfo{author}{Millinger, M.}, \bibinfo{author}{Vaubaillon, J.},
  \bibinfo{author}{Alius, W.}, \bibinfo{author}{Camodeca, G.},
  \bibinfo{author}{Hein, F.}, \bibinfo{author}{Laskar, J.},
  \bibinfo{author}{Gastineau, M.}, \bibinfo{author}{Fienga, A.},
  \bibinfo{author}{Schwarzkopf, G.J.}, \bibinfo{author}{Herzog, J.},
  \bibinfo{author}{Gutsche, K.}, \bibinfo{author}{Skuppin, N.},
  \bibinfo{author}{Srama, R.}, \bibinfo{year}{2019}.
\newblock \bibinfo{title}{{{IMEM2}}: A meteoroid environment model for the
  inner solar system}.
\newblock \bibinfo{journal}{Astronomy \& Astrophysics} \bibinfo{volume}{628},
  \bibinfo{pages}{A109}.
\newblock \DOIprefix\doi{10.1051/0004-6361/201834892}.
\bibitem[{Sommer et~al.(2020)Sommer, Yano and Srama}]{Sommer2020effects}
\bibinfo{author}{Sommer, M.}, \bibinfo{author}{Yano, H.},
  \bibinfo{author}{Srama, R.}, \bibinfo{year}{2020}.
\newblock \bibinfo{title}{Effects of neighbouring planets on the formation of
  resonant dust rings in the inner {{Solar System}}}.
\newblock \bibinfo{journal}{Astronomy \& Astrophysics} \bibinfo{volume}{635},
  \bibinfo{pages}{A10}.
\newblock \DOIprefix\doi{10.1051/0004-6361/201936676}.
\bibitem[{Srama et~al.(2004)Srama, Ahrens, Altobelli, Auer, Bradley, Burton,
  Dikarev, Economou, Fechtig, G{\"o}rlich, Grande, Graps, Gr{\"u}n, Havnes,
  Helfert, Horanyi, Igenbergs, Jessberger, Johnson, Kempf, Krivov, Kr{\"u}ger,
  {Mocker-Ahlreep}, {Moragas-Klostermeyer}, Lamy, Landgraf, Linkert, Linkert,
  Lura, McDonnell, M{\"o}hlmann, Morfill, M{\"u}ller, Roy, Sch{\"a}fer,
  Schlotzhauer, Schwehm, Spahn, St{\"u}big, Svestka, Tschernjawski, Tuzzolino,
  W{\"a}sch and Zook}]{Srama2004cassini}
\bibinfo{author}{Srama, R.}, \bibinfo{author}{Ahrens, T.J.},
  \bibinfo{author}{Altobelli, N.}, \bibinfo{author}{Auer, S.},
  \bibinfo{author}{Bradley, J.G.}, \bibinfo{author}{Burton, M.},
  \bibinfo{author}{Dikarev, V.V.}, \bibinfo{author}{Economou, T.},
  \bibinfo{author}{Fechtig, H.}, \bibinfo{author}{G{\"o}rlich, M.},
  \bibinfo{author}{Grande, M.}, \bibinfo{author}{Graps, A.},
  \bibinfo{author}{Gr{\"u}n, E.}, \bibinfo{author}{Havnes, O.},
  \bibinfo{author}{Helfert, S.}, \bibinfo{author}{Horanyi, M.},
  \bibinfo{author}{Igenbergs, E.}, \bibinfo{author}{Jessberger, E.K.},
  \bibinfo{author}{Johnson, T.V.}, \bibinfo{author}{Kempf, S.},
  \bibinfo{author}{Krivov, A.V.}, \bibinfo{author}{Kr{\"u}ger, H.},
  \bibinfo{author}{{Mocker-Ahlreep}, A.},
  \bibinfo{author}{{Moragas-Klostermeyer}, G.}, \bibinfo{author}{Lamy, P.},
  \bibinfo{author}{Landgraf, M.}, \bibinfo{author}{Linkert, D.},
  \bibinfo{author}{Linkert, G.}, \bibinfo{author}{Lura, F.},
  \bibinfo{author}{McDonnell, J.A.M.}, \bibinfo{author}{M{\"o}hlmann, D.},
  \bibinfo{author}{Morfill, G.E.}, \bibinfo{author}{M{\"u}ller, M.},
  \bibinfo{author}{Roy, M.}, \bibinfo{author}{Sch{\"a}fer, G.},
  \bibinfo{author}{Schlotzhauer, G.}, \bibinfo{author}{Schwehm, G.H.},
  \bibinfo{author}{Spahn, F.}, \bibinfo{author}{St{\"u}big, M.},
  \bibinfo{author}{Svestka, J.}, \bibinfo{author}{Tschernjawski, V.},
  \bibinfo{author}{Tuzzolino, A.J.}, \bibinfo{author}{W{\"a}sch, R.},
  \bibinfo{author}{Zook, H.A.}, \bibinfo{year}{2004}.
\newblock \bibinfo{title}{The {{Cassini Cosmic Dust Analyzer}}}.
\newblock \bibinfo{journal}{Space Science Reviews} \bibinfo{volume}{114},
  \bibinfo{pages}{465--518}.
\newblock \DOIprefix\doi{10.1007/s11214-004-1435-z}.
\bibitem[{Staubach et~al.(1997)Staubach, Gr{\"u}n and
  Jehn}]{Staubach1997meteoroid}
\bibinfo{author}{Staubach, P.}, \bibinfo{author}{Gr{\"u}n, E.},
  \bibinfo{author}{Jehn, R.}, \bibinfo{year}{1997}.
\newblock \bibinfo{title}{The meteoroid environment near {{Earth}}}.
\newblock \bibinfo{journal}{Advances in Space Research} \bibinfo{volume}{19},
  \bibinfo{pages}{301--308}.
\newblock \DOIprefix\doi{10.1016/S0273-1177(97)00017-3}.
\bibitem[{Stenborg et~al.(2021)Stenborg, Howard, Hess and
  Gallagher}]{Stenborg2021psp}
\bibinfo{author}{Stenborg, G.}, \bibinfo{author}{Howard, R.A.},
  \bibinfo{author}{Hess, P.}, \bibinfo{author}{Gallagher, B.},
  \bibinfo{year}{2021}.
\newblock \bibinfo{title}{{{PSP}}/{{WISPR}} observations of dust density
  depletion near the {{Sun}}: {{I}}. {{Remote}} observations to 8
  {{{\emph{R}}}} {\textsubscript{{$\odot$}}} from an observer between 0.13 and
  0.35 {{AU}}}.
\newblock \bibinfo{journal}{Astronomy \& Astrophysics} \bibinfo{volume}{650},
  \bibinfo{pages}{A28}.
\newblock \DOIprefix\doi{10.1051/0004-6361/202039284}.
\bibitem[{Stenborg et~al.(2022)Stenborg, Howard, Vourlidas and
  Gallagher}]{Stenborg2022psp}
\bibinfo{author}{Stenborg, G.}, \bibinfo{author}{Howard, R.A.},
  \bibinfo{author}{Vourlidas, A.}, \bibinfo{author}{Gallagher, B.},
  \bibinfo{year}{2022}.
\newblock \bibinfo{title}{{{PSP}}/{{WISPR Observations}} of {{Dust Density
  Depletion}} near the {{Sun}}. {{II}}. {{New Insights}} from within the
  {{Depletion Zone}}}.
\newblock \bibinfo{journal}{The Astrophysical Journal} \bibinfo{volume}{932},
  \bibinfo{pages}{75}.
\newblock \DOIprefix\doi{10.3847/1538-4357/ac6b36}.
\bibitem[{Sterken(2022)}]{Sterken2022dolphin}
\bibinfo{author}{Sterken, V.}, \bibinfo{year}{2022}.
\newblock \bibinfo{title}{The {{DOLPHIN}} mission and unique opportunities in
  2030 to probe the dust-heliosphere interactions}, in:
  \bibinfo{booktitle}{44th {{COSPAR Scientific Assembly}}}, p.
  \bibinfo{pages}{1016}.
\bibitem[{Sterken et~al.(2012)Sterken, Altobelli, Kempf, Schwehm, Srama and
  Gr{\"u}n}]{Sterken2012flow}
\bibinfo{author}{Sterken, V.J.}, \bibinfo{author}{Altobelli, N.},
  \bibinfo{author}{Kempf, S.}, \bibinfo{author}{Schwehm, G.},
  \bibinfo{author}{Srama, R.}, \bibinfo{author}{Gr{\"u}n, E.},
  \bibinfo{year}{2012}.
\newblock \bibinfo{title}{The flow of interstellar dust into the solar system}.
\newblock \bibinfo{journal}{Astronomy \& Astrophysics} \bibinfo{volume}{538},
  \bibinfo{pages}{A102}.
\newblock \DOIprefix\doi{10.1051/0004-6361/201117119}.
\bibitem[{Sternovsky et~al.(2022)Sternovsky, Horanyi, Ayari, Kempf, Mikula,
  Hillier, Postberg and Srama}]{Sternovsky2022measuring}
\bibinfo{author}{Sternovsky, Z.}, \bibinfo{author}{Horanyi, M.},
  \bibinfo{author}{Ayari, E.}, \bibinfo{author}{Kempf, S.},
  \bibinfo{author}{Mikula, B.}, \bibinfo{author}{Hillier, J.},
  \bibinfo{author}{Postberg, F.}, \bibinfo{author}{Srama, R.},
  \bibinfo{year}{2022}.
\newblock \bibinfo{title}{Measuring the composition of interstellar and
  interplanetary dust particles with the {{IDEX}} instrument onboard the
  {{IMAP}} mission}, in: \bibinfo{booktitle}{44th {{COSPAR Scientific
  Assembly}}}, p. \bibinfo{pages}{1206}.
\bibitem[{Strub et~al.(2019)Strub, Sterken, Soja, Kr{\"u}ger, Gr{\"u}n and
  Srama}]{Strub2019heliospheric}
\bibinfo{author}{Strub, P.}, \bibinfo{author}{Sterken, V.J.},
  \bibinfo{author}{Soja, R.}, \bibinfo{author}{Kr{\"u}ger, H.},
  \bibinfo{author}{Gr{\"u}n, E.}, \bibinfo{author}{Srama, R.},
  \bibinfo{year}{2019}.
\newblock \bibinfo{title}{Heliospheric modulation of the interstellar dust flow
  on to {{Earth}}}.
\newblock \bibinfo{journal}{Astronomy \& Astrophysics} \bibinfo{volume}{621},
  \bibinfo{pages}{A54}.
\newblock \DOIprefix\doi{10.1051/0004-6361/201832644}.
\bibitem[{Sulzer(2004)}]{Sulzer2004meteoroid}
\bibinfo{author}{Sulzer, M.P.}, \bibinfo{year}{2004}.
\newblock \bibinfo{title}{Meteoroid velocity distribution derived from head
  echo data collected at {{Arecibo}} during regular world day observations}.
\newblock \bibinfo{journal}{Atmospheric Chemistry and Physics}
  \bibinfo{volume}{4}, \bibinfo{pages}{947--954}.
\newblock \DOIprefix\doi{10.5194/acp-4-947-2004}.
\bibitem[{Svedhem et~al.(2000)Svedhem, Drolshagen, Gr{\"u}n, Grafodatsky and
  Prokopiev}]{Svedhem2000new}
\bibinfo{author}{Svedhem, H.}, \bibinfo{author}{Drolshagen, G.},
  \bibinfo{author}{Gr{\"u}n, E.}, \bibinfo{author}{Grafodatsky, O.},
  \bibinfo{author}{Prokopiev, U.}, \bibinfo{year}{2000}.
\newblock \bibinfo{title}{New results from in situ measurements of {{Cosmic
  Dust}} \textemdash{} {{Data}} from the {{GORID}} experiment}.
\newblock \bibinfo{journal}{Advances in Space Research} \bibinfo{volume}{25},
  \bibinfo{pages}{309--314}.
\newblock \DOIprefix\doi{10.1016/S0273-1177(99)00951-5}.
\bibitem[{Szalay et~al.(2020)Szalay, Pokorn{\'y}, Bale, Christian, Goetz,
  Goodrich, Hill, Kuchner, Larsen, Malaspina, McComas, Mitchell, Page and
  Schwadron}]{Szalay2020nearsun}
\bibinfo{author}{Szalay, J.R.}, \bibinfo{author}{Pokorn{\'y}, P.},
  \bibinfo{author}{Bale, S.D.}, \bibinfo{author}{Christian, E.R.},
  \bibinfo{author}{Goetz, K.}, \bibinfo{author}{Goodrich, K.},
  \bibinfo{author}{Hill, M.E.}, \bibinfo{author}{Kuchner, M.},
  \bibinfo{author}{Larsen, R.}, \bibinfo{author}{Malaspina, D.},
  \bibinfo{author}{McComas, D.J.}, \bibinfo{author}{Mitchell, D.},
  \bibinfo{author}{Page, B.}, \bibinfo{author}{Schwadron, N.},
  \bibinfo{year}{2020}.
\newblock \bibinfo{title}{The {{Near-Sun Dust Environment}}: {{Initial
  Observations}} from {{{\emph{Parker Solar Probe}}}}}.
\newblock \bibinfo{journal}{The Astrophysical Journal Supplement Series}
  \bibinfo{volume}{246}, \bibinfo{pages}{27}.
\newblock \DOIprefix\doi{10.3847/1538-4365/ab50c1}.
\bibitem[{Szalay et~al.(2021)Szalay, Pokorn{\'y}, Malaspina, Pusack, Bale,
  Battams, Gasque, Goetz, Kr{\"u}ger, McComas, Schwadron and
  Strub}]{Szalay2021collisional}
\bibinfo{author}{Szalay, J.R.}, \bibinfo{author}{Pokorn{\'y}, P.},
  \bibinfo{author}{Malaspina, D.M.}, \bibinfo{author}{Pusack, A.},
  \bibinfo{author}{Bale, S.D.}, \bibinfo{author}{Battams, K.},
  \bibinfo{author}{Gasque, L.C.}, \bibinfo{author}{Goetz, K.},
  \bibinfo{author}{Kr{\"u}ger, H.}, \bibinfo{author}{McComas, D.J.},
  \bibinfo{author}{Schwadron, N.A.}, \bibinfo{author}{Strub, P.},
  \bibinfo{year}{2021}.
\newblock \bibinfo{title}{Collisional {{Evolution}} of the {{Inner Zodiacal
  Cloud}}}.
\newblock \bibinfo{journal}{The Planetary Science Journal} \bibinfo{volume}{2},
  \bibinfo{pages}{185}.
\newblock \DOIprefix\doi{10.3847/PSJ/abf928}.
\bibitem[{Wallis and Hassan(1985)}]{Wallis1985stochastic}
\bibinfo{author}{Wallis, M.K.}, \bibinfo{author}{Hassan, M.H.A.},
  \bibinfo{year}{1985}.
\newblock \bibinfo{title}{Stochastic diffusion of interplanetary dust grains
  orbiting under {{Poynting-Robertson}} forces}.
\newblock \bibinfo{journal}{Astronomy and Astrophysics} \bibinfo{volume}{151},
  \bibinfo{pages}{435--441}.
\bibitem[{Wehry et~al.(2004)Wehry, Kr{\"u}ger and Gr{\"u}n}]{Wehry2004analysis}
\bibinfo{author}{Wehry, A.}, \bibinfo{author}{Kr{\"u}ger, H.},
  \bibinfo{author}{Gr{\"u}n, E.}, \bibinfo{year}{2004}.
\newblock \bibinfo{title}{Analysis of {{Ulysses}} data: {{Radiation}} pressure
  effects on dust particles}.
\newblock \bibinfo{journal}{Astronomy \& Astrophysics} \bibinfo{volume}{419},
  \bibinfo{pages}{1169--1174}.
\newblock \DOIprefix\doi{10.1051/0004-6361:20035613}.
\bibitem[{Wehry and Mann(1999)}]{Wehry1999identification}
\bibinfo{author}{Wehry, A.}, \bibinfo{author}{Mann, I.}, \bibinfo{year}{1999}.
\newblock \bibinfo{title}{Identification of {$\beta$}-meteoroids from
  measurements of the dust detector onboard the {{ULYSSES}} spacecraft}.
\newblock \bibinfo{journal}{Astronomy \& Astrophysics} \bibinfo{volume}{341},
  \bibinfo{pages}{296--303}.
\bibitem[{Weidenschilling(1978)}]{Weidenschilling1978distribution}
\bibinfo{author}{Weidenschilling, S.J.}, \bibinfo{year}{1978}.
\newblock \bibinfo{title}{The distribution of orbits of cosmic dust particles
  detected by {{Pioneers}} 8 and 9}.
\newblock \bibinfo{journal}{Geophysical Research Letters} \bibinfo{volume}{5},
  \bibinfo{pages}{606--608}.
\newblock \DOIprefix\doi{10.1029/GL005i007p00606}.
\bibitem[{Weidenschilling and Jackson(1993)}]{Weidenschilling1993orbital}
\bibinfo{author}{Weidenschilling, S.J.}, \bibinfo{author}{Jackson, A.A.},
  \bibinfo{year}{1993}.
\newblock \bibinfo{title}{Orbital {{Resonances}} and {{Poynting-Robertson
  Drag}}}.
\newblock \bibinfo{journal}{Icarus} \bibinfo{volume}{104},
  \bibinfo{pages}{244--254}.
\newblock \DOIprefix\doi{10.1006/icar.1993.1099}.
\bibitem[{Wilck and Mann(1996)}]{Wilck1996radiation}
\bibinfo{author}{Wilck, M.}, \bibinfo{author}{Mann, I.}, \bibinfo{year}{1996}.
\newblock \bibinfo{title}{Radiation pressure forces on ``typical''
  interplanetary dust grains}.
\newblock \bibinfo{journal}{Planetary and Space Science} \bibinfo{volume}{44},
  \bibinfo{pages}{493--499}.
\newblock \DOIprefix\doi{10.1016/0032-0633(95)00151-4}.
\bibitem[{Wolf et~al.(1976)Wolf, Rhee and Berg}]{Wolf1976orbital}
\bibinfo{author}{Wolf, H.}, \bibinfo{author}{Rhee, J.}, \bibinfo{author}{Berg,
  O.E.}, \bibinfo{year}{1976}.
\newblock \bibinfo{title}{Orbital elements of dust particles intercepted by
  pioneers 8 and 9}, in: \bibinfo{booktitle}{{{Proc.}} of the {{IAU Colloq.}}
  31: Interplanetary {{Dust}} and {{Zodiacal Light}}}, pp.
  \bibinfo{pages}{165--169}.
\newblock \DOIprefix\doi{10.1007/3-540-07615-8_478}.
\bibitem[{Wood et~al.(2015)Wood, Malaspina, Andersson and
  Horanyi}]{Wood2015hypervelocity}
\bibinfo{author}{Wood, S.R.}, \bibinfo{author}{Malaspina, D.M.},
  \bibinfo{author}{Andersson, L.}, \bibinfo{author}{Horanyi, M.},
  \bibinfo{year}{2015}.
\newblock \bibinfo{title}{Hypervelocity dust impacts on the {{Wind}}
  spacecraft: {{Correlations}} between {{Ulysses}} and {{Wind}} interstellar
  dust detections}.
\newblock \bibinfo{journal}{Journal of Geophysical Research: Space Physics}
  \bibinfo{volume}{120}, \bibinfo{pages}{7121--7129}.
\newblock \DOIprefix\doi{10.1002/2015JA021463}.
\bibitem[{Wozniakiewicz et~al.(2021)Wozniakiewicz, Bridges, Burchell, Carey,
  Carpenter, Della~Corte, Dignam, Genge, Hicks, Hilchenbach, Hillier, Kearsley,
  Kr{\"u}ger, Merouane, Palomba, Postberg, Schmidt, Srama, Trieloff,
  {van-Ginneken} and Sterken}]{Wozniakiewicz2021cosmic}
\bibinfo{author}{Wozniakiewicz, P.J.}, \bibinfo{author}{Bridges, J.},
  \bibinfo{author}{Burchell, M.J.}, \bibinfo{author}{Carey, W.},
  \bibinfo{author}{Carpenter, J.}, \bibinfo{author}{Della~Corte, V.},
  \bibinfo{author}{Dignam, A.}, \bibinfo{author}{Genge, M.J.},
  \bibinfo{author}{Hicks, L.}, \bibinfo{author}{Hilchenbach, M.},
  \bibinfo{author}{Hillier, J.}, \bibinfo{author}{Kearsley, A.T.},
  \bibinfo{author}{Kr{\"u}ger, H.}, \bibinfo{author}{Merouane, S.},
  \bibinfo{author}{Palomba, E.}, \bibinfo{author}{Postberg, F.},
  \bibinfo{author}{Schmidt, J.}, \bibinfo{author}{Srama, R.},
  \bibinfo{author}{Trieloff, M.}, \bibinfo{author}{{van-Ginneken}, M.},
  \bibinfo{author}{Sterken, V.J.}, \bibinfo{year}{2021}.
\newblock \bibinfo{title}{A cosmic dust detection suite for the deep space
  {{Gateway}}}.
\newblock \bibinfo{journal}{Advances in Space Research} \bibinfo{volume}{68},
  \bibinfo{pages}{85--104}.
\newblock \DOIprefix\doi{10.1016/j.asr.2021.04.002}.
\bibitem[{Wyatt and Whipple(1950)}]{Wyatt1950poyntingrobertson}
\bibinfo{author}{Wyatt, S.P.}, \bibinfo{author}{Whipple, F.L.},
  \bibinfo{year}{1950}.
\newblock \bibinfo{title}{The {{Poynting-Robertson}} effect on meteor orbits}.
\newblock \bibinfo{journal}{The Astrophysical Journal} \bibinfo{volume}{111},
  \bibinfo{pages}{134--141}.
\newblock \DOIprefix\doi{10.1086/145244}.
\bibitem[{Zaslavsky et~al.(2021)Zaslavsky, Mann, Soucek, Czechowski, P{\'i}{\v
  s}a, Vaverka, {Meyer-Vernet}, Maksimovic, Lorf{\`e}vre, Issautier, Babic,
  Bale, Morooka, Vecchio, Chust, Khotyaintsev, Krasnoselskikh, Kretzschmar,
  Plettemeier, Steller, {\v S}tver{\'a}k, Tr{\'a}vn{\'i}{\v c}ek and
  Vaivads}]{Zaslavsky2021first}
\bibinfo{author}{Zaslavsky, A.}, \bibinfo{author}{Mann, I.},
  \bibinfo{author}{Soucek, J.}, \bibinfo{author}{Czechowski, A.},
  \bibinfo{author}{P{\'i}{\v s}a, D.}, \bibinfo{author}{Vaverka, J.},
  \bibinfo{author}{{Meyer-Vernet}, N.}, \bibinfo{author}{Maksimovic, M.},
  \bibinfo{author}{Lorf{\`e}vre, E.}, \bibinfo{author}{Issautier, K.},
  \bibinfo{author}{Babic, K.R.}, \bibinfo{author}{Bale, S.D.},
  \bibinfo{author}{Morooka, M.}, \bibinfo{author}{Vecchio, A.},
  \bibinfo{author}{Chust, T.}, \bibinfo{author}{Khotyaintsev, Y.},
  \bibinfo{author}{Krasnoselskikh, V.}, \bibinfo{author}{Kretzschmar, M.},
  \bibinfo{author}{Plettemeier, D.}, \bibinfo{author}{Steller, M.},
  \bibinfo{author}{{\v S}tver{\'a}k, {\v S}.},
  \bibinfo{author}{Tr{\'a}vn{\'i}{\v c}ek, P.}, \bibinfo{author}{Vaivads, A.},
  \bibinfo{year}{2021}.
\newblock \bibinfo{title}{First dust measurements with the {{Solar Orbiter
  Radio}} and {{Plasma Wave}} instrument}.
\newblock \bibinfo{journal}{Astronomy \& Astrophysics} \bibinfo{volume}{656},
  \bibinfo{pages}{A30}.
\newblock \DOIprefix\doi{10.1051/0004-6361/202140969}.
\bibitem[{Zaslavsky et~al.(2012)Zaslavsky, {Meyer-Vernet}, Mann, Czechowski,
  Issautier, Le~Chat, Pantellini, Goetz, Maksimovic, Bale and
  Kasper}]{Zaslavsky2012interplanetary}
\bibinfo{author}{Zaslavsky, A.}, \bibinfo{author}{{Meyer-Vernet}, N.},
  \bibinfo{author}{Mann, I.}, \bibinfo{author}{Czechowski, A.},
  \bibinfo{author}{Issautier, K.}, \bibinfo{author}{Le~Chat, G.},
  \bibinfo{author}{Pantellini, F.}, \bibinfo{author}{Goetz, K.},
  \bibinfo{author}{Maksimovic, M.}, \bibinfo{author}{Bale, S.D.},
  \bibinfo{author}{Kasper, J.C.}, \bibinfo{year}{2012}.
\newblock \bibinfo{title}{Interplanetary dust detection by radio antennas:
  {{Mass}} calibration and fluxes measured by {{STEREO}}/{{WAVES}}}.
\newblock \bibinfo{journal}{Journal of Geophysical Research: Space Physics}
  \bibinfo{volume}{117}.
\newblock \DOIprefix\doi{10.1029/2011JA017480}.
\bibitem[{Zook and Berg(1975)}]{Zook1975source}
\bibinfo{author}{Zook, H.A.}, \bibinfo{author}{Berg, O.E.},
  \bibinfo{year}{1975}.
\newblock \bibinfo{title}{A source for hyperbolic cosmic dust particles}.
\newblock \bibinfo{journal}{Planetary and Space Science} \bibinfo{volume}{23},
  \bibinfo{pages}{183--203}.
\newblock \DOIprefix\doi{10.1016/0032-0633(75)90078-1}.

\end{thebibliography}

\end{document}